\RequirePackage{fix-cm}

\documentclass[natbib]{svjour3}          
\usepackage[left=1in,top=1in, right=1in, bottom=1in]{geometry}

\smartqed  
\usepackage{graphicx}
\usepackage{ltxtable, tabularx, longtable}
\usepackage{geometry}
\usepackage{multirow}
\usepackage{rotating}
\usepackage{multicol}
\usepackage{array}
\usepackage{color}
\usepackage{mathptmx}
\usepackage{enumerate}
\usepackage{wrapfig}
\usepackage{epstopdf}
\usepackage{amsmath}
\usepackage{bigstrut}
\usepackage{lscape}
\usepackage{pifont}
\usepackage[titletoc,title]{appendix}
\newcommand{\cmark}{\ding{51}}%
\newcommand{\xmark}{\ding{55}}%
\begin{document}

\title{Automatic generation of analysis class diagrams from use case specifications
}
\author{Jitendra Singh Thakur \and Atul Gupta 
}
\institute{Jitendra Singh Thakur \at
              Indian Institute of Information Technology Design and Manufacturing Jabalpur, India.\\							
              \email{jsthakur@iiitdmj.ac.in}           
           \and
           Atul Gupta \at
             Indian Institute of Information Technology Design and Manufacturing Jabalpur, India.\\
							\email{atul@iiitdmj.ac.in}           
}
\date{July 29, 2016}
\maketitle
\begin{abstract}
In object oriented software development, the analysis modeling is concerned with the task of identifying problem level objects along with the relationships between them from software requirements. The software requirements are usually written in some natural language, and the analysis modeling is normally performed by experienced human analysts. The huge gap between the software requirements which are unstructured texts and analysis models which are usually structured UML diagrams, along with human slip-ups inevitably makes the transformation process error prone. The automation of this process can help in reducing the errors in the transformation. In this paper we propose a tool supported approach for automated transformation of use case specifications documented in English language into analysis class diagrams. The approach works in four steps. It first takes the textual specification of a use case as input, and then using a natural language parser generates type dependencies and parts of speech tags for each sentence in the specification. Then, it identifies the sentence structure of each sentence using a set of comprehensive sentence structure rules. Next, it applies a set of transformation rules on the type dependencies and parts of speech tags of the sentences to discover the problem level objects and the relationships between them. Finally, it generates and visualizes the analysis class diagram. We conducted a controlled experiment to compare the correctness, completeness and redundancy of the analysis class diagrams generated by our approach with those generated by the existing automated approaches. The results showed that the analysis class diagrams generated by our approach were more correct, more complete, and less redundant than those generated by the other approaches.
\end{abstract}
\section{Introduction}
The model driven object oriented software development process~\citep{kleppe2003mda,kuznetsov2007uml} can be visualized as a sequence of transformation steps where the first set of transformations is about constructing analysis models also called Platform Independent Models (PIM) from textual requirements. The the second set of transformations is about developing design models also called Platform Specific Models (PSM) from the analysis models. And the third set of transformations is about converting the design models into code. Understandably, an essential requirement and the underlying assumption here is that these transformations are loss-less for the requirements to be successfully transformed into a software product that will meet the user requirements. However, this is normally not the case as each of these transformations deals with a complex problem of identifying relevant model elements, their properties, and the contextual relationships between them. The complexity of the problem along with the human slip-ups~\citep{norman1981categorization} make these transformations error prone, non-repeatable and grossly influenced by the skills of these developers.

Here we focus on analysis modeling activity (also referred as object oriented analysis) which is about discovering the domain classes, attributes, operations and the relationships between the domain classes from the vocabulary of the problem domain to obtain the analysis models~\citep{booch2010object}. The Object Management Group's (OMG) Model Driven Architecture (MDA)\footnote{http://www.omg.org/mda/} refers these models as the Platform Independent Models (PIM) as they represent the functionality, behavior and structure of the system at the problem level. These models do not contain the platform specific details about the implementation of the system~\citep{kuznetsov2007uml}. They represent different viewpoints of a problem domain, so as to allow the developers to talk and reason about the domain objects in order to enhance their understanding of the problem and avoid potential pitfalls~\citep{mike2005OOAD}.

The analysis modeling is usually carried by human analysts. This activity requires the analysts to read and analyse typically several hundred pages of software requirement specifications which involves signiﬁcant efforts and time. Moreover, the chances of this transformation to be ‘lossy’ are even more prominent as it requires understanding the user requirements documented as natural language (NL) texts and mapping these textual specifications to domain models like the analysis class diagrams. Naturally, the varying skills of human analysts, their understanding of domain knowledge and of the mapping process make the transformation non-repeatable. The different analysts may derive different analysis models~\citep{harmain2003cm,deeptimahanti2009automated} which may capture the domain information incompletely and may also be inconsistent with the problem specification. The problem gets even worse when it comes to keep the domain models in sync with changing requirements.\\

Sensing the problem, there have been some efforts in the direction of framing semi-automated and automated approaches that can help to generate the analysis models from textual specifications. The semi-automated approaches assist the user in deriving the analysis models but most of them are highly dependent on the user skills for identifying the elements of the analysis models~\citep{overmyer2001conceptual,harmain2003cm,samarasinghe2005generating}. On the other hand the automated approaches~\citep{harmain2003cm,liu2004natural,ilieva2006models,popescu2008reducing,simula12} though do not require human interventions, but have some serious issues such as failing to identify the major elements of the analysis models, incomplete transformation of requirements into analysis models, identification of analysis model elements which are semantically incorrect, redundancy in the generated models and presence of many unconnected components in the models.\\

In this paper we propose an automated approach to generate the analysis class diagrams from use case specifications that overcomes many of the shortcomings of the existing approaches. The approach first parses the sentences of the input use case specification (UCS) using the Stanford NL parser\footnote{http://nlp.stanford.edu/software/lex-parser.shtml} to generate parts of speech tags (POS-tags) and type dependencies (TDs) (POS-tags represent the tagging of each word in a sentence with parts of speech such as noun, pronoun, verb, adjective, adverb etc. TDs~\citep{de2006generating, de2008stanford} represent grammatical relationships between the words of a sentence. POS tags and TDs are discussed in more detail in Section~\ref{sec:NLPModels}). Then it applies the proposed set of comprehensive sentence structure rules on TDs and POS-tags of the sentences to identify their sentence structures. We have framed these comprehensive sentence structure rules based on the twenty five verb patterns proposed by A.S.Hornby (known for various achievements in linguistic and literature)~\citep{hanks2008lexical} in Oxford Advanced Learner’s Dictionary of Current English~\citep{OALD1974english,OALD2000english} and in~\cite{hornby1975english}. After identifying the sentence structures, the approach then applies the proposed set of comprehensive transformation rules on TDs and POS-tags to identify the potential elements for the generation of analysis class diagram. The transformation rules take into account the sentence structure of the sentences as well as the syntactic and semantic relationships between the words of the sentences to precisely identify these elements. The approach finally generates the analysis class diagram and using the GraphViz API\footnote{http://www.graphviz.org/, https://github.com/jabbalaci/graphviz-java-api} visualizes the generated diagram. We have prototyped the proposed approach in a GUI based tool named \textsl{AutoAMG (Automatic Analysis Model Generator)}.

As a validation of the proposed approach, we report the outcome of a controlled experiment conducted to compare the analysis class diagrams generated by our approach with those generated by the other two existing approaches - one proposed by~\cite{popescu2008reducing} and the other by~\cite{simula12,yue2015atoucan}. The approaches were compared on the basis of correctness, completeness and redundancy of the analysis class diagrams generated by them for the forty UCSs taken from various software engineering books~\citep{mike2005OOAD,gomaa2011software,doug2008use,rosenberg2007use,bruegge1999object} and research works~\citep{liu2004natural,popescu2008reducing,simula12,yue2015atoucan,deeptimahanti2011semi}.  The experiment design was a complete block design in which forty subjects evaluated the analysis class diagrams obtained from the three approaches. Each subject was randomly given one UCS, the analysis class diagrams obtained by the three approaches for the given UCS and a set of questionnaires for each class diagram. The questionnaires were based on the quality measures for analysis class diagrams presented in Section~\ref{subsec:Metrics}. For statistical analysis of the data obtained from answers to the questionnaires collected from the subjects, we first applied \textsl{Kolmogorov-Smirnov} test~\citep{massey1951kolmogorov} and found that the data was non-normal. Then we applied the \textsl{Friedman} test (a non parametric test)~\citep{corder2009nonparametric,sheskin2003handbook} on the data that showed significant differences between the analysis class diagrams generated using the three approach. Next we applied post hoc \textsl{Friedman} tests on the data for pairwise comparison of the analysis class diagrams generated by the three approaches to find whether the class diagrams obtained from one of the approach are better in terms of correctness completeness and redundancy than those obtained from the other approaches. The results clearly showed that the analysis class diagrams generated by the proposed approach were significantly better than those generated by other two existing approaches in terms of the correctness, completeness and redundancy. Specifically, the results for the forty UCSs showed that the analysis class diagrams generated by our approach were 46\% more correct, 55\% more complete and 31\% less redundant than those generated by Popescu et al. approach, and were 33\% more correct, 31\% more complete and 31\%  less redundant than those generated by Yue et al. approach. \\
\\
\noindent The main contributions of this paper are as follows:
\begin{enumerate}
	\item We propose an automated approach to generate analysis class diagrams from UCSs.
	\item We framed a comprehensive set of sentence structure rules to identify the sentence structures of the sentences, and a comprehensive set of transformation rules to identify the elements of the analysis class diagrams.
	\item We developed a tool support \textsl{AutoAMG} for the above stated automation.	
	\item We report on a controlled experiment that we have conducted with 40 subjects and 40 UCSs to compare the correctness, completeness and redundancy of the analysis class diagrams generated by the proposed approach with those generated by other two exiting automated approaches.	
\end{enumerate}
The rest of the paper is organized as follows. Section~\ref{sec:Background} presents the background of the proposed work. Section~\ref{sec:MotivationProblemDescription} presents the motivation and the problem description. Section~\ref{sec:ProposedApproach} presents the proposed approach. Section~\ref{sec:ToolSupport} presents the tool support that we have developed for the automation. Section~\ref{sec:ExperimentalStudy} presents the details of the experiment conducted for the evaluation of the proposed approach. Section~\ref{sec:Discussion} discusses the strengths and limitations of the approach. Section~\ref{sec:RelatedWork} presents the related work in the literature. Section~\ref{sec:Conclusion} presents the conclusion and future directions of the present work.
\vspace{-2ex}
\section{Background}
\label{sec:Background}
The earliest attempt for analysis of textual requirements to identify classes and objects was made by~\cite{abbott1983program}, in which he proposed extracting the classes and their operations using noun-verb analysis of textual description of the problem. The Nouns found in the text represent candidate classes and the verbs represent the candidate operations. The idea of simple noun-verb textual analysis was useful but suffers from dilemma of enlisting incorrect candidate objects and operations as any noun can be verbed, and any verb can be nouned~\citep{booch2010object}.

Another technique to identify the potential classes proposed by~\cite{arango1989domain}, in which an analyst consults domain experts and identifies the candidate objects, operations, and relationships from the vocabulary that the domain experts use while speaking about the given problem. A domain expert may be a user of the system or any person who is familiar with all the details of the given system.

\cite{shlaer1988object} and~\cite{yourdon1990object} suggested other ways of identifying the prospective classes and objects from sources such as perceptible things (ATM card, customer record, sensors etc), roles (roles played by the people who interact with the system), events (an event that must be remembered), sentence structures (``is a" and ``part of" relationships), other systems (other system with which this system interacts), devices (with which the system interacts) and organizational units (companies, departments, divisions etc. that the system must keep tract of) discussed in the problem description.

\cite{beck1989laboratory} suggested an idea of using simple cards that they named CRC (Class Responsibility Collaborator) cards for teaching the concepts of objects to novice programmers, and for introducing complicated existing designs to experienced programmers. CRC cards became an effective method for analyzing use case scenarios by a team of analysts to obtain analysis models.\\

The advancements in the field of natural language processing motivated researchers to automate the process of analysis modeling. \cite{mich1996nl},~\cite{mich2002nl} proposed a CASE tool \textsl{NL-OOPS (Natural Language - Object Oriented Production System)} for identifying classes and relationships using a semantic network of words of a natural language processing system \textsl{LOLITA }\textsl{(Large-scale Object-based Language Interactor, Translator and Analyser)}. The tool requires user intervention for deleting the extra nodes representing spurious classes, and to set the level of details for class hierarchy.

\cite{overmyer2001conceptual} proposed another semi automated approach supported by a tool named Linguistic Assistant for Domain Analysis (LIDA) that uses POS tags generated by NL parser to present the user with list of nouns, adjectives and verbs from which user can identify and mark the candidate classes, attributes and the methods. A Similar approach is proposed by~\cite{harmain2003cm} named CM-Builder1, but there approach also assigns frequency of reference in the text to each class which helps the user to identify the candidate classes.\\

Further efforts have been made to remove the human intervention and to fully automate the transformation process. \cite{liu2004natural} proposed an automatic approach named UCDA for generating analysis class diagrams using POS tags generated by a NL parser from UCS written using restricted grammar. Their approach processes the sentences based on the classification of sentences as transitive, intransitive, ditransitive, intensive, complex transitive, prepositional and non-finite given in~\cite{roberts1956patterns}. The approach was dependent on using a glossary for identifying classes, it fails to identify attributes, relationship names, aggregation relationships and generalization relationships. \cite{ilieva2006models} proposed an approach to generate domain models from unrestricted NL requirements. The approach was based on using a semantic network of words created from the POS-tags generated by parser. The semantic network is then transformed into the domain model. The domain model confine only the identified classes and the relationship between them, it lacks the attributes, operations, relationship names and relationship types (association, generalization and aggregation).

\cite{popescu2008reducing} proposed a different approach whose main objective was to identify inconsistencies in the requirement specifications but with the help of automatically generated domain models. The approach processes the requirement specifications using the constraining grammar proposed in~\cite{juristo2000use}.The approach uses link types generated by link grammar parser to extract the domain elements from such sentences in the requirement specification. The generated domain models can then be examined by the user to identify the inconsistencies. The analysis models obtained are highly incomplete with many unconnected components.

Recently,~\cite{simula12,yue2015atoucan} proposed an another automated approach for deriving analysis models from UCSs using parse trees and POS tags generated by a NL parser. The approach requires the input UCSs to be written using twenty six restrictions rules proposed in their previous work. Their approach processes the sentences on the basis of sentence structures formed using five basic English sentence patterns given in~\cite{greenbaum1996oxford}. The approach does not recognize domain objects and attributes which are documented as a group of words (nouns). The generated class diagrams have many unconnected components. The approach dumps most of the operations in a single control class, whereas most of the other classes are assigned no operations at all.
\vspace{-2ex}
\section{Motivation and problem description}
\label{sec:MotivationProblemDescription}
The analysis modeling is usually performed by human analysts in which the human analysts read the textual requirements to identify the elements for the generation of analysis class diagram. The analysis class diagram consists of the identified domain classes (entity, boundary and control) their attributes and operations, and the relationships (association, generalization and aggregation) between the domain classes. The huge gap between the software requirements and analysis models, along with human slip-ups~\citep{norman1981categorization} inevitably make the transformation of the requirements into analysis models error prone. Because of the varying skills of human analysts, their understanding of domain knowledge and of the mapping process, the different analysts may derive different analysis models from the same set of requirements~\citep{harmain2003cm,deeptimahanti2009automated}, which makes the transformation non-repeatable. The problem gets even worse when the domain models are to be kept in sync with the changing requirements. The automation of this process can help in reducing these problems.

As there are many possible ways in which a same concept or thing can be expressed in a natural language such as English, the relevant elements to construct the analysis model are embedded at different positions in different sentences of the requirement specifications. Hence to extract the relevant elements from the text, the two foremost requirement for an automated approach are 1) It should be able to recognize the structures of the sentences in the specification, 2) It should be able to accurately obtain the syntactic and semantic relationships between the words in the sentences. An ideal automated approach should be able to interpret all the possible sentences of the natural language without any restrictions. The set of sentence patterns that an automated approach can interpret is referred in the paper as language model. The stronger the language model utilized by an approach makes the approach more useful and more appealing. But, the issues associated with the natural language such as ambiguity, variety of sentence types, anaphora (or pronoun) resolution problem and inconsistency~\citep{kamsties2000taming,nuseibeh2001making,fabbrini2001linguistic,yang2010extending} makes the interpretation of all the possible sentences hard for an automated approach. Therefore to overcome the problems associated with natural languages, the automated approaches impose restrictions on free usage of natural language to specify the requirements, but the requirements analysts (or use case authors) don't want too much restrictions (too much restrictions prevent them in expressing their ideas). \\

As discussed in the last section there have been a few efforts in the direction of framing semi-automated and automated approaches to derive the analysis class diagrams from textual specifications. The major drawback of the semi automated approaches~\citep{mich1996nl,mich2002nl,overmyer2001conceptual,harmain2003cm} was their high dependency on the user to identify the elements for generating analysis class diagrams. The automated approaches~\citep{liu2004natural,ilieva2006models} though do not require any human intervention but they have some serious issues such as the approaches do not identify the major elements of the analysis class diagrams (Table~\ref{tab:ClassDiagramElementsExistingAppraoches}) and require glossary files (files containing list of domain specific terms) to identify the classes. The automated approaches proposed by~\cite{popescu2008reducing} and~\cite{simula12,yue2015atoucan} identify the major elements for generating analysis class diagrams (Table~\ref{tab:ClassDiagramElementsExistingAppraoches}) without using any glossary file to identify the domain classes but they too have some serious issues which are listed in Table~\ref{tab:IssuesExistingApproaches} along with the possible causes.

\begin{table}[htb]
\vspace{-1ex}
  \centering
	\tiny
  \caption{Analysis class diagram elements identified by the existing automated approaches}
    \begin{tabular}{lp{1cm}lcccccccc}
    \hline
    \multicolumn{1}{l}{\multirow{3}[6]{*}{\textbf{Approach}}} & \multicolumn{1}{p{1cm}}{\multirow{3}[6]{1cm}{\textbf{NLP constructs used}}} & \multicolumn{1}{c}{\multirow{3}[6]{*}{\textbf{Input}}} & \multicolumn{8}{c}{\textbf{Class Diagram Elements Identified (Output) }} \bigstrut\\
\cline{4-11}    \multicolumn{1}{c}{} & \multicolumn{1}{c}{} & \multicolumn{1}{c}{} & \multicolumn{4}{c}{\textbf{Classes }} & \multicolumn{4}{c}{\textbf{ Relationships }} \bigstrut\\
\cline{4-6} \cline{8-11}   \multicolumn{1}{c}{} & \multicolumn{1}{c}{} & \multicolumn{1}{c}{} & \textbf{Class Name } & \textbf{Attributes } & \textbf{Operations } & & \textbf{Relationship Name } & \textbf{Association} & \textbf{Aggregation } & \textbf{Generalization } \bigstrut\\
    \hline
   \bigstrut
    \cite{liu2004natural}  & POS tags  & Restricted NL Text  & \cmark     & \xmark     & \cmark     & & \xmark     & \cmark     & \xmark     & \xmark  \\ \bigstrut
    \cite{ilieva2006models} & POS tags  & Unrestricted NL Text  & \cmark     & \xmark     & \xmark     & & \xmark     & \cmark     & \xmark     & \xmark  \\ \bigstrut
    \cite{popescu2008reducing} & Link types & Restricted NL Text  & \cmark     & \cmark     & \cmark     & & \cmark     & \cmark     & \cmark     & \cmark  \\ \bigstrut
    \cite{simula12,yue2015atoucan} & Parse tree & Restricted NL Text  & \cmark     & \cmark     & \cmark     & & \cmark     & \cmark     & \xmark     & \cmark~~* \\
    \hline
		\multicolumn{11}{l}{* can not identify generalization relationship from text but obtains generalization relationship between control classes of UCSs directly from generalization field of the UCS}
    \end{tabular}%
  \label{tab:ClassDiagramElementsExistingAppraoches}%
\vspace{-1ex}
\end{table}%

\begin{table}[htb]
  \centering
	\scriptsize
  \caption{Issues in existing automated approaches}
    \begin{tabular}{p{1cm}p{6cm}p{8cm}}
  \hline
    \multicolumn{1}{p{1cm}}{\textbf{Approach}} & \multicolumn{1}{l}{\textbf{Issues}} & \multicolumn{1}{p{8cm}}{\textbf{Possible causes of the Issues}} \bigstrut\\
    \hline
    \multicolumn{1}{p{1cm}}{\multirow{4}[2]{1cm}{A1~\cite{popescu2008reducing}}} & \multicolumn{1}{p{6cm}}{1) The analysis class diagrams generated by the approach contains many isolated classes and unconnected components (chuncks of classes and relationships).} & \multicolumn{1}{p{8cm}}{\multirow{3}[1]{8cm}{1) The language model (set of sentence patterns used by the approach to interpret the sentences) which is based on the five sentence patterns proposed in~\cite{juristo2000use}, is unable to interpret the variety of simple sentences in English like those involving infinitive (e.g. ``The system prompts to enter the password", ``The system commands the motor to start"), present participle (e.g. ``The system prints the receipt showing transaction number and date"), past participle (e.g. ``The system validates the record entered by the customer"), gerund (e.g. ``The system starts printing the document"), etc. }} \bigstrut[t]\\
    \multicolumn{1}{l}{} & \multicolumn{1}{p{6cm}}{2) The class diagrams lacks several relevant classes and relationships (domain classes and relationships which should be be deduced from the problem specifications) } & \multicolumn{1}{p{8cm}}{} \\
    \multicolumn{1}{l}{} & \multicolumn{1}{p{6cm}}{3) A number of identified domain classes are difficult to be semantically termed as domain classes as they seems to violate the encapsulation principle.}
		& \multicolumn{1}{p{8cm}}{} \\
    \multicolumn{1}{l}{} & \multicolumn{1}{p{6cm}}{4) The class diagrams contain some redundant classes and relationships.} & \multicolumn{1}{p{8cm}}{2) Inability of the transformation rules to extract all the relevant elements from the sentences.} \bigstrut[b]\\
    \hline
    \multicolumn{1}{p{1cm}}{\multirow{7}[2]{1cm}{A2~\cite{simula12,yue2015atoucan}}} & \multicolumn{1}{p{6cm}}{1) The analysis class diagrams generated by the approach contains some isolated classes and unconnected components (chuncks of classes and relationships).} & \multicolumn{1}{p{8cm}}{\multirow{4}[1]{8cm}{1) The approach uses parse trees to identify class diagram elements. As a parse tree represents only the syntactic structure of a sentence (it does not represent the semantic relationships between the words of the sentence), the approach is unable to disambiguate the extraction of relevant elements from the text.
}} \bigstrut[t]\\
    \multicolumn{1}{l}{} & \multicolumn{1}{p{6cm}}{2) Same as Issue 2 of approach A1} & \multicolumn{1}{p{8cm}}{} \\
    \multicolumn{1}{l}{} & \multicolumn{1}{p{6cm}}{3) Same as Issue 3 of approach A1} & \multicolumn{1}{p{8cm}}{\multirow{2}[0]{8cm}{2) The language model (set of sentence patterns used by the approach to interpret the sentences) is not comprehensive enough to interpret common complex sentences like those involving that clause (e.g. ``The system check that the password is correct"), conjunctive clause (e.g. ``The system stops the motor when the tank is full", ``The motor stops when the tank is full.") etc.}} \\
    \multicolumn{1}{l}{} & \multicolumn{1}{p{6cm}}{4) Same as Issue 4 of approach A1} & \multicolumn{1}{p{8cm}}{} \\
    \multicolumn{1}{l}{} & \multicolumn{1}{p{6cm}}{5) The approach fails to recognize domain classes which are documented as a group of words (or nouns) in the specifications.} & \multicolumn{1}{p{8cm}}{} \\
    \multicolumn{1}{l}{} &   \multicolumn{1}{p{6cm}}{6) The approach assigns most of the operations in a single control class, whereas most of the other classes are assigned no operations at all. Such classes are called smells (defective classes) in~\cite{arendt2010uml}.} &  \multicolumn{1}{p{8cm}}{3) The transformation rules identify the entity classes only from those noun phrases in the sentences that either contains a single noun (e.g. ``customer"), or a single noun with a determiner (e.g. ``The customer") , or a possessive noun (e.g. ``customer's address") (Rule B1.1 and B1.2~\cite{simula12,yue2015atoucan}).} \\
    \multicolumn{1}{l}{} &  \multicolumn{1}{p{6cm}}{7) The approach is unable to identify aggregation relationships. The identification of generalization relationships require human intervention~\citep{simula12,yue2015atoucan}.} &  \multicolumn{1}{p{8cm}}{4) No transformation rules to extract aggregation and generalization relationships from the sentences in UCS.} \bigstrut[b]\\
    \hline
		\end{tabular}%
  \label{tab:IssuesExistingApproaches}%
	\vspace{-2ex}
\end{table}%

In this paper we aim to propose a tool supported automated approach that can generate analysis class diagrams which are more correct, more complete and less redundant than those generated by existing automated approaches.
The proposed approach takes as input the software requirements documented as use case specifications (UCSs) with a few restrictions shown in Table~\ref{tab:RestrictionRules}. To handle the diverse set of sentences in the specifications the approach systematically process the sentences on the basis of their sentence structures identified using the proposed comprehensive sentence structure rules. To correctly extract potential classes, their attributes, operations and the relationships, the approach uses the proposed set of comprehensive transformation rules. We have framed these transformation rules to precisely extract the relevant elements of analysis class diagram from the text through various drills on hundreds of sentences of each sentence pattern. For extracting the relevant elements from the text, the transformation rules take into account the sentence structure of the sentences as well as the syntactic and semantic relationships between the words of the sentences. For finding these relationships between the words of the sentences, the transformation rules use type dependencies (TDs) and part of speech tags (POS-tags) generated using Stanford Natural Language Parser API\footnote{http://nlp.stanford.edu/software/lex-parser.shtml}. The approach finally visualizes the generated analysis class diagram in place with the help of GraphViz API.

We conducted a controlled experiment to compare the analysis class diagrams generated by the proposed approach with those generated by two existing automated approaches~\citep{popescu2008reducing} and~\citep{simula12,yue2015atoucan} for the forty UCSs obtained from various software engineering books and research papers. In the experimental forty subjects evaluated the analysis class diagrams generated by the three approaches for the forty UCSs in terms of correctness, completeness and redundancy. The results showed that the analysis class diagrams generated by our approach were significantly more correct, more complete, and less redundant than those generated by the other approaches.
\vspace{-2ex}
\section{Proposed approach}
\label{sec:ProposedApproach}
This section describes how the proposed approach obtains the analysis class diagrams from the input use case specifications. Section~\ref{sec:BuildingBlocks} describes the artifacts which are the building blocks of the proposed approach. Then Section~\ref{sec:WorkingOfProposedApproach} step by step describes how the input UCS is read and parsed, and how the various elements of analysis class diagram are identified from the UCS.
\vspace{-2ex}
\subsection{Building blocks of the proposed approach}
\label{sec:BuildingBlocks}
This section presents the artifacts that are the backbone of the proposed approach. Section~\ref{sec:UCSTemplate} presents the UCS template used as input in the approach. Section~\ref{sec:MetaModels} presents the use case meta model used to store the elements of the input UCS, and class diagram meta model used to store the class diagram elements identified by the approach. Section~\ref{sec:NLPModels} presents the Natural Language Processing (NLP) constructs (POS-tags and TDs) used by the approach to analyse the textual specifications, and to extract the class diagram elements from the text. Section~\ref{sec:LanguageModel} presents the language model used by the approach, and the restriction rules to be used for documenting the UCS.

\subsubsection{Use case model}
\label{sec:UCSTemplate}
\begin{figure}[htbp]
\vspace{-1ex}
\centering
\includegraphics[width=2in]{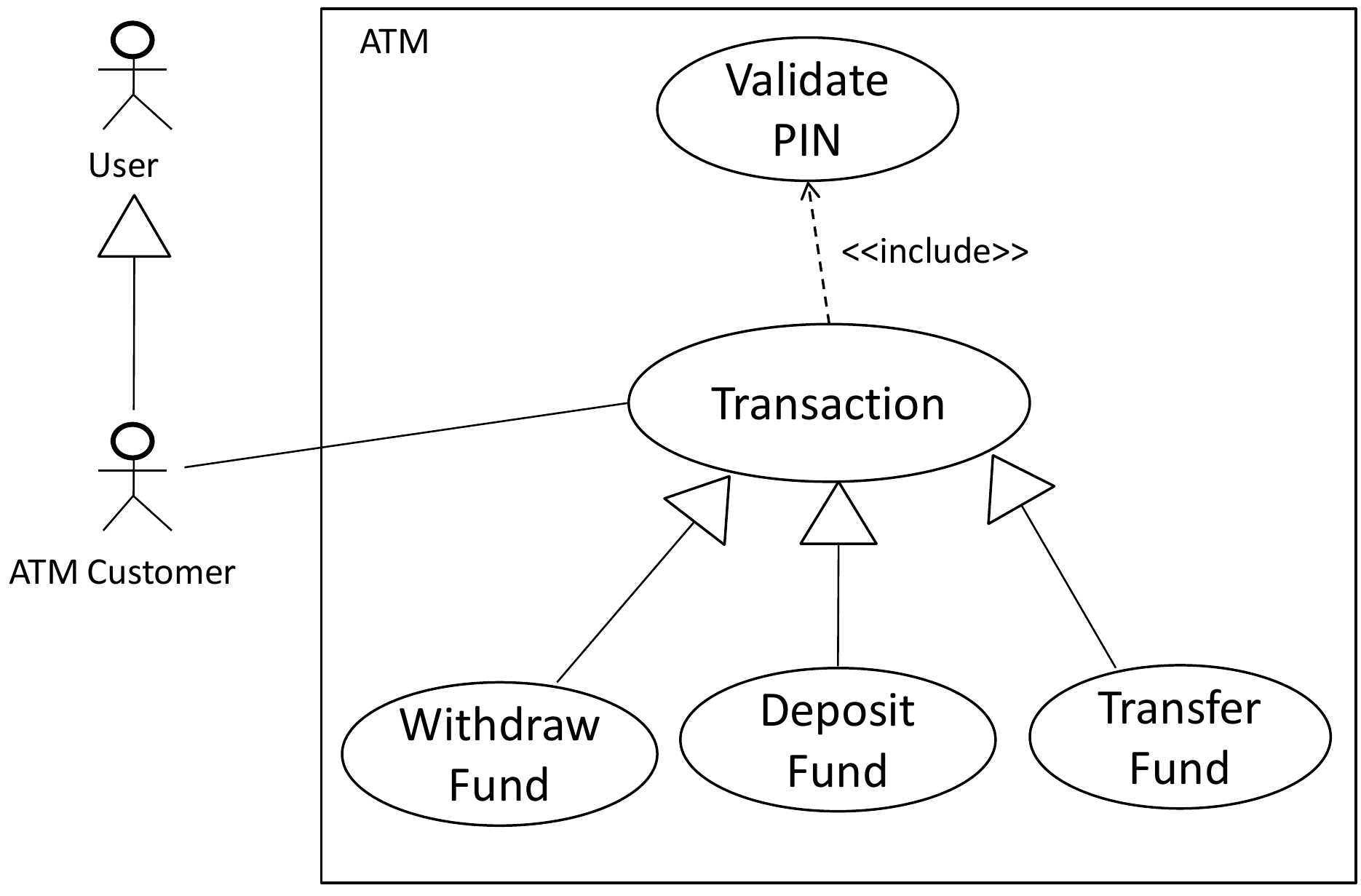}
\caption{Partial use cases of ATM system}
\label{fig:UseCaseATM}
\vspace{-1ex}
\end{figure}

Use case diagrams (Example Figure~\ref{fig:UseCaseATM}) along with their textual specifications (Example Table~\ref{tab:UseCaseWithdrawFund}) are commonly used for documenting the functional requirements of a system. A use case specification (UCS) defines the functional requirements of a system in terms of a sequence of actions preformed by the system and the actor(s) of the system that provides the desired function for the actor(s).

Our approach takes as input the functional requirements documented as UCS (Example Table~\ref{tab:UseCaseWithdrawFund}). The presented UCS template contains the most common elements (name, description, actor, basic flow and alternate flow)~\citep{siqueira2011essential} that are used for documenting the UCS. A part from these essential elements the template also contains a few more elements viz. parent actor name to specify actor generalization relationships which are shown in a use case diagram, parent use case name to specify use case generalization relationships which are shown in a use case diagram, and sub flow to specify the steps that can be performed in parallel with some steps in main flow (basic flow). Each flow step in our use case template can contain a pre-condition and a post-condition. The element alternate flow is divided into two sub elements viz. specific alternate flow and global alternate flow as done in~\cite{yue2013facilitating} template, where the specific alternate flows specifies the steps which are alternatives to some specific step in the UCS, and the global alternate flow specifies the steps whose scope is all the other steps in the UCS (e.g. If a user press cancel or select cancel, when some other step is being carried out in the UCS.)

\begin{table}[htbp]
\vspace{-1ex}
  \centering
	\scriptsize
  \caption{Use case specification (UCS): Withdraw fund taken from~\cite{simula12,yue2015atoucan}, but a few modifications done to fit our template)}
    \begin{tabular}{llllll}
    \hline
    \multicolumn{1}{p{.9in}}{\textbf{Use Case Name :-}} & \multicolumn{2}{p{2.2in}}{Withdraw Fund} & \multicolumn{1}{p{.35in}}{\textbf{}} & \multicolumn{1}{p{.35in}}{} & \multicolumn{1}{p{.9in}}{} \bigstrut \\
		\hline
    \multicolumn{1}{p{.9in}}{\textbf{Main System Name :-}} & \multicolumn{2}{l}{ATM} & \multicolumn{2}{p{.65in}}{\textbf{Parent Use Case Name :-}} & \multicolumn{1}{l}{Transaction} \\
\hline
    \multicolumn{1}{p{.9in}}{\textbf{Description :-}} & \multicolumn{5}{p{4.9in}}{Represents steps to withdraw cash from ATM.} \\
  \hline
    \multicolumn{1}{p{.9in}}{\textbf{Constraints :-}} & \multicolumn{5}{l}{The system must process the transaction within 20 seconds} \\
    \hline
    \multicolumn{6}{c}{\textbf{Actors}} \\
    \hline
    \multicolumn{1}{r}{\textbf{Actor No.}} & \textbf{Actor Type} & \textbf{Actor Name } & \multicolumn{3}{l}{\textbf{Parent Actor Name}} \\
    \hline
    \multicolumn{1}{r}{\textbf{1}} & Primary & ATM Customer & \multicolumn{3}{l}{User} \\

    \multicolumn{1}{r}{\textbf{2}} & Secondary &        & \multicolumn{3}{c}{} \\
   \hline
  \multicolumn{6}{c}{\textbf{Main Flow Steps}} \\ \hline
    \multicolumn{1}{p{.9in}}{\textbf{Pre Condition/Guard}} & \multicolumn{1}{p{.35in}}{\textbf{FlowId}} &  \multicolumn{1}{p{2in}}{\textbf{Step}} & \multicolumn{1}{p{.35in}}{\textbf{Sub FlowId}} & \multicolumn{1}{p{.35in}}{\textbf{Alternate FlowId}} & \multicolumn{1}{p{1in}}{\textbf{Post Condition}} \\ \hline
    \multicolumn{1}{p{.9in}}{ATM customer has inserted an ATM card in the card reader} & \multicolumn{1}{p{.35in}}{M1} &  \multicolumn{1}{p{2in}}{INCLUDE USE CASE Validate PIN.} & \multicolumn{1}{p{.35in}}{} & \multicolumn{1}{l}{} & \multicolumn{1}{l}{} \\
    \multicolumn{1}{p{.9in}}{} & \multicolumn{1}{p{.35in}}{M2} &  \multicolumn{1}{p{2.1in}}{ATM customer selects Withdrawal.} & \multicolumn{1}{l}{} & \multicolumn{1}{l}{} & \multicolumn{1}{l}{} \\
    \multicolumn{1}{p{.9in}}{} & \multicolumn{1}{p{.35in}}{M3} &  \multicolumn{1}{p{2.1in}}{ATM customer enters the withdrawal amount.} & \multicolumn{1}{l}{} & \multicolumn{1}{l}{} & \multicolumn{1}{l}{} \\
    \multicolumn{1}{p{.9in}}{} & \multicolumn{1}{p{.35in}}{M4} &  \multicolumn{1}{p{2.1in}}{ATM customer selects account number.} & \multicolumn{1}{l}{} & \multicolumn{1}{l}{} & \multicolumn{1}{l}{} \\
    \multicolumn{1}{p{.9in}}{} & \multicolumn{1}{p{.35in}}{M5} &  \multicolumn{1}{p{2.1in}}{The system validates that the account number is valid.} & \multicolumn{1}{l}{} & \multicolumn{1}{l}{A1} & \multicolumn{1}{l}{} \\
    \multicolumn{1}{p{.9in}}{} & \multicolumn{1}{p{.35in}}{M6} &  \multicolumn{1}{p{2.1in}}{The system validates that ATM customer has enough funds in the account.} & \multicolumn{1}{l}{} & \multicolumn{1}{l}{A1} & \multicolumn{1}{l}{} \\
    \multicolumn{1}{p{.9in}}{} & \multicolumn{1}{p{.35in}}{M7} &  \multicolumn{1}{p{2.1in}}{The system validates that the withdrawal amount does not exceed the daily limit of the account.} & \multicolumn{1}{l}{} & \multicolumn{1}{l}{A1} & \multicolumn{1}{l}{} \\
    \multicolumn{1}{p{.9in}}{} & \multicolumn{1}{p{.35in}}{M8} &  \multicolumn{1}{p{2.1in}}{The system validates that the ATM has enough funds.} & \multicolumn{1}{l}{} & \multicolumn{1}{l}{A1-A2} & \multicolumn{1}{l}{} \\
    \multicolumn{1}{p{.9in}}{} & \multicolumn{1}{p{.35in}}{M9} &  \multicolumn{1}{p{2.1in}}{The system dispenses the cash amount.} & \multicolumn{1}{l}{} & \multicolumn{1}{l}{} & \multicolumn{1}{l}{} \\
    \multicolumn{1}{p{.9in}}{} & \multicolumn{1}{p{.35in}}{M10} &  \multicolumn{1}{p{2.1in}}{The system prints a receipt showing transaction number, transaction type, amount withdrawn, and account balance.} & \multicolumn{1}{l}{S1-S2} & \multicolumn{1}{l}{} & \multicolumn{1}{p{1in}}{ATM customer funds have been withdrawn.} \\ \hline
    \multicolumn{6}{c}{\textbf{Sub Flow Steps}} \\ \hline
    \multicolumn{1}{l}{} & \multicolumn{1}{p{.35in}}{S1} &  \multicolumn{1}{p{2in}}{The system ejects the ATM card.} & \multicolumn{1}{l}{} & \multicolumn{1}{l}{} & \multicolumn{1}{l}{} \\
		           & S2     & The system displays the Welcome message. &        &        &  \\
    \hline
    \multicolumn{6}{c}{\textbf{Specific Alternate Flow Steps}} \\ \hline
            & A1     & The system displays an apology message. & S1     &        & \multicolumn{1}{p{1in}}{ATM customer funds have not been withdrawn.} \\

           & A2     & The system shuts down. &        &        & \multicolumn{1}{p{1in}}{The system is shut down.} \\
    \hline
    \multicolumn{6}{c}{\textbf{Global Alternate Flow Steps}} \\ \hline
        \multicolumn{1}{p{.9in}}{IF ATM customer enters Cancel}  & GA1    & The system cancels the transaction. & S1-S2  &        & \multicolumn{1}{p{1in}}{ATM customer funds have not been withdrawn.} \bigstrut[t]\\
		\hline
    \end{tabular}%
 \label{tab:UseCaseWithdrawFund}
\vspace{-2ex}
\end{table}%

\subsubsection{Metamodels}
\label{sec:MetaModels}
A metamodel describes how various elements in a model are arranged, related and constrained~\citep{bezivin2006model}. Metamodels allow the model transformation tools to effectively apply the transformation operations on the models~\citep{siqueira2011essential} (A model is an instance of a metamodel).

\begin{figure}[htb]
\centering
\includegraphics[width=\linewidth]{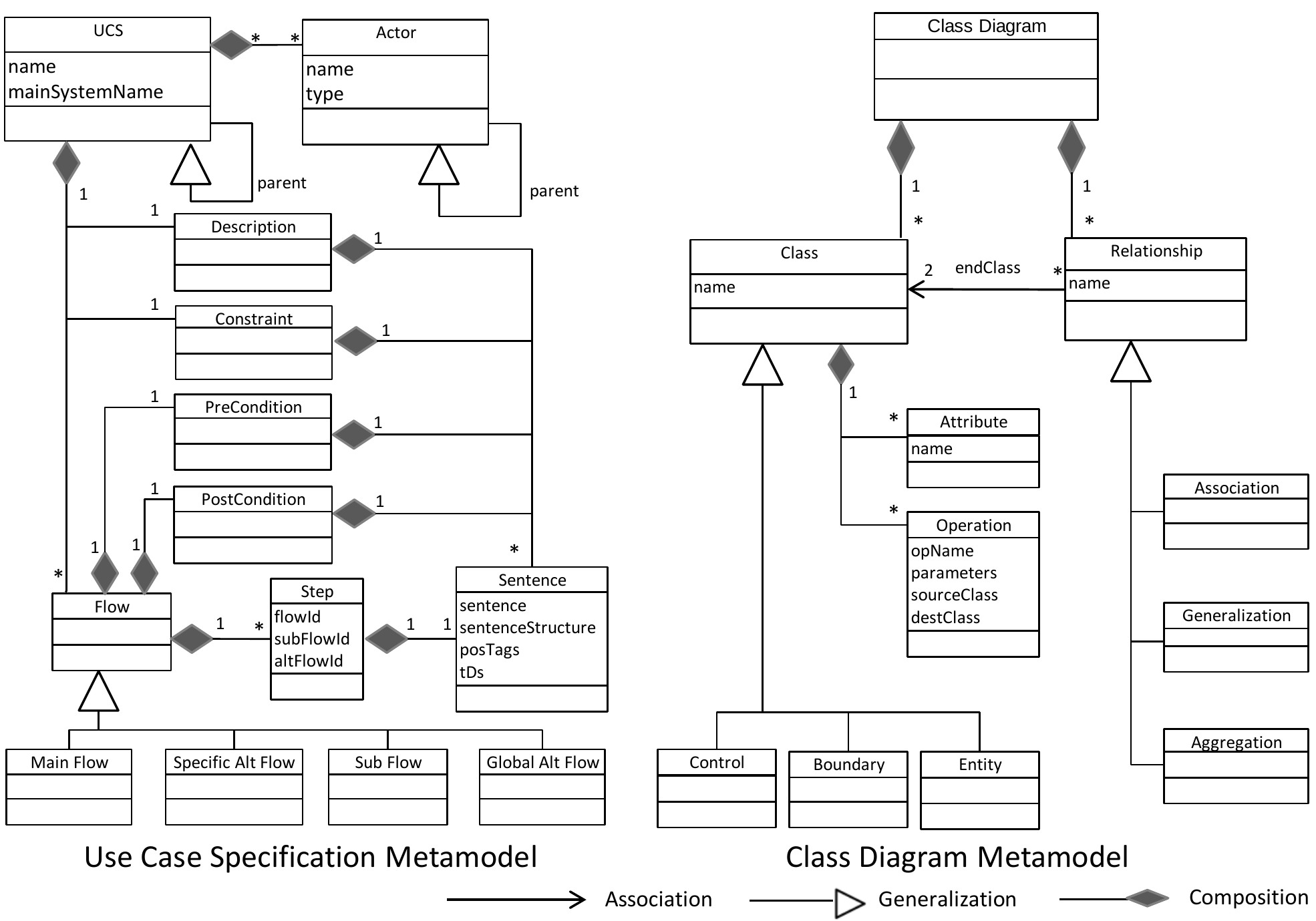}
\caption{Metamodels used in the approach}
\label{fig:UCSandCDMetaModel}
\end{figure}

Our approach stores the elements of the input UCS in an instance of UCS metamodel, and the elements of output class diagram in an instance of class diagram metamodel (Figure~\ref{fig:UCSandCDMetaModel}). As our approach generates analysis models (analysis class diagrams) which are platform independent models so the class diagram metamodel used used in our approach is different from those in the literature (including OMG's metamodel) which contain some platform specific details required for creating design models.

\subsubsection{NLP models}
\label{sec:NLPModels}
The approach generates two NLP constructs i) Parts of speech tags (POS-tags) and ii) Type dependencies (TDs) from the sentences in the UCS using the Stanford NL parser API. The approach then uses these NLP constructs first to identify the sentence structure of the sentences, then to extract the class diagram elements from the text.\\
\\
i) \textbf{Parts of speech tags (POS-tags)}: The POS-tags\footnote{http://nlp.stanford.edu/software/tagger.shtml} refer to the words in a sentence tagged (or annotated) with parts of speech, such as noun, pronoun, verb, adjective, adverb, etc. When a sentence is given as input to the parser, it analyses the sentence, and assigns each word in the sentence with a POS-tag taken from a set of 36 such POS-tags\footnote{http://www.comp.leeds.ac.uk/amalgam/tagsets/upenn.html}. The output sentence generated by the parser is same as the input sentence, but each word in the output sentence is now tagged with a POS-tag.
\begin{example}
\label{ex:POSTags}
If a sentence: \textit{``The bank sends the customer an sms."} is given as input to the parser then the POS tagged sentence generated by the parser as output is:\\
\\
\textit{``[The\textcolor[rgb]{0.65,0.16,0}{/}\textcolor[rgb]{0,0,1}{DT}, bank\textcolor[rgb]{0.65,0.16,0}{/}\textcolor[rgb]{0,0,1}{NNP}, sends\textcolor[rgb]{0.65,0.16,0}{/}\textcolor[rgb]{0,0,1}{VBZ}, the\textcolor[rgb]{0.65,0.16,0}{/}\textcolor[rgb]{0,0,1}{DT}, customer\textcolor[rgb]{0.65,0.16,0}{/}\textcolor[rgb]{0,0,1}{NN}, an\textcolor[rgb]{0.65,0.16,0}{/}\textcolor[rgb]{0,0,1}{DT}, sms\textcolor[rgb]{0.65,0.16,0}{/}\textcolor[rgb]{0,0,1}{NN}, .\textcolor[rgb]{0.65,0.16,0}{/}\textcolor[rgb]{0,0,1}{.}]"}\\
\\
Where, the tag \textcolor[rgb]{0.65,0.16,0}{/}\textcolor[rgb]{0,0,1}{DT} after a word denote that the word is determiner, \textcolor[rgb]{0.65,0.16,0}{/}\textcolor[rgb]{0,0,1}{NNP} denote that the word is a singular proper noun, \textcolor[rgb]{0.65,0.16,0}{/}\textcolor[rgb]{0,0,1}{NN} denote that the word is a singular noun, and \textcolor[rgb]{0.65,0.16,0}{/}\textcolor[rgb]{0,0,1}{VBZ} denote that the word is a 3rd person singular present tense verb.\\
\end{example}
\noindent ii) \textbf{Type dependencies (TDs)}: A type dependency (TD) represents grammatical dependency relationship (bi-lexical asymmetrical relationship) between the words of a sentence. The TDs can be used to obtain the semantic relationships between the words in a sentence~\citep{de2008stanford}, so they are used in our approach to disambiguate the extraction of relevant elements from the sentences. Our approach uses Stanford Parser APIs version 2.0.4 to generate TDs from the sentences. The Stanford TDs~\citep{de2006generating, de2008stanford} are based on Lexical-Functional Grammar~\citep{bresnan2001lexical}. The dependency relationships and the naming schemes are inherent of two representations~\citep{carroll1999corpus,king2003parc} that also follow the Lexical-Functional Grammar. The present Stanford typed dependencies\footnote{http://nlp.stanford.edu/software /dependencies\_manual.pdf} set can identify 53 such grammatical relationships. The parser generates a TD as a triplet structure \textit{tdName(head, dependent)}, where \textsl{tdName} represents the name of the dependency, \textsl{head} represents the head word and \textsl{dependent} represents the dependent word. More formally \textsl{tdName} depict that the \textsl{dependent} word is related to the \textsl{head} word by the dependency \textsl{tdName}.

\begin{example} \label{ex:TDs}
For the sentence \textit{``The bank sends the customer an sms."}, the TDs generated by the parser are given below.  \\
\\
\noindent \textit{TDs = [\textcolor[rgb]{0,0,1}{det}(\textcolor[rgb]{0.24,0.7,0.44}{bank}-2, \textcolor[rgb]{0.65,0.16,0}{The}-1), \textcolor[rgb]{0,0,1}{nsubj}(\textcolor[rgb]{0.24,0.7,0.44}{sends}-3, \textcolor[rgb]{0.65,0.16,0}{bank}-2), \textcolor[rgb]{0,0,1}{root}(\textcolor[rgb]{0.24,0.7,0.44}{ROOT}-0, \textcolor[rgb]{0.65,0.16,0}{sends}-3), \textcolor[rgb]{0,0,1}{det}(\textcolor[rgb]{0.24,0.7,0.44}{customer}-5, \textcolor[rgb]{0.65,0.16,0}{the}-4), \textcolor[rgb]{0,0,1}{iobj}(\textcolor[rgb]{0.24,0.7,0.44}{sends}-3, \textcolor[rgb]{0.65,0.16,0}{customer}-5), \textcolor[rgb]{0,0,1}{det}(\textcolor[rgb]{0.24,0.7,0.44}{sms}-7, \textcolor[rgb]{0.65,0.16,0}{an}-6), \textcolor[rgb]{0,0,1}{dobj}(\textcolor[rgb]{0.24,0.7,0.44}{sends}-3, \textcolor[rgb]{0.65,0.16,0}{sms}-7)]}\\

The description of the TDs are shown in Table~\ref{tab:TDs}. The last column in this table presents the graphical view of some important dependencies in the sentence, and how these TDs help to disambiguate the extraction of desired elements of the analysis class diagram from the sentence. The TD \textcolor[rgb]{0,0,1}{root}(\textcolor[rgb]{0.24,0.7,0.44}{ROOT}-0, \textcolor[rgb]{0.65,0.16,0}{sends}-3) depict that the main verb in the sentence is ``sends", the subject of the verb is ``bank", so a class can be created with name ``bank". The direct object of the verb is ``sms" which is receiving the action ``sends"' so an another class can be created with name ``sms", ``sends" can be added as an operation in it, and an association relationship with name ``sends" can be created between the class ``bank" and the class ``sms". The indirect object of the verb is ``customer" which is indirectly receiving the action ``sends" so it can be added as parameter to the to the operation ``sends" in the ``sms" class.
\end{example}
\begin{table}[htb]
  \centering
	\scriptsize
  \caption{TDs generated by Stanford parser for sentence: \textit{``The bank sends the customer an sms."} \label{tab:POSandTD}}{%
    \begin{tabular}{llllp{0.38\textwidth}}
    \hline
     \multicolumn{3}{c}{\textbf{Type dependencies (TDs)}} & \multirow{2}[4]{.24\textwidth}{\textbf{Depicting semantic relationship between the words in the sentence}} & \multirow{1}[1]{0.38\textwidth}{\textbf{Disambiguating extraction of relevant elements}} \bigstrut\\
     \cline{1-3} \cline{5-5}    \textsl{\textbf{TDName}} & \textsl{\textbf{Head}} & \textsl{\textbf{Dependent}} &       & \hspace{-3ex} \multirow{4}[0]{0.38\textwidth}{\includegraphics[width=0.42\textwidth]{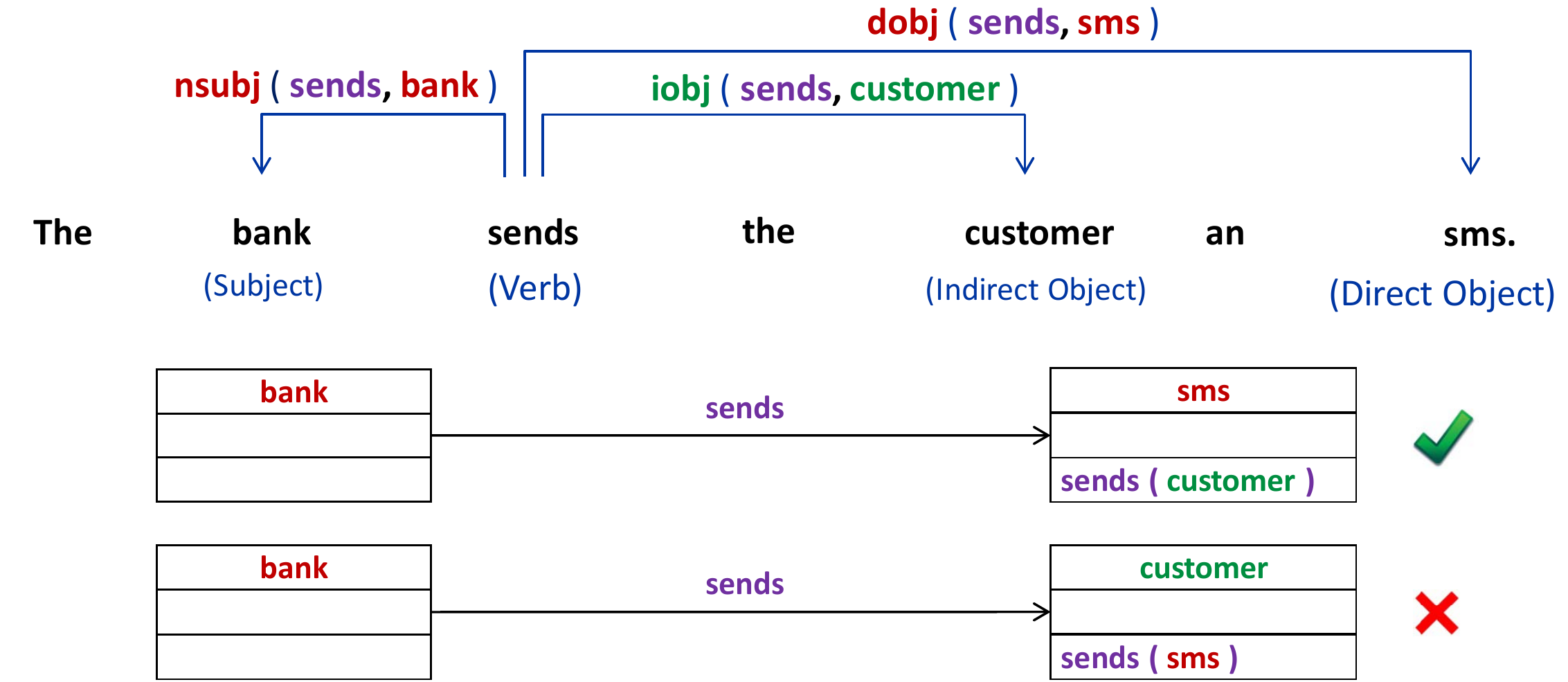}} \bigstrut\\
    \cline{1-4}	\textit{det} & Bank-2 & The-1  & ``The" is determiner of ``Bank" & \multirow{1}[1]{.33\textwidth}{}{} \bigstrut\\
      \textit{nsubj} & sends-3 & Bank-2 & ``Bank" is subject of ``sends"  & \multirow{1}[1]{.33\textwidth}{} \bigstrut\\
    \textit{root} & ROOT-0 & sends-3 & ``sends" is root of the sentence &  \multirow{1}[1]{.33\textwidth}{} \bigstrut\\
    \textit{det} & customer-5 & the-4  & ``the" is determiner of ``customer" &  \multirow{1}[1]{.38\textwidth}{}\bigstrut\\
     \textit{iobj} & sends-3 & customer-5 & ``customer" is indirect object of ``sends" & \multirow{1}[1]{.33\textwidth}{} \bigstrut\\
    \textit{det} & sms-7  & an-6   & ``an" is determiner of ``sms" &
		\multirow{1}[1]{.38\textwidth}{} \bigstrut\\
     \textit{dobj} & sends-3 & sms-7  & ``sms" is direct object of ``sends" &  \multirow{1}[1]{.4\textwidth}{} \\
    \hline
    \end{tabular}}%
  \label{tab:TDs}%

\end{table}%

\subsubsection{Language model}
\label{sec:LanguageModel}
As a same concept or thing can be expressed in many ways (using sentences with different sentence structures) in a natural language like English, the essential elements for generating analysis class diagrams are embedded at different places in different sentences. To correctly interpret the sentences for extracting the essential elements, the structures of the sentences should to be identified. \\
\\
\begin{table}[htb]
	\centering
 \scriptsize
   \caption{Sentence structure rules}
     \begin{tabular}{p{.5cm}p{2.7cm}p{2cm}p{1.9cm}p{6.4cm}}
    \hline
    \textbf{Rule \#} & \textbf{Antecedent (If the sentence contains TDs:)} & \textbf{Consequent (then the identified sentence structure is:)} & \textbf{Example sentence} & \textbf{Type dependencies of Example sentence} \bigstrut\\
    \hline
   \textbf{SSR1} & \textit{\textbf{nsubj}(A,B), \textbf{iobj}(A,C), \textbf{dobj}(A,D)} & SVIODO (Subject-Verb-IndirectObject-DirectObject) & The system sends the user an email. & \textit{[det(system-2, The-1), \textbf{nsubj}(sends-3, system-2), root(ROOT-0, sends-3), det(user-5, the-4), \textbf{iobj}(sends-3, user-5), det(email-7, an-6), \textbf{dobj}(sends-3, email-7)]}
	\bigstrut\\		
		\textbf{SSR2} & \textit{\textbf{nsubj}(A,B), \textbf{dobj}(A,C), \textbf{complm}(D,E), \textbf{nsubj}(D,F)} & SVDOThatClause (Subject-Verb-DirectObject-ThatClause) & The system informs the user that the battery is full & \textit{[det(system-2, The-1), \textbf{nsubj}(informs-3, system-2), root(ROOT-0, informs-3), det(user-5, the-4), \textbf{dobj}(informs-3, user-5), \textbf{complm}(full-10, that-6), det(battery-8, the-7), \textbf{nsubj}(full-10, battery-8), cop(full-10, is-9), ccomp(informs-3, full-10)]}
		\bigstrut\\
		\textbf{SSR3} & \textit{\textbf{nsubj}(A,B), \textbf{complm}(C,D), \textbf{nsubj}(C,E)} & SVThatClause (Subject-Verb-ThatClause) & The system validates that the password is correct & \textit{[det(system-2, The-1), \textbf{nsubj}(validates-3, system-2), root(ROOT-0, validates-3), \textbf{complm}(correct-8, that-4), det(password-6, the-5), \textbf{nsubj}(correct-8, password-6), cop(correct-8, is-7), ccomp(validates-3, correct-8)]} 		\bigstrut\\
		 \textbf{SSR4} & \textit{\textbf{nsubj}(A,B), \textbf{dobj}(A,C), \textbf{neg}(D,E), \textbf{aux}(D,F), \textbf{infmod}(C,D)} & SVDONotToInf (Subject-Verb-DirectObject-Not-To-Infinitive) & The system warns the user not to restart the system. & \textit{[det(system-2, The-1), \textbf{nsubj}(warns-3, system-2), root(ROOT-0, warns-3), det(user-5, the-4), \textbf{dobj}(warns-3, user-5), \textbf{neg}(restart-8, not-6), \textbf{aux}(restart-8, to-7), \textbf{infmod}(user-5, restart-8), det(system-10, the-9), dobj(restart-8, system-10)]}
				\bigstrut\\
		 \textbf{SSR5} & \textit{\textbf{nsubj}(A,B), \textbf{neg}(C,D), \textbf{aux}(C,E), \textbf{xcomp}(A,C), \textbf{dobj}(C,F)} & SVNotToInf (Subject-Verb-Not-To-Infinitive) & The customer selects not to fill the tank & \textit{[det(customer-2, The-1), \textbf{nsubj}(selects-3, customer-2), root(ROOT-0, selects-3), \textbf{neg}(fill-6, not-4), \textbf{aux}(fill-6, to-5), \textbf{xcomp}(selects-3, fill-6), det(tank-8, the-7), \textbf{dobj}(fill-6, tank-8)]}
		\bigstrut\\
		 \textbf{SSR6} & \textit{\textbf{nsubj}(A,B), \textbf{nsubj}(C,D), \textbf{aux}(C,E), \textbf{cop}(C,F), \textbf{xcomp}(A,C)} & SVDOtobeComp (Subject-Verb-DirectObject-to-be-Complement) & The system marks the errors to be red. & \textit{[det(system-2, The-1), \textbf{nsubj}(marks-3, system-2), root(ROOT-0, marks-3), det(errors-5, the-4), \textbf{nsubj}(red-8, errors-5), \textbf{aux}(red-8, to-6), \textbf{cop}(red-8, be-7), \textbf{xcomp}(marks-3, red-8)]}
		\bigstrut\\
		\textbf{SSR7} & \textit{\textbf{nsubj}(A,B), \textbf{dobj}(A,C), \textbf{aux}(D,E), \textbf{infmod}(C,D)} & SVDOToInf (Subject-Verb-DirectObject-To-Infinitive) & The system commands the motor to start. & \textit{[det(system-2, The-1), \textbf{nsubj}(commands-3, system-2), root(ROOT-0, commands-3), det(motor-5, the-4), \textbf{dobj}(commands-3, motor-5), \textbf{aux}(start-7, to-6), \textbf{infmod}(motor-5, start-7)]}
		\bigstrut\\
		 \textbf{SSR8} & \textit{\textbf{nsubj}(A,B), \textbf{dobj}(A,C), \textbf{partmod}(C,D) and \textbf{POS-tag(D)}==``VBG"} & SVDOPresentPart (Subject-Verb-DirectObject-PresentParticiple) & The system keeps the user waiting. & \textit{[det(system-2, The-1), \textbf{nsubj}(keeps-3, system-2), root(ROOT-0, keeps-3), det(user-5, the-4), \textbf{dobj}(keeps-3, user-5), \textbf{partmod}(user-5, \textbf{waiting}-6)].}
		\bigstrut\\
		\textbf{SSR9} & \textit{\textbf{nsubj}(A,B), \textbf{dobj}(A,C), \textbf{partmod}(C,D) and \textbf{POS-tag(D)}==``VBN"} & SVDOPastPart (Subject-Verb-DirectObject-PastParticiple) & The system validates the record entered by the customer. & \textit{[det(system-2, The-1), \textbf{nsubj}(validates-3, system-2), root(ROOT-0, validates-3), det(record-5, the-4), \textbf{dobj}(validates-3, record-5), \textbf{partmod}(record-5, \textbf{entered}-6), prep(entered-6, by-7), det(customer-9, the-8), pobj(by-7, customer-9)]}
		\bigstrut\\
	\textbf{SSR10} & \textit{\textbf{nsubj}(A,B), \textbf{nsubj}(C,D), \textbf{xcomp}(A,C) and \textbf{POS-tag(C)}==``JJ"} & SVDOAdj (Subject-Verb-DirectObject-Adjective-Complement) & The system keeps the door open & \textit{[det(system-2, The-1), \textbf{nsubj}(keeps-3, system-2), root(ROOT-0, keeps-3), det(door-5, the-4), \textbf{nsubj}(open-6, door-5), \textbf{xcomp}(keeps-3, \textbf{open}-6)]} \bigstrut\\
    \hline
    \end{tabular}%
   \label{tab:SSRules}
\end{table}%

\begin{table}[htb]
  \centering
	\scriptsize
  \caption{Restriction rules for documenting use case specifications}
    \begin{tabular}{p{.3cm} p{7.5cm} p{7cm}}
    \hline
   \textbf{Rule\#} & \textbf{Rule description} & \textbf{Rationale} \\
   \hline
    \textbf{1} & Use simple sentences to write the steps, except for a few complex sentences such as sentences expressing validation or check (example: “The system validates that the password is correct.”) and sentences specifying conditions (example: “If the ATM card is invalid, the system ejects the card.”) & In order to reduce ambiguity~\citep{kamsties2000taming, wiegers2013software}, various authors in literature recommend the use of simple sentences for documenting the UCSs. The sentence structure rules used by the approach presently are able to process the simple sentences and a few complex sentences. \\
    \textbf{2} & Do not use the pronouns. & The approach avoid the use of pronouns for documenting the UCS to avoid the errors in pronoun resolution as the state of the art approaches for pronoun resolution problem have precisions between 80-90\%~\citep{lee2011stanford}. \\
    \textbf{3} & Use consistent names for things, concepts etc. (i.e. use of different names to represent same thing or concept at different places in UCS must be avoided) & The approach recommend using consistent names for things and concepts in order to avoid the identification of multiple classes representing same thing or concept. \\
    \textbf{4} & Use``system" or use case name to refer to the system under development. & To avoid the identification of multiple classes representing the system. \\
    \textbf{5} & Use keywords: &  \\
          & – INCLUDE to specify include relationship with other UCS & The approach creates an association relationship named INCLUDE between the control classes of the two UCSs. \\
          & – EXTEND to specify extend relationship with other UCS & The approach creates an association relationship named EXTEND between the control classes of the two UCSs. \\
          & – RESUME to specify resume or return of control to a specific step in the UCS & The approach creates a resumeStep() operation in the control class of the UCS. This can further help in the generation of template code which could be one of the future directions of the proposed approach. \\
          & – REPEAT to specify the repeated execution of some steps in the UCS. & The approach creates a repeatSteps() operation in the control class of the UCS. This can further help in the generation of template code which could one of the future directions of the proposed approach. \\
\hline
				\multicolumn{3}{l}{For types of sentences in English please refer Appendix~C} \bigstrut
		\end{tabular}%
  \label{tab:RestrictionRules}%
\end{table}%

\noindent\textbf{Language model:} We define the language model of an approach as the set of sentence patterns that an approach can interpret.

The language model of our approach consists of a comprehensive set of sentence patterns that include all the simple sentences (including sentences containing participles, infinitives and gerunds) and a few complex sentences (sentences specifying conditions and the sentences containing that clause and conjunctive clause) written in English. A simple sentence contains only one independent clause. A complex sentence contains one independent clause and one or more dependent clauses. (A clause consists a subject or noun phrase and a predicate or verb phrase. An independent clause is a clause that do not depend on any other clause to expresses a complete thought. A dependent clause is a clause which is dependent on some other clause or dependent clause to express a complete thought. For more details on types of sentences please refer~\ref{app:TypesOfSentnecesInEnglish}.)

Our language model is constructed using the twenty five verb patterns proposed by A. S. Hornby, known for various achievements in linguistic and literature~\citep{hanks2008lexical}, in Oxford Advanced Learner’s Dictionary of Current English~\citep{OALD1974english,OALD2000english} and in~\cite{hornby1975english}. These verb patterns define the twenty five possible ways in which a verb phrase in a sentence can be written. Using all these verb patterns, we framed the comprehensive set of sentence structure rules to recognize the sentences structure of the sentences. These rules use TDs of the sentences to identify the sentence structures. The first ten sentence structure rules are presented in Table~\ref{tab:SSRules}, for complete set of rules please refer~\ref{app:SentenceStructureRules}. These rules are presented in \textsl{Antecedent-Consequent} format and they are ordered. To identify the sentence structure of a sentence, the approach one by one checks that all TDs shown in \textsl{Antecedent} part of the rule are found in the TDs of the given sentence, if this is true then the approach gets the sentence structure from \textsl{Consequent} part of the rule. \textit{(Note: The rules SSR8-SSR11 uses POS-tags along with TDs to identify the sentence structures. The rules SSR30-SSR33 identify sentence structures through keyword matching.)}

The language model along with the need to avoid ambiguities~\citep{kamsties2000taming, wiegers2013software} in the sentences, enforce the UCSs to be written in English language using a few restriction rules. The restriction rules along with their rationale are presented in Table~\ref{tab:RestrictionRules}.

The following example shows how the approach identifies the sentence structures.

\begin{example} Here we show how the approach identifies the sentence structure of a given sentence using the rules presented in Table~\ref{tab:SSRules}. For sentence say S=``The system commands the motor to start.", the TDs generated by the parser are:\\
\textsl{[det(system-2, The-1), \textbf{nsubj}(commands-3, system-2), root(ROOT-0, commands-3), det(motor-5, the-4), \textbf{dobj}(commands-3, motor-5), \textbf{aux}(start-7, to-6), \textbf{infmod}(motor-5, start-7)]
} \\
 \\
Here, for each rule shown in Table~\ref{tab:SSRules} the approach one by one checks whether the TDs present in the \textsl{Antecedent} part of the rule are found in the TDs of the sentence. All the TDs in the \textsl{Antecedent} part of the rule \textbf{SSR7} are found in the TDs of the sentence, hence the approach gets the sentence structure as \textbf{SVDOToInf} from \textsl{Consequent} part of the rule.
\end{example}
\subsection{Working of the approach}
\label{sec:WorkingOfProposedApproach}
\begin{figure}[htbp]
\centering
\includegraphics[width=\linewidth]{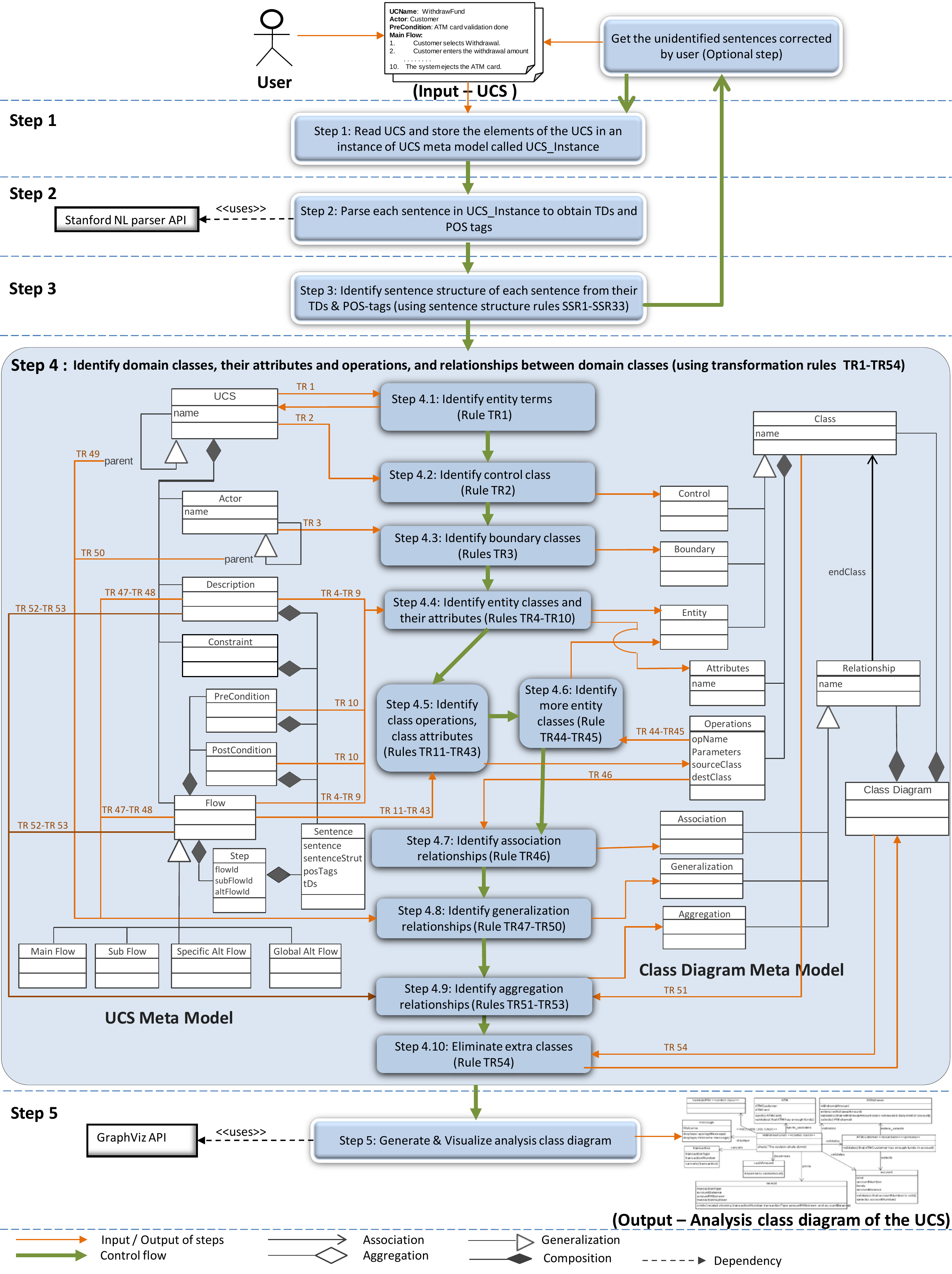}
\caption{Schematic diagram of the proposed approach}
\label{fig:ActivityDiagramOfProposedApproach}
\end{figure}

The proposed approach works in five steps (Figure~\ref{fig:ActivityDiagramOfProposedApproach}). The approach first reads the UCS, and stores the elements of the UCS into an instance of UCS metamodel that we call UCS\_Instance (Step~1). It parses the sentences in the UCS\_Instance using Stanford NL Parser API to generate TDs and POS-tags (Step~2). It then applies the comprehensive set of proposed sentence structure rules on TDs and POS-tags to identify sentence structures of the sentences (Step~3). Then it applies the proposed transformation rules on the TDs and POS-tags to identify problem level classes, attributes, operations, the relationships between the classes (Step~4). Finally, it generates and visualizes the analysis class diagram (Step~5). The terms used in the transformation rules are shown in Table~\ref{tab:Terminology}. The details of the steps are presented as follows: \\

\begin{table}[htb]
  \centering
	\scriptsize
   \caption{Terminology used in the transformation rules}
    \begin{tabular}{llll}
    \textbf{UCS\_Instance} & an instance of UCS metamodel (Figure~\ref{fig:UCSandCDMetaModel}) & \textbf{op.Name} & name of the operation \\
    \textbf{ClassDiagram\_Instance} & an instance of the class diagram metamodel (Figure~\ref{fig:UCSandCDMetaModel}) & \textbf{op.SourceEntityTerm} & entity term that calls the operation \\
    \textbf{createClass( )} & create new class, and add it to ClassDiagram\_Instance & \textbf{op.DestEntityTerm} & entity term that hosts the operation \\
    \textbf{class.addAttribute()} & add attribute to class if already not present & \textbf{op.Para} & operation parameter \\
    \textbf{class.addOperation()} & add operation to class if already not present & \textbf{`='} & assignment operator \\
    \textbf{createRelationship( )} & create new relationship, and add it to ClassDiagram\_Instance & \textbf{`=='} & relational equality operation \\
    \textbf{op} & a meta object to store identified operation &       &  \\
    \end{tabular}%
  \label{tab:Terminology}%
\end{table}%

\noindent \textbf{\textsl{Step 1: Read UCS and generate UCS\_Instance}}\\
\\
This step reads the UCS (Example Figure~\ref{tab:UseCaseWithdrawFund}) and then stores the data from all the fields of the UCS into the respective fields in an instance of UCS metamodel (Example Figure~\ref{fig:UCSandCDMetaModel}) called UCS\_Instance. \\
\\
\noindent \textbf{\textsl{Step 2: Parse sentences in UCS\_Instance to obtain their TDs and POS tags}}\\
\\
This step reads the sentences of each section (Main Flow, Sub Flow, Alternate Flow, Global Alternate Flow, Pre-Condition, Post-Condition and Description section) of the UCS from UCS\_Instance, and parses each sentence using Stanford NL Parser API to generate POS-tags and TDs. The TDs and POS-tags of the sentences are then used in Step 3 to identify the sentence structure of the sentences, and in Step 4 to identify the elements of the analysis class diagram.

\begin{example}
\label{ex:Step2}
For the sentence =``ATM customer enters the withdrawal amount." of the UCS WithdrawFund (Table~\ref{tab:UseCaseWithdrawFund}), the TDs and POS-tags generated by the are:\\
\\
TDs=[nn(customer-2, ATM-1), nsubj(enters-3, customer-2), root(ROOT-0, enters-3), det(amount-6, the-4), nn(amount-6, withdrawal-5), dobj(enters-3, amount-6)]
\\
POS-tags=[ATM/NNP, customer/NN, enters/VBZ, the/DT, withdrawal/NN, amount/NN, ./.]
\end{example}

\noindent \textbf{\textsl{Step 3: Identify sentence structures of the sentences}}\\
 \\
As a same concept or thing can be expressed in many ways, the essential elements for generating analysis class diagrams are embedded at different places in different sentences. To correctly interpret the sentences for extracting the essential elements, the structures of the sentences are to be identified. The approach uses the proposed set of comprehensive sentence structure rules SSR1-SSR33 to identify the sentence structures of the sentences as described in Section~\ref{sec:LanguageModel} (Table~\ref{tab:SSRules} presents first ten sentence structure rules, for full rule set please refer~\ref{app:SentenceStructureRules}). These rules are presented in \textsl{Antecedent-Consequent} format and they are ordered. To identify the sentence structure of a sentence, the approach one by one checks that all TDs shown in \textsl{Antecedent} part of the rule are found in the TDs of the given sentence, if this is true then the approach gets the sentence structure from \textsl{Consequent} part of the rule.
If sentence structures of some sentences cannot be identified then those sentences are marked and presented to the user for modification. The user can modify sentences or may choose to continue without modifying. If user chooses to continue without modifying the unidentified sentences then the unidentified sentences are skipped by the transformation process. The identified sentence structure of the sentences along with the TDs and POS-tags identified in previous step are then used in Step~4.5 to disambiguate the identification of class operations and/or class attributes. The identified operations are in turn used disambiguate the identification of entity classes in Step~4.6, and association relationships in Step~4.7.

\begin{example}
\label{ex:Step3}
As described in Section~\ref{sec:LanguageModel}, to identify the sentence structure of the sentence = ``ATM customer enters the withdrawal amount.", the approach applies the sentence structure rule \textbf{SSR27} (shown in Appendix A) on TDs and POS-tags of the sentence (shown in Example~\ref{ex:Step2}) to identify the sentence structure of the sentence as \textbf{SVDO}.
\end{example}
\noindent \textbf{\textsl{Step 4: Identify elements of analysis class diagram}}\\
 \\
The nouns in the sentences are the prospects of classes and attributes, and the verbs are the prospects of operations and relationships. But every noun may not be a class or attribute and every verb may not be an operation or relationship. To disambiguate the process of identifying the right classes, attributes, operations and relationships of the analysis class diagram from the text, our approach uses various heuristics (the proposed transformation rules TR1-TR54) that in turn uses the syntactic and semantics relationships between the words in the sentences obtained from TDs and POStags. The following steps present how the approach identifies the elements of analysis class diagram.
\\

\noindent \textbf{\textsl{Step 4.1: Identify entity terms}}
\\
\\
The noun phrases in the sentences may contain single nouns (Example: Transaction, Withdrawal) or a group of two or more consecutive nouns (group of nouns representing a single term, Example: Transaction number, ATM Card, ATM Card PIN number). These single nouns or group of two or more consecutive nouns are called entity terms. The entity terms are the prospects of potential classes or attributes. This step combines two or more consecutive nouns in the sentences to identify the entity terms (transformation rule-TR1). This is done by concatenating two or more consecutive words in the sentences whose POS-tag starts with ``NN". The TDs and POS-tags of the sentence are updated to reflect the changes (concatenation) done if any. A similar method for entity term extraction is used in an approach for entity disambiguation proposed by~\cite{misra2013entity}.\\
\\
\textbf{\textsl{Rule-TR1}}: For each sentence in the UCS\\
\setlength\parindent{55pt}
\indent Concatenate two or more consecutive words in the sentences whose POS-tags starts with ``NN". \\
\indent Update the TDs and POS-tags of the sentences to reflect the changes\\
\setlength\parindent{40pt}
\indent EndFor\\
\indent Scan POS-tags of each sentence, and store all the nouns in a set named \textsl{setOfEntityTerms}.\\
\begin{example}
\label{ex:Step4.1}
For sentence ``ATM customer enters the ATM Card Pin Number.", the POS-tags generated by the parser are: \textsl{[ATM/NNP, customer/NN, enters/VBZ, the/DT, ATM/NNP, Card/NNP, Pin/NNP, Number/NNP, ./.]}

Here ``ATM" and ``customer" are the two consecutive words whose POS-tags start with ``NN", hence they are concatenated using the transformation rule TR1 to get ``ATMcustomer" representing a single entity term. Similarly, the four consecutive words ``ATM", ``Card", ``Pin" and ``Number" are also concatenated using rule TR1 to get ``ATMCardPinNumber" representing another single entity term. \\
\\
The modified sentence is : ``ATMcustomer enters the ATMCardPinNumber."\\
The updated TDs are: \textsl{[nsubj(enters-2, ATMcustomer-1), root(ROOT-0, enters-2), det(ATMCardPinNumber-4, the-3), dobj(enters-2, ATMCardPinNumber-4)]}\\
The updated POS-tags are: \textsl{[ATMcustomer/NNP, enters/VBZ, the/DT, ATMCardPinNumber/NN, ./.]}\\
\textsl{setOfEntityTerms} =\{``ATMcustomer", ``ATMCardPinNumber"\}
\\
\\
Similarly, when two more sentences, ``ATM customer selects Withdrawal." and ``ATM customer enters withdrawal amount.", are processed, new entity terms are identified and the setOfEntityTerms is updated as:\\
\textsl{setOfEntityTerms} =\{``ATMcustomer", ``ATMCardPinNumber", ``Withdrawal", ``withdrawalamount"\}
\end{example}
\noindent\textbf{\textsl{Step 4.2: Identify control class} \textsl{(Rules TR2)}} \\
\\
This step creates a control class from the name of the UCS stored in UCS\_Instance (transformation rule-TR2).
\\ \\
\textbf{\textsl{Rule-TR2}}: \textsl{createClass}(UCS.name,``${<<}$control class${>>}$"); \\
For one UCS only one control class is created and from now all the references to the word ``System" and to the name of UCS in the sentences refer to this control class.
\begin{example}
From UCS shown in Table~\ref{tab:UseCaseWithdrawFund} the approach creates the control class named \textsl{``WithdrawFunds \textless \textless control class \textgreater \textgreater"}, where the stereotype \textsl{``\textless \textless control class \textgreater \textgreater"} denotes that the class is control class.
\end{example}

\noindent\textbf{\textsl{Step 4.3: Identify boundary class(es)} \textsl{(Rules TR3)}} \\
\\
This step creates a boundary class for each actor of the UCS stored in UCS\_Instance (transformation rule-TR3).
\\
\\
\textbf{\textsl{Rule-TR3}}: For each actor of the UCS stored in UCS\_Instance, \textsl{createClass}(actor.name,``${<<}$boundary class${>>}$");
\begin{example}
From UCS shown in Table~\ref{tab:UseCaseWithdrawFund} the approach creates the boundary class named \textsl{``ATMcustomer \textless \textless boundary \textgreater \textgreater \textless \textless primary \textgreater \textgreater"}, where the stereotype \textsl{``\textless \textless boundary \textgreater \textgreater \textless \textless primary \textgreater \textgreater"} denotes that the class is boundary class of primary actor.
\end{example}

\noindent\textbf{\textsl{Step 4.4: Identify entity classes and their attributes} \textsl{(Rules TR4-TR10)}}\\ \\
The entity terms identified in Step~4.1 are the prospects of classes or attributes. But every entity term may not be a class or an attribute. The step applies heuristics (transformation rules TR4-TR10) on the identified entity terms and TDs of the sentences to identify the entity classes and attributes. Rule TR4 is given below, for rules TR5-TR10 please refer~\ref{app:TransformationRules} \\
\\
\textbf{\textsl{Rule-TR4}}:
For every two entity terms t1 and t2 in \textsl{setOfEntityTerms}\\
\setlength\parindent{55pt}
\indent If (t2 startsWith t1) AND (t2$\neq$t1) then\\
\setlength\parindent{65pt}
\indent class = \textsl{createClass}(t1,``${<<}$entity class${>>}$");  class.addAttribute(t2); \\
\setlength\parindent{55pt}
\indent EndIf\\
EndFor

\begin{example}
\label{ex:EntityClass}
The \textsl{setOfEntityTerms} =\{``ATMcustomer", ``ATMCardPinNumber", ``Withdrawal", ``withdrawalamount"\}, shown in Example~\ref{ex:Step4.1}, contains two entity terms say t1=``Withdrawal" and  t2=``withdrawalAmount"\\
\begin{tabular}{p{5.2in}p{1in}}
Here the entity term t2 starts with entity term t1 hence by rule TR4 ``Withdrawal" is identified as a class and ``withdrawalAmount" is added as an attribute to class ``Withdrawal" & \multirow{1}[10]{1in}{\includegraphics[width=.8in]{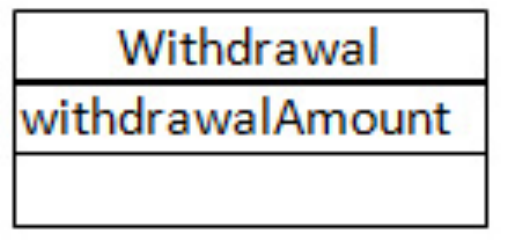} }\\
 &  \\
 \end{tabular}
 \end{example}

\noindent \textbf{\textsl{Step 4.5: Identify class operations and more class attributes} \textsl{(Rules TR11-TR43)}}\\ \\
As the flow sentences (sentences in the main flow, sub flow and alternate flow sections of UCS) specifies the sequence of actions preformed by the system and the actor(s) of the system, the verbs representing these actions can be used to identify class operations. For each operation, three things are to be identified: i) the operation, ii) the source entity term that calls this operation and iii) the destination entity term that hosts this operation.

The verbs in the flow sentences are the prospects of class operations, but not all the verbs in a sentence represent valid operations. This step applies the heuristics (or transformation rules TR11-TR43) to identify the operations, the source entity terms and the destination entity terms from the flow sentences stored in UCS\_Instance. These heuristics (rules) uses the sentence structure of the sentences, and the syntactic and semantic relationships between the words in the sentences depicted by TD to disambiguate the process of identifying these elements. The transformation rules TR11-TR20 are presented in Table~\ref{tab:TRules}, for full rule set please refer~\ref{app:TransformationRules}. These rules are presented in \textsl{Antecedent-Consequent} format. The Antecedent part or If part (Column A) contains the sentence structure of the sentences to be matched for the rules to be fired, the Consequent part or then part has two columns (Column B and column C) where, column B presents the TDs of the sentence to be used for identifying operations/attributes and column C defines how the approach identifies class operations and attributes from these TDs when the rule is fired. To identify the class operations and attributes from a given sentence, the approach one by one compares the sentence structure of the given sentence with that present in column A, if match is found then the approach uses the TDs present in column B to identify the operations/attributes as shown in column C. These identified operations are further used in Step 4.6 for identifying more entity classes, and in Step 4.7 for identifying association relationships between the classes.

\begin{table}[htb]
	\scriptsize
  \caption{Transformation rules to identify operations and attributes of classes}
    \begin{tabular}{|c|p{2cm}|p{5.5cm}|p{6cm}|}
		\hline 	
     & \textbf{Antecedent ( If A)} & \multicolumn{2}{l|}{\textbf{Consequent ( then use B to identify C )}} \bigstrut\\
		\cline{2-4} \textbf{Rule \#}  & \textbf{A (Sentence structure of sentence is:)}  & \textbf{B (TDs of the sentence:)} & \textbf{C (operations/attributes:)} \bigstrut\\
    \hline
		   \textbf{TR11} & SVIODO & \textit{\textbf{nsubj}(A,B), \textbf{iobj}(A,C), \textbf{dobj}(A,D)} & op.SourceEntityTerm=B, op.DestEntityTerm=D, op.name=A \bigstrut\\
    \hline
		\textbf{TR12} & SVDOThatClause & \textit{\textbf{nsubj}(A,B), \textbf{dobj}(A,C), \textbf{complm}(D,E), \textbf{nsubj}(D,F)} & op.SourceEntityTerm=B, op.DestEntityTerm=C, op.name=A \bigstrut\\
    \hline
		\textbf{TR13} & SVThatClause & \textit{\textbf{nsubj}(A,B), \textbf{complm}(C,D), \textbf{nsubj}(C,E)} & op.SourceEntityTerm=B, op.DestEntityTerm=E, op.name=A \bigstrut\\
    \hline
		\textbf{TR14} & SVDONotToInf & \textit{\textbf{nsubj}(A,B), \textbf{dobj}(A,C), \textbf{neg}(D,E), \textbf{aux}(D,F), \textbf{infmod}(C,D)} & op.SourceEntityTerm=B, op.DestEntityTerm=C, op.name=A \bigstrut\\
    \hline
		 \textbf{TR15} & SVNotToInf & \textit{\textbf{nsubj}(A,B), \textbf{neg}(C,D), \textbf{aux}(C,E), \textbf{xcomp}(A,C), \textbf{dobj}(C,F)} & op.SourceEntityTerm=B, op.DestEntityTerm=F, op.name=A \bigstrut\\
    \hline
		 \textbf{TR16} & SVDOtobeComp & \textit{\textbf{nsubj}(A,B), \textbf{nsubj}(C,D), \textbf{aux}(C,E), \textbf{cop}(C,F), \textbf{xcomp}(A,C)} & op.SourceEntityTerm=B, op.DestEntityTerm=D, op.name=A \bigstrut\\
    \hline
		 \multirow{4}[8]{*}{\textbf{TR17}} & \multirow{4}[8]{*}{SVDOToInf} & \multirow{4}[8]{4cm}{\textit{\textbf{nsubj}(A,B), \textbf{dobj}(A,C), \textbf{aux}(D,E), \textbf{infmod}(C,D)}} & op.SourceEntityTerm=B, op.DestEntityTerm=D, op.name=A \bigstrut\\
           &        &        & If (TDs of the sentence contains TD dobj(D,F)) then  \bigstrut\\
           &        &        & op.SourceEntityTerm2=C, op.DestEntityTerm2=F, op.name2=D  \bigstrut\\
          &        &        & EndIf \bigstrut\\
    \hline
		\multirow{7}[14]{*}{\textbf{TR18}} & \multirow{7}[14]{*}{SVDOPresentPart} & \multirow{7}[14]{*}{\textit {\textbf{nsubj}(A,B), \textbf{dobj}(A,C), \textbf{partmod}(C,D) \textbf{dobj}(D,E)}} & op.SourceEntityTerm=B, op.DestEntityTerm=D, op.name=A
 \bigstrut\\
          &        &        & If (TDs of the sentence contains TD dobj(D,E)) then \bigstrut\\
           &        &        & op.DestEntityTerm.addAttribute(E) \bigstrut\\
           &        &        & For each TD=conj(X,Y) and (X==E) after dobj(D,E)  \bigstrut\\
           &        &        & destClass.addAttribute(Y) \bigstrut\\
           &        &        & EndFor \bigstrut\\
          &        &        & EndIf  \bigstrut\\
    \hline
		\textbf{TR19} & SVDOPastPart & \textit{\textbf{nsubj}(A,B), \textbf{dobj}(A,C), \textbf{partmod}(C,D)} & op.SourceEntityTerm=B, op.DestEntityTerm=C, op.name=A \bigstrut\\
    \hline
		\textbf{TR20} & SVDOAdj & \textit{\textbf{nsubj}(A,B), \textbf{nsubj}(C,D), \textbf{xcomp}(A,C)} & op.SourceEntityTerm=B, op.DestEntityTerm=D, op.name=A \bigstrut\\
    \hline
		\multicolumn{4}{l}{\textsl{Note: ``op" is a meta object used to store the identified operation}} \bigstrut\\
    \end{tabular}%
  \label{tab:TRules}%
\end{table}%

\begin{example}
\label{ex:ExampleTR10}
For sentence=``\textsl{ATM customer enters the withdrawal amount.}" shown in Example~\ref{ex:Step2} the sentence structure as identified in Example~\ref{ex:Step3} is \textbf{SVDO}, which that matches with the \textsl{Antecedent} part of rule TR37 (presented in Appendix~B), hence the approach applies the transformation rule TR37 on the TDs \textsl{nsubj(enters-2,ATMcustomer-1), dobj(enters-2,withdrawalAmount-4)} of the sentence to identify following operation:\\
\textsl{op.SourceEntityTerm}=\textsl{``ATMcustomer", \textsl{op.DestEntityTerm}=``withdrawalAmount", op.name=``enters"}\\
We can see that \textsl{``ATM"} and \textsl{``customer"} are consecutive nouns so they had already been concatenated to represent single entity term \textsl{``ATMcustomer"} using Rule TR1 in Step 4.1. Similar is the case for \textsl{``withdrawal"} and \textsl{``amount"}.
\end{example}

\noindent \textbf{\textsl{Step 4.6: Identify more entity classes} \textsl{(Rule TR44-TR45)}}\\
\\
The source entity terms and destination entity terms of the operations identified in previous step are used in this step to identify more entity classes. A source entity term is the caller of the identified operation hence it is the prospect of only a class (if a class for the source entity term does not exist in \textsl{ClassDiagram\_Instance}, a new entity class is created for it, and is added to \textsl{ClassDiagram\_Instance} by rule TR44). Whereas, a destination entity term has two possibilities: it may either be an attribute of an existing class in which the identified operation is to be hosted or may be a prospect of a class (if a class for the destination entity term exists in \textsl{ClassDiagram\_Instance} then the identified operation is hosted in that class, otherwise a new entity class is created, the identified operations is hosted in that class, and the class is added to \textsl{ClassDiagram\_Instance}) (rule TR45). \\
\\
\textbf{\textsl{Rule-TR44}}: If op.SourceEntityTerm is not present in \textsl{ClassDiagram\_Instance} then \textsl{createClass}(op.SourceEntityTerm,``${<<}$entity class${>>}$");
\begin{example}
 To the identified operation shown in Example~\ref{ex:ExampleTR10}\\
\label{ex:OpSourceClassEntity}
\begin{tabular}{p{5.2in}p{1in}}
 The approach apply rule TR44 to create a new entity class ``ATMcustomer" and add it to \textsl{ClassDiagram\_Instance}  & \multirow{1}[10]{1in}{\includegraphics[width=.8in]{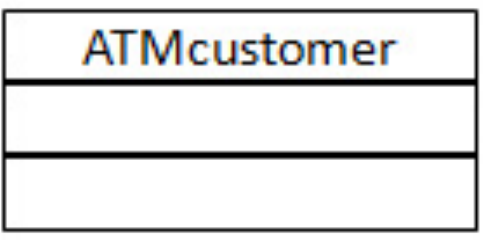} }\\
 &  \\
 \end{tabular}
\end{example}
\noindent \textbf{\textsl{Rule-TR45}}: For each class C in ClassDiagram\_Instance \\		
\setlength\parindent{65pt}
\indent If (op.DestEntityTerm.name==C.name)AND(C does not contain operation op.name(op.Para)) then\\
\setlength\parindent{80pt}
 \indent	 C.addOperation(op.name(op.Para));\\		
\setlength\parindent{65pt}
\indent EndIf\\
\setlength\parindent{55pt}
\indent EndFor\\
\indent If no such class is found then \\
\setlength\parindent{65pt}
\indent For each class C in ClassDiagram\_Instance \\
\setlength\parindent{80pt}
\indent If(op.DestEntityTerm.name==a.name for some attribute a of class C)AND\\
\setlength\parindent{85pt}
\indent(C does not contains operation op.name(op.Para)) then\\
\setlength\parindent{95pt}
\indent	  C.addOperation(op.name(op.Para));\\	
\setlength\parindent{80pt}	
\indent		EndIf\\		
\setlength\parindent{65pt}	
\indent EndFor\\
\setlength\parindent{55pt}
\indent EndIf \\
\indent If no such class is found then \\
\setlength\parindent{65pt}	
\indent	C=createClass(op.DestEntityTerm.name,``${<<}$entity class${>>}$"); C.addOperation(op.name(op.Para));\\
\setlength\parindent{55pt}
\indent EndIf

\begin{example}
 To the same operation identified in Example~\ref{ex:ExampleTR10} \\
\begin{tabular}{p{5.2in}p{1in}}
the approach applies rule TR45 to add the operation ``enters(withdrawalAmount)" to class=``Withdrawal", since op.DestEntityTerm=``withdrawalAmount" is found as an attribute of class=``Withdrawal" & \multirow{1}[10]{1in}{\includegraphics[width=.9in]{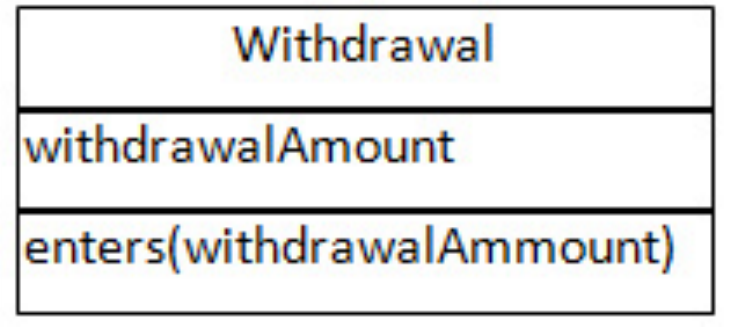} }\\
 &  \\
 \end{tabular}
\end{example}
\noindent \textbf{\textsl{Step 4.7: Identify association relationships} \textsl{(Rule TR46)}}\\ \\
From the source entity term, the destination entity term and the operation name of each operation identified in step~4.5, the approach identifies association relationship between the class representing source entity term and the class representing destination entity term. The navigability of the relationship is set from the class representing the source entity term to the class representing the destination entity term. (transformation rule TR-46)

For two classes say class A and class B, if class A calls operations op1 of class B then an association relationship named op1 is created between the class A and Class B. If class A also calls another operation op2 of class B then op2 is appended to the name of the association relationship to get the new name of the association relationship as op1op2 (Note: a new relationship op2 is not created between class A and class B, because this will result in redundant relationships between the classes i.e. two association relationship of navigability from class A to class B). The appending of such operation name to association relationship names will be helpful during the modification that may be done automatically or manually at later stage. Suppose at later stage we remove the operation op1 from class B, then this will require to remove the relationship named op1 between class A and class B, this can be done easily by removing the op1 from the name of association relationship between class A and class B to get the new relationship name as op2. If the relationships had not been renamed this way then the removal of the operation op1 from class B will also remove the only association relationship with name op1 between the two classes.
\\
\\
\textbf{\textsl{Rule-TR46}}: For each relationship r in \textsl{ClassDiagram\_Instance} \\
\setlength\parindent{55pt}
\indent If(op.SourceEntityTerm==r.class1 and op.DestEntityTerm==r.class2)AND(r.name does not contains op.name) \\
\setlength\parindent{80pt}
\indent     append op.name to r.name\\
\setlength\parindent{55pt}
\indent		EndIf \\
\setlength\parindent{45pt}
\indent EndFor \\
\indent If (no such relationship found) then \\
\setlength\parindent{55pt}
\indent For each class c in ClassDiagram\_Instance \\
\setlength\parindent{65pt}
\indent If(op.DestEntityTerm==c.Name)\\
\setlength\parindent{80pt}
\indent     rName=op.name; createRelationship(op.SourceEntityTerm, c, rName, ``association");\\
\setlength\parindent{65pt}
\indent		EndIf \\
\setlength\parindent{55pt}
\indent EndFor \\
\indent If(no such class is found) then\\
\setlength\parindent{65pt}
\indent For each class c in ClassDiagram\_Instance \\
\setlength\parindent{75pt}
\indent If(op.DestEntityTerm==a.Name for some attribute a in class c)\\
\setlength\parindent{85pt}
\indent rName=op.name; createRelationship(op.SourceEntityTerm, c, rName, ``association");\\
\setlength\parindent{75pt}
\indent		EndIf \\
\setlength\parindent{65pt}
\indent EndFor \\
\setlength\parindent{55pt}
\indent EndIf\\
\setlength\parindent{45pt}
\indent EndIf\\
(Note: the operation createRelationship(), sets the navigability from the class representing the sourceEntityTerm to the class representing destEntityTerm when it creates association relationship between the classes.)

\begin{example}
To the operation op = ``enters", op.sourceEntityTerm = ``ATMcustomer" and op.destEntityTerm = ``withdrawlAmount"  shown in Example~\ref{ex:ExampleTR10} of Step~4.5, the approach applies rule TR46\\
\begin{tabular}{p{4.2in}p{2in}}
to create anassociation relationship r with r.name = `enters", r.endClass1 = ``ATMcustomer", r.endClass2 = ``Withdrawal" (since op.DestEntityTerm = ``withdrawalAmount" is found as an attribute of class = ``Withdrawal") and with navigability from ``ATMcustomer" to ``Withdrawal"  & \multirow{1}[10]{2in}{\includegraphics[width=1.9in]{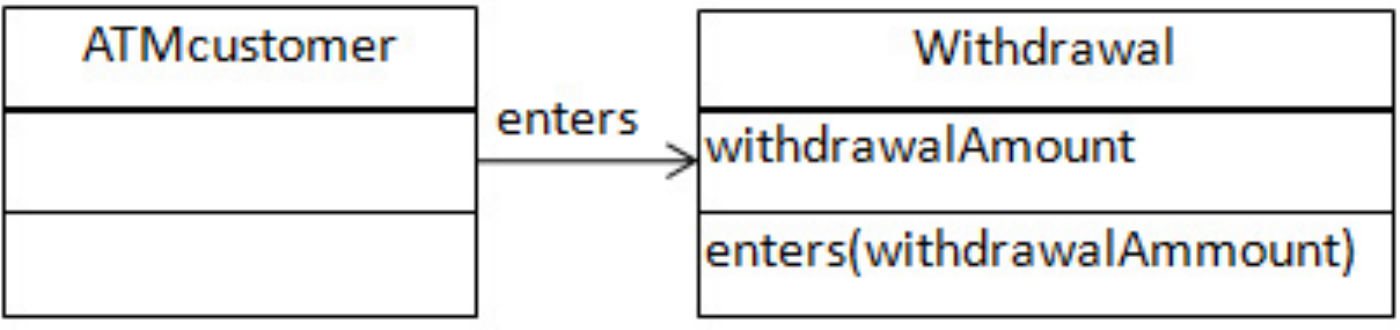} }\\
 &  \\
 \end{tabular}
\end{example}
\noindent \textbf{\textsl{Step 4.8: Identify generalization relationships} \textsl{(Rules TR47-TR50)}}\\
 \\
The sentences in flows and description sections of UCS are scanned and the sentences containing sub strings of types ``is a", ``kind of", and all their synonyms are used to identify generalization relationships. We call such sub strings as Generalization Sub String (\textit{GenSubString}). We categorise the sentences containing \textit{GenSubString} into two kinds:
\begin{enumerate}
\vspace{2ex}
\item \textit{\textit{Child-\textit{GenSubString}-Parent}} sentences: The sentences containing sub strings ``is a", ``type of", ``kind of", and all their synonyms are referred as \textit{\textit{Child-\textit{GenSubString}-Parent} sentences} because in these sentences the child class/classes is/are present on the left of \textit{GenSubString} and the parent class is present on the right of \textit{GenSubString}. Rule TR47 is used the identify generalization relationships from these sentences.
\vspace{2ex}
\item \textit{\textit{Parent-\textit{GenSubString}-Child}} sentences: The sentences containing sub strings ``parent of", ``categorized into", ``has types", ``of types", ``classified into", ``classified as" and all their synonyms are referred as \textit{Parent-\textit{GenSubString}-Child} sentences because in these sentences the parent class is present on the left of \textit{GenSubString} and the child class/classes is/are present on the right of \textit{GenSubString}. Rule TR48 presented in Appendix~B is used the identify generalization relationships from these sentences.\\
\end{enumerate}
\textbf{\textsl{Rule-TR47}}: For each sentence of type \textit{Child-\textit{GenSubString}-Parent}, the POS-tags of the sentence are scanned and \\  \setlength\parindent{60pt}
\indent    parentClass=createClass(noun nr on the right of \textit{GenSubString},``${<<}$entity class${>>}$");\\
\indent		 For each noun nl on the left of \textit{GenSubString} \\ \setlength\parindent{70pt}
\indent    childClass=createClass(nl,``${<<}$entity class${>>}$");\\
\indent		 createRelationship(parentClass,childClass,``generalization");\\
  \setlength\parindent{60pt}
\indent    EndFor\\
\setlength\parindent{40pt}
\indent EndFor		
\\
\\
\begin{example} For sentence \textsl{``The withdrawal, deposit, transfer and query are types of transaction."} POS-tags generated by the parser are: \\
\\
\textsl{[The/DT, withdrawal/NN, ,/,, deposit/NN, ,/,, transfer/NN, and/CC, query/NN, are/VBP, types/NNS, of/IN, transaction/NNS, ./.]}\\
\\
\begin{tabular}{p{3.8in}p{2.5in}}
\noindent As this sentence contains \textit{GenSubString}=``types of" hence it is \textit{Child-\textit{GenSubString}-Parent} sentence, therefore rule TR47 is applied. From POS-tags the nouns to the left of \textit{GenSubString} \textsl{``types of"} are \textsl{withdrawal}, \textsl{deposit}, \textsl{transfer} and \textsl{query}, a child class is created for each of these nouns. And the noun to the right of \textit{GenSubString} \textsl{``types of"} is \textsl{transaction}, a parent class is created for this noun. Generalization relationship is established between the identified child classes (\textsl{withdrawal, deposit, transfer and query}) and the identified parent class (\textsl{transaction})  & \multirow{1}[10]{2.5in}{\includegraphics[width=2.5in]{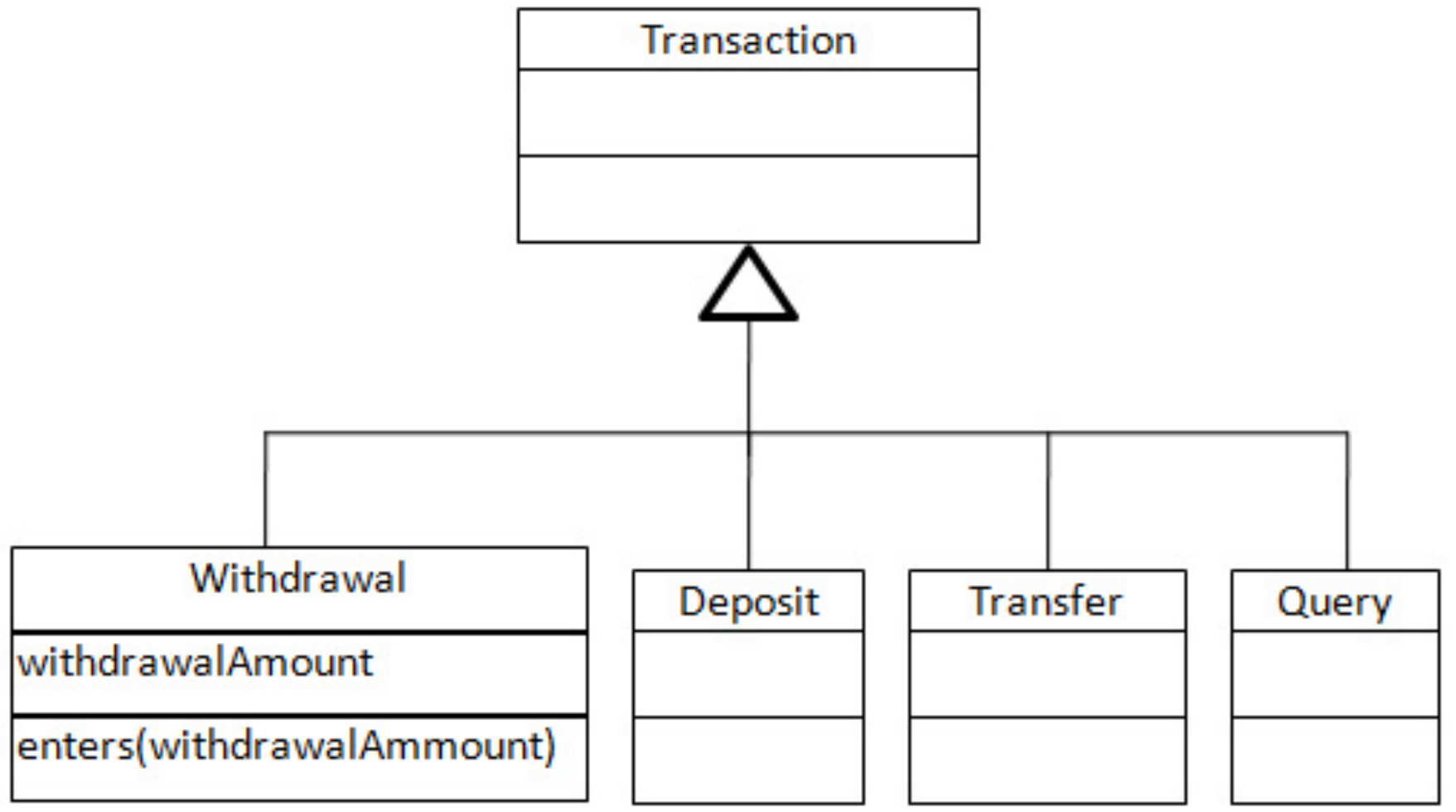} }\\
 &  \\
 \end{tabular}
\end{example}
\noindent \textbf{\textsl{Step 4.9: Identify aggregation relationships} \textsl{(Rules TR51-TR53)}}\\ \\
An aggregation relationship is a part-whole relationship between two classes, in which the part class is contained in the whole class (or part class is present as an attribute inside the whole class). The approach identifies aggregation relationships using two tactics: 1) When in ClassDiagram\_Instance, a class say C1 is present as an attribute in another class say C2, then class C1 is recognized as part class and C2 is recognized as whole class, and aggregation relationship is created between the part class and whole class (transformation rule TR51 shown below). 2) The sentences in flows and description sections of UCS containing sub strings such as ``part of", ``consists of", ``contains" and all their synonyms are the prospects of aggregation relationships. The transformation rules TR52-TR53 (given is Appendix~B) identify aggregation relationships from such sentences.\\
\\
\textbf{\textsl{Rule-TR51}}: For each of the two classes c1 and c2 in \textsl{ClassDiagram\_Instance}\\		
\setlength\parindent{60pt}
\indent       If c2 is attribute of c1 then\\		
\setlength\parindent{80pt}	
\indent             wholeClass=c1;partClass=c2;\\
\indent createRelationship(wholeClass,partClass,``aggregation"); \\			
\setlength\parindent{60pt}
\indent       EndIf\\		
\setlength\parindent{40pt}	
\indent EndFor
\begin{example} Let \textsl{ClassDiagram\_Instance} contains two classes say c1=``Book" and c2=``BookDetail" then by rule TR51, the approach creates an aggregation relationship between wholeClass=``Book" and the partClass=``BookDetail" \\
\begin{tabular}{p{4.2in}p{2in}}
 then by rule TR51, the approach creates an aggregation relationship between wholeClass=``Book" and the partClass=``BookDetail" & \multirow{1}[10]{2in}{\includegraphics[width=1.9in]{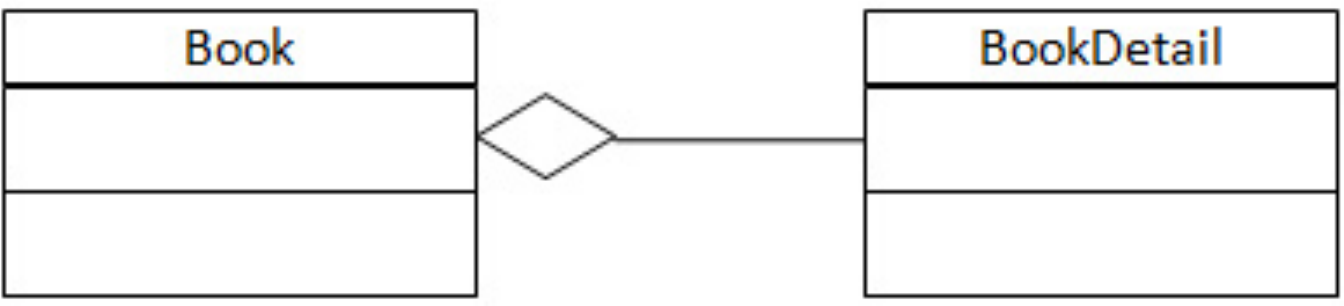} }\\
 &  \\
 \end{tabular}
\end{example}
\noindent \textbf{\textsl{Step 4.10: Eliminate extra classes} \textsl{(Rule TR54)}}\\
 \\
Every noun or entity term may not be a class, if a noun or an entity term is identified as a class by the approach, then it must participate in relationship with some other class (or classes) in the class diagram. Hence, if some classes are not participating in any relationships then those classes are extra classes and needs to be removed from the class diagram. These extra classes are removed by applying rule TR54.\\
\\
\textbf{\textsl{Rule-TR54}}: For each class c in \textsl{ClassDiagram\_Instance}, if c is not present as EndClass of any relationship r in \textsl{ClassDiagram\_Instance} then delete c from \textsl{ClassDiagram\_Instance}.\\
\\
\textbf{\textsl{Step 5: Generate and visualize the class diagram}}\\
 \\
The identified classes and relationships are then used to generate DOT language commands (commands that are used to create and visualize various graph elements such as lines, rectangles, triangles etc. in GraphViz. The class diagram is then visualized using GraphViz API.

\section{Tool support developed}
\label{sec:ToolSupport}
We have implemented the proposed approach in a prototype tool named Automatic Analysis Model Generator (AutoAMG) in Java 1.6 using Eclipse Indigo IDE release 3.7.0.
AutoAMG uses Stanford NLP parser APIs version 2.0.4\footnote{http://nlp.stanford.edu/software/lex-parser.shtml} for parsing the sentences. It uses Apache POI API 3.9\footnote{http://poi.apache.org/} for reading the UCSs written in MS Excel files. For in-place visualizing of the generated class diagrams it uses Graphviz\footnote{http://www.graphviz.org/}. It also uses a Java API\footnote{https://github.com/jabbalaci/graphviz-java-api} to call dot (GraphViz) from Java program. The current implementation supports the input of one UCS at a time, and the generation of analysis class diagram for one UCS at a time. The interactions of the given UCS with other UCSs, as specified in the given UCS using INCLUDE, EXTEND keywords and Parent Use Case Name field of the UCS, are shown in the generated analysis class diagram with the help of INCLUDE, EXTEND and generalization relationships between the control class of given UCS and the control classes of the other UCSs respectively. The tool can be easily extended take a set of UCS for a given problem as input, and to generate the analysis class diagram for the problem.
\\
\\
\noindent \textbf{ User interface provided by AutoAMG}: AutoAMG provides a GUI interface, the menu bar provides options to:
\begin{enumerate}
	\item Browse and select a UCS file for the input: Once the UCS file is selected it is read and parsed by the tool. The sentence structures of the sentences in UCS are identified and the UCS is displayed in a tab at center of screen. The sentences whose sentence structures cannot be identified are marked by the tool. The user can either modify those sentences or can continue by selecting \textsl{Generate and Visualize Analysis Class Diagram } from the menu bar.
	\item Generate and visualize analysis class diagram: When user selects this options the tool automatically generates the analysis class diagram and displays it in a new tab at the center of screen.
	\item Export class diagram as JPEG image: The user can browse to select location for storing the class diagram image file.
	\item Export class diagram as XMI file: The user can browse to select location for storing the class diagram XMI file. The XMI file can be imported in an open source tool (e.g. ArgoUML\footnote{http://argouml.tigris.org/}) for modifying the class diagram if required.\\
\end{enumerate}

\section{Experimental study}
\label{sec:ExperimentalStudy}
This section reports the controlled experiment that we conducted for comparing the analysis class diagrams generated by the proposed approach with those generated by the two existing approaches, one proposed by~\cite{popescu2008reducing} and the other proposed by~\cite{simula12,yue2015atoucan}. In the experiment the analysis class diagrams generated by the three approaches for forty UCSs were evaluated by forty subjects on the basis of correctness, completeness and redundancy of the analysis class diagrams generated by them. The procedure followed for conducting the experiment were based the guidelines for experimental studies presented in~\citep{wohlin2003empirical,sjoberg2005survey,wohlin2012experimentation}

Here, we present the details of the experiment. Section~\ref{subsec:GoalResearchQuestions} presents the goal of the experiment and the research question. Section~\ref{subsec:Metrics} presents the definition of the metrics used for assessing the correctness, completeness and redundancy of the analysis class diagrams. Section~\ref{subsec:HypothesisFormulation} states the hypothesis based on the research questions. Section~\ref{subsec:Variables} describes the variables used. Section~\ref{subsec:SubjectObject} presents subject and object selection. Section~\ref{subsec:ExperimentalDesign} introduces the experimental design. Section~\ref{subsec:ResultsAnalysis} presents the study results and describes the analysis procedure. Section~\ref{subsec:ValidityConsideration} discusses the validity considerations.

\subsection{Goal and research question}
\label{subsec:GoalResearchQuestions}
The objective of the experiment was to compare the quality (correctness, completeness and redundancy) of the analysis class diagrams generated by the three approaches viz. the proposed approach,~\cite{popescu2008reducing} and~\cite{simula12,yue2015atoucan}. Hence we framed following research question:
\begin{itemize}
\item \textbf{RQ}: Are the three approaches differ significantly in terms of the quality (class diagram correctness ($CD_{cr}$), class diagram completeness ($CD_{cm}$) and  class diagram redundancy ($CD_{rd}$)) of the analysis class diagrams generated by them? If yes, then what are the significant differences between them in terms of the quality?\\
\end{itemize}

\subsection{Metrics used to assess the quality}
\label{subsec:Metrics}
 The metrics used for assessing the quality of the analysis class diagrams along with how to use them for the assessment are as follows:
\begin{enumerate}
	  \item Class diagram correctness ($CD_{cr}$): Class diagram correctness of an analysis class diagram is found in terms of Average class correctness ($AvgC_{cr}$) and Average relationship correctness ($AvgR_{cr}$). The Average class correctness is the average of the Class correctness ($C_{cr}$) of all the classes in the class diagram. The Average relationship correctness is the average of the Relationship correctness ($R_{cr}$) of all the relationships in the class diagram.\\		
			$CD_{cr}$ = ( $AvgC_{cr}$ + $AvgR_{cr}$ ) / 2
		\begin{itemize}
				\item Class correctness ($C_{cr}$): is the proportion of the correctness of various class elements like class name, class stereotype, class attributes and class operations.\\
				$C_{cr}$ = ($C_{cr1}$ + $C_{cr2}$ + $C_{cr3}$ + $C_{cr4}$ + $C_{cr5}$) / 5,
				So to calculate the correctness of a class, we need to find the values of variables $C_{cr1}$, $C_{cr2}$, $C_{cr3}$, $C_{cr4}$ and $C_{cr5}$, which are found as follows:
				\begin{itemize}
				\item	Correctly identified as class ($C_{cr1}$) =1, if the identified class represents some significant concept or thing of the problem domain for which a separate class is required in the analysis class diagram, 0 otherwise.
				\item Correctly named ($C_{cr2}$) = 1, if the name assigned to the class is correct, 0 otherwise.
        \item Correctly stereotyped ($C_{cr3}$) =1, if a correct stereotype (\textless \textless entity \textgreater \textgreater, \textless \textless boundary \textgreater \textgreater or \textless \textless control \textgreater \textgreater) is assigned to the class, 0 otherwise.
				\item Proportion of correctly identified attributes ($C_{cr4}$) = No. of correctly identified attributes in the class /
Total no. of identified attributes in the class, if the Total no. of identified attributes in the class \textgreater 0, 0 otherwise.
				\item Proportion of correctly identified operations ($C_{cr5}$) = No. of correctly identified operations in the class / Total no. of identified operations in the class, if the Total no. of identified operations in the class \textgreater 0, 0 otherwise.
        \end{itemize}
				\item Relationship correctness ($R_{cr}$): is the proportion of the correctness of various relationship elements like relationship name, relationship type (association, generalization or aggregation), navigability, End-Class1 and End-Class2 (ends of the relationships).\\
				$R_{cr}$ = ($R_{cr1}$ + $R_{cr2}$ + $R_{cr3}$ + $R_{cr4}$ + $R_{cr5}$ + $R_{cr6}$) / 6,
			  So to calculate Relationship correctness of a relationship between two classes, we need to find the values of the variables $R_{cr1}$, $R_{cr2}$, $R_{cr3}$, $R_{cr4}$, $R_{cr5}$ and $R_{cr6}$, which can be found as follows:
				\begin{itemize}
					\item Correctly assigned End-Class1 ($R_{cr1}$) = 1, if the relationship end point one is correctly identified, 0 otherwise.
					\item Correctly assigned End-Class2 ($R_{cr2}$) = 1, if the relationship end point two is correctly identified, 0 otherwise.
					\item Correctly identified as relationship ($R_{cr3}$) = 1, if identified relationship represents some significant relationship which is required between the two end classes in the analysis class diagram, 0 otherwise (more than one relationship of same relationship type and navigability between the two same end classes are also considered as incorrect relationships)
					\item Correctly named ($R_{cr4}$) = 1, if the relationship name is correctly identified, 0 otherwise.
					\item Correctly identified relationship type ($R_{cr5}$) = 1, if the relationship type (association, generalization or aggregation) is correctly identified for the relationship, 0 otherwise.
					\item Correctly assigned navigability ($R_{cr6}$) =1, if the navigability of the relationship is correctly identified, 0 otherwise.\\
				\end{itemize}
		\end{itemize}
		\item Class diagram completeness ($CD_{cm}$):	Class diagram completeness of an analysis class diagram is measured as the average of Class operation completeness ($CO_{cm}$) and Relationship completeness ($R_{cm}$)\\		
		$CD_{cm}$ = ( $CO_{cm}$ + $R_{cm}$ ) / 2
		\begin{itemize}
			\item Class operation completeness ($CO_{cm}$): Class operation completeness is measured in terms of the No. of sentences in the functional requirements whose functionalities are assigned as operations to some class (or classes) in the class diagram ($N_{sf}$). \\
			$CO_{cm}$ = $N_{sf}$ / $N_{s}$, where $N_{s}$ is the total number of sentences in the functional requirements
			\item Relationship completeness ($R_{cm}$): Relationship completeness is measured in terms of the No. of separate groups of classes and relationships in the class diagram ($N_{sg}$).\\
			  $R_{cm}$ = 1-($N_{sg}$-1)/($N_{c}$-1),  if $N_{r}$\textgreater0, 0 otherwise (where $N_{c}$ is the total number of classes in the class diagram and $N_{r}$ is the total number of relationships in the class diagram) \\
		\end{itemize}		
			\item Class diagram redundancy ($CD_{rd}$): Class diagram redundancy of an analysis class diagram is measured as the average of Class redundancy ($C_{rd}$) and Relationship redundancy ($R_{rd}$).\\  	
			$CD_{rd}$ = ( $C_{rd}$ + $R_{rd}$ ) / 2
		\begin{itemize}
			\item Class redundancy ($C_{rd}$) : Class redundancy is found in terms of No. of redundant or extra classes in the class diagram ($N_{rc}$). A class is considered as a redundant class if it does not participate in any relationship with other classes in the class diagram or if it is an incorrectly identified class. Redundant classes are extra classes that are identified by an approach (those classes that are not needed). The extra classes can be of two types: i) the class that do not participate in any relationship to other class/classes in the class diagram. ii) An incorrect class (a class which was identified by the approach due to some mistake) is also an extra class (this class should not be identified by the approach hence it is extra or redundant class) \\			
			 $C_{rd}$ = $N_{rc}$ / $N_{c}$,  if $N_{c}$ \textgreater 0, 0 otherwise (where, $N_{c}$ = Total no. of classes in the class diagram)
			\item Relationship redundancy ($R_{rd}$): Relationship redundancy is found in terms of No. of redundant or extra relationships in the class diagram ($N_{rr}$). More than one relationships of same relationship type and navigability between the two same end classes are considered as redundant relationships. An incorrect relationship is also considered as redundant relationship because it is identified by the approach my mistake so it is also extra or redundant relationship.\\			
			 $R_{rd}$ = $N_{rr}$ / $N_{r}$,  if $N_{r}$ \textgreater 0, 0 otherwise (where, $N_{r}$ = Total no. of relationships in the class diagram)
			\end{itemize}
	\end{enumerate}

\subsection{Hypothesis formulation}
\label{subsec:HypothesisFormulation}
We hypothesize that there are differences between the quality (correctness, completeness and redundancy) of the analysis class diagrams generated by the three approaches (Popescu et al. approach, Yue et al. approach and Our approach)(Table~\ref{tab:HypothesisFormulation}).

\begin{table}[htb]
  \centering
	\scriptsize
  \caption{Hypothesis formulation}
    \begin{tabular}{p{1.2cm}p{7.2cm}p{6.2cm}}
    \hline
    \textbf{Hypothesis} & \textbf{Null hypothesis} & \textbf{Alternate hypothesis} \bigstrut \\
    \hline
    Hypothesis~1 & H1-0: There are no significant differences between the correctness of the analysis class diagrams generated by the three approaches.  & H1-1: The three approaches differ significantly in terms of the correctness of the analysis class diagrams generated by them. \\
          & $CD_{cr}$(Popescu et al.) = $CD_{cr}$(Yue et al.) = $CD_{cr}$(Our approach) & \bigstrut \\
    Hypothesis~2 & H2-0: There are no significant differences between the completeness of the analysis class diagrams generated by the three approaches.  & H2-1: The three approaches differ significantly in terms of the completeness of the analysis class diagrams generated by them. \\
          & $CD_{cm}$(Popescu et al.) = $CD_{cm}$(Yue et al.) = $CD_{cm}$(Our approach) & \bigstrut \\
    Hypothesis~3 & H3-0: There are no significant differences between the redundancy in the analysis class diagrams generated by the three approaches.  & H3-1: The three approaches differ significantly in terms of redundancy of the analysis class diagrams generated by them. \bigstrut \\
          & $CD_{rd}$(Popescu et al.) = $CD_{rd}$(Yue et al.) = $CD_{rd}$(Our approach) &   \\
    \hline
    \end{tabular}%
  \label{tab:HypothesisFormulation}%
\end{table}%

\subsection{Variables}
\label{subsec:Variables}
Following are the independent and the dependent variables of the experimental study
\begin{enumerate}
	\item Independent variables: The analysis class diagrams generated using the three approaches.
	\item Dependent variables: Class diagram correctness ($CD_{cr}$),  Class diagram completeness ($CD_{cm}$) and Class diagram redundancy ($CD_{rd}$) of the analysis class diagrams generated by the three approaches.
\end{enumerate}

\subsection{Subject and object selection}
\label{subsec:SubjectObject}
\begin{figure}[htb]
\centering
\includegraphics[width=\linewidth,height=135mm]{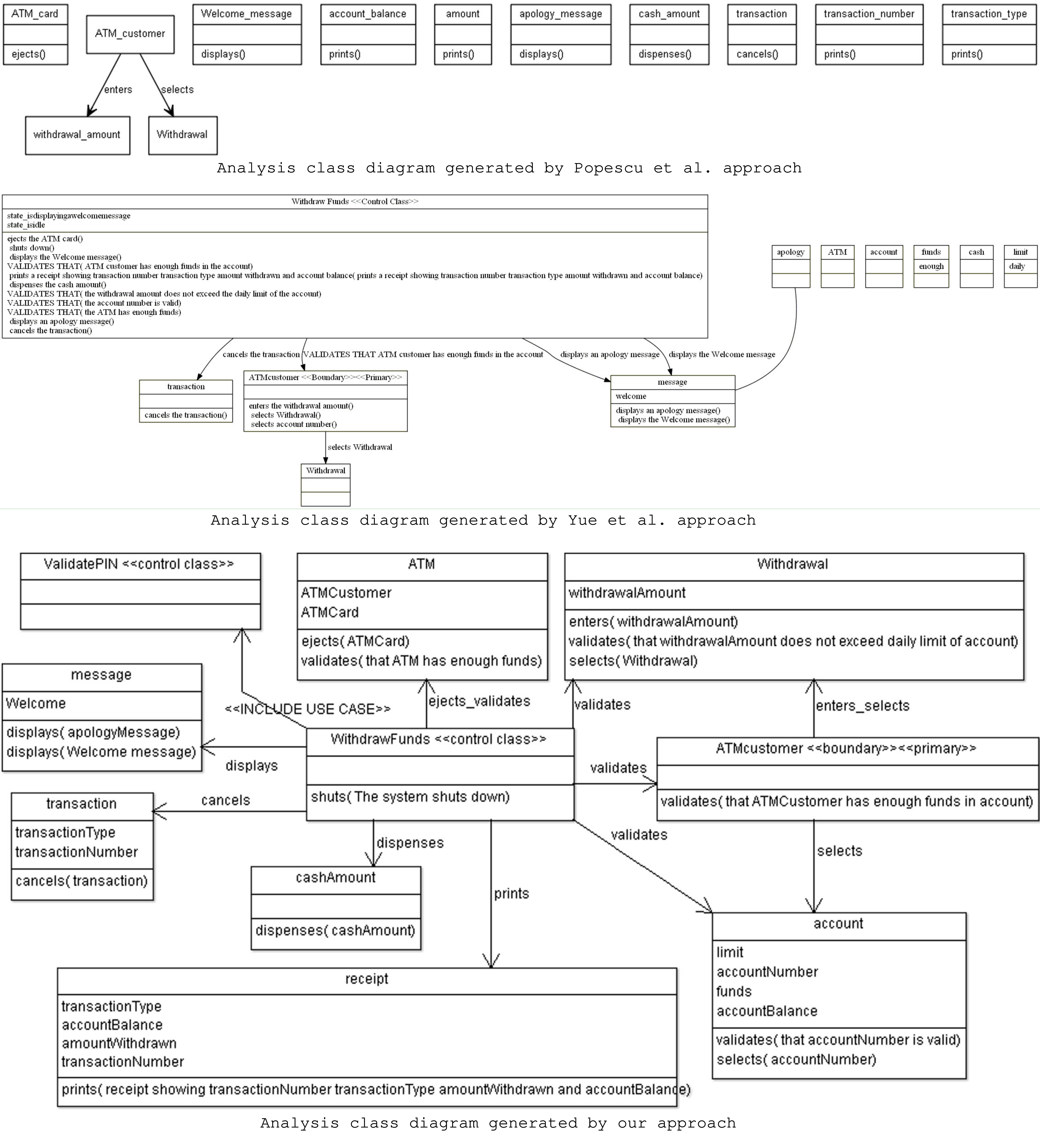}
\caption{Analysis class diagrams generated by Popescu et al. approach, Yue et al. approach and our approach for the UCS Withdraw Fund}
\label{fig:CDWithdrawFundXYZ}
\end{figure}

\noindent \textbf{\emph{Subject selection}}: The subjects (or participants) in the experiment were the forty students of Computer Science and Engineering discipline from Indian Institute of Information Technology Design \& Manufacturing Jabalpur, India. The subjects were a mix of Ph.D. scholars,  M.Tech. final year students and B.Tech final year students. All the subjects had done a basic course in \textsl{Software Engineering} as well as either \textsl{Object-Oriented Software Engineering} course or \textsl{Object-Oriented Analysis and Design} course, both of which required a course project which needed them to develop object-oriented analysis and design models for some specific problems. So the subjects were already familiar with object oriented analysis and modeling. Additionally, we conducted a training session of two hours to brush up their analysis modeling concepts. The training was not biased to any approach evaluated in the experimental study. The training given to the subject was very general in which the concepts for identifying the objects, their attributes and operations, and the relationships between the objects were based on the heuristics proposed by Abbott and Grady Booch et al. The concepts and understanding of the subjects were then tested through two different exercises each of two hours. In first exercise they were given a UCS and were asked to create an analysis class diagram for the UCS. In the second exercise they were given a UCS with its corresponding analysis class diagram, and were told to fill the same set of questionnaire for quality measures of analysis class diagrams, as used in the experiment. The answers to the questionnaires were checked and the mistakes done by the students were told to them, so that they don't make the same mistakes in the experiment.\\
\\
\textbf{\emph{Object selection}}: The objects were the analysis class diagrams generated by the three approaches viz. \cite{popescu2008reducing},~\cite{simula12,yue2015atoucan} and our approach for forty UCSs. We took these forty UCSs from various software engineering books~\citep{booch2010object,mike2005OOAD,gomaa2011software,doug2008use,rosenberg2007use,bruegge1999object} and research works~\citep{liu2004natural,popescu2008reducing,simula12,yue2015atoucan,deeptimahanti2011semi}. Figure~\ref{fig:CDWithdrawFundXYZ} presents the analysis class diagrams which were generated using the three approaches for one of the UCS named \textsl{Withdraw Fund} shown in Table~\ref{tab:UseCaseWithdrawFund}.

\subsection{Experimental design}
\label{subsec:ExperimentalDesign}
The experiment design was a complete block design in which each subject evaluated the analysis class diagrams of all the three approaches for a UCS provided to him/her. In all there were forty subjects and forty UCSs, each subject was given one UCS, the class diagrams obtained from the three approaches for the given UCS and a set of questionnaires with each class diagram. These experimental materials were randomly distributed to the subjects. The blocking variable was the order in which the questionnaire for the approaches to be answered, so to nullify the ordering effect we asked the subjects to answer the questionnaires for the three approaches in the orders presented in Table~\ref{tab:OrderOfEvaluation}. In the experiment the subjects first read the UCS, then examining the analysis class diagrams with respect to the UCS answered to the given questionnaires regarding the correctness, completeness and redundancy of the analysis class diagrams. The questionnaires were based on the quality measures for evaluating correctness, completeness and redundancy of analysis class diagrams presented in Section~\ref{subsec:Metrics}\\
\begin{table}[htb]
  \centering
  \caption{Order of the evaluation for the class diagrams of the three approaches to nullify the effect of blocking variable - Ordering effect}
    \begin{tabular}{ccccc}
    \hline
    \multirow{2}[4]{*}{\textbf{Students seat \#}} & \multirow{2}[4]{*}{\textbf{Number of students}} & \multicolumn{3}{c}{\textbf{Order of evaluation}} \bigstrut\\
\cline{3-5}          &       & \textbf{First} & \textbf{Second} & \textbf{Third} \bigstrut\\
    \hline
    1 to 7 & 7     & Popescu et al. & Yue et al. & Our approach \bigstrut[t]\\
    8 to 14 & 7     & Popescu et al. & Our approach & Yue et al. \\
    15 to 21 & 7     & Yue et al. & Popescu et al. & Our approach \\
    21 to 28 & 7     & Yue et al. & Our approach & Popescu et al. \\
    29 to 34 & 6     & Our approach & Popescu et al. & Yue et al. \\
    35 to 40 & 6     & Our approach & Yue et al. & Popescu et al. \\
		\hline
    \end{tabular}%
	
  \label{tab:OrderOfEvaluation}%
\end{table}%

\noindent \textbf{\emph{Instrumentation/Experimental-material}}: The experiment material provided to each subject was:\\
1. A use case specification (UCS). (A sample UCS is shown in Table~\ref{tab:UseCaseWithdrawFund})\\
2. The analysis class diagrams generated by the three approaches for the given UCS. (The samples of analysis class diagrams generated by the three approaches for the UCS Withdraw Fund is shown in Figure~\ref{fig:CDWithdrawFundXYZ})\\
3. Questionnaires for evaluating correctness, completeness and redundancy of analysis class diagrams of each approach. The questionnaires were based on the class diagram quality measures presented in Section~\ref{subsec:Metrics}. The sample questionnaires for correctness, completeness and redundancy of the analysis class diagram generated by one of the approach for UCS Withdraw Fund are given in Appendix~\ref{sec:sampleQuestionaires}.

\subsection{Execution of the experiment}
The experimental material (a UCS, three analysis class diagrams and three set of questionnaires one for each class diagram) were randomly distributed to the subjects. The subjects were clearly explained the steps for reading the UCS, examining the analysis class diagrams and filling the answers to the questionnaires. We provided the subjects maximum of three hours time for completing the experiment (In a prior mock experiment with four UCSs of different sizes taken from the 40 UCSs used in the experimental study, we found that the time needed to answer the questionnaires was 1 hour 20 minutes for the UCS of smallest size and 2 hours 45 minutes for the UCS of largest size. So we decided the maximum time of 3 hours for the experiment.). Each subject had to answer the questionnaires for all the three analysis class diagrams provided to him/her. In the experiment each subject first read the UCS provided to him/her, then following the order shown in Table~\ref{tab:OrderOfEvaluation} examined the analysis class diagrams and filled the answers to the given questionnaires. The subjects were not allowed to talk and share the answers with each other during the experiment. As soon as a subject  completed his/her experiment, the answers to questionnaire were collected from him/her, and the subject is allowed to move out.
	
\subsection{Results \& Analysis}
\label{subsec:ResultsAnalysis}
\begin{figure}[htb]
\centering
\includegraphics[width=\linewidth]{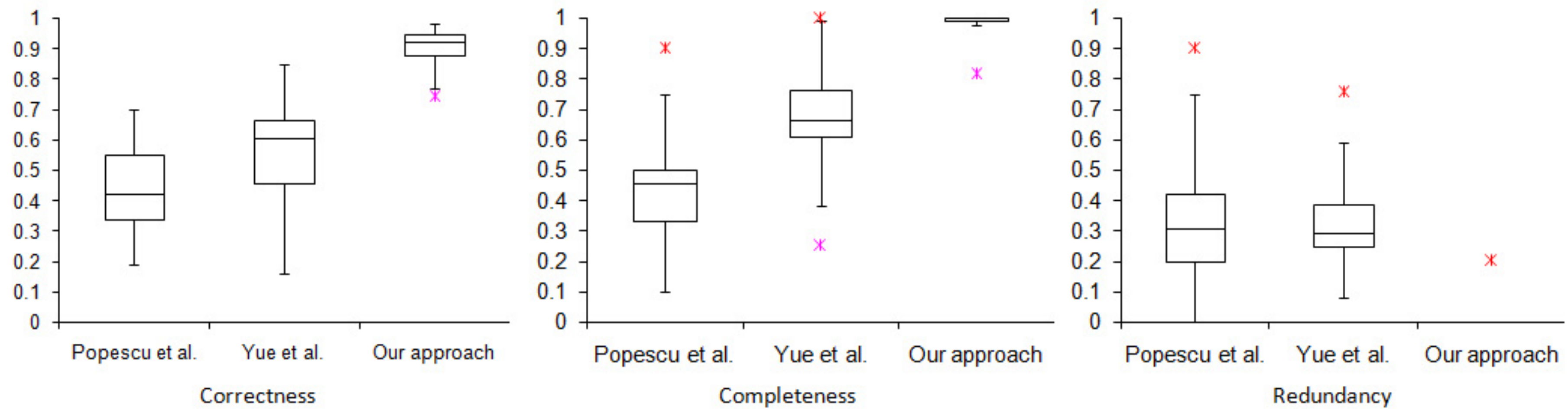}
\caption{Box plot for the correctness, completeness and redundancy of the analysis class diagrams - showing the IQR, whiskers (1.5*IQR) , and max/min outliers.}
\label{fig:BoxPlot}
\end{figure}

This section analyzes the data set obtained from the experiment. Table~\ref{tab:ExperimentResults} shows the data set for the correctness, completeness and redundancy of the analysis class diagrams of the three approaches for each UCS. This data set is obtained from the answers to the questionnaires collected from the subjects. The variability of the data for the correctness, completeness and redundancy of the analysis class diagrams generated by the approaches for the forty UCSs are shown using box plots in Figure~\ref{fig:BoxPlot}.\\
\begin{table}[htb]

	\scriptsize
   \caption{The correctness, completeness and redundancy of the analysis class diagrams generated by the three approaches for the forty UCSs.}
    \begin{tabular}{clccccccccccc}
		\hline
   \multicolumn{1}{c}{\multirow{2}[4]{*}{\textbf{S\#}}} & \multicolumn{1}{l}{\multirow{2}[4]{*}{\textbf{UseCaseName}}} & \multicolumn{3}{c}{\textbf{CD Correctness ($CD_{cr}$)}}&  & \multicolumn{3}{c}{\textbf{CD Completeness ($CD_{cm}$)}} & & \multicolumn{3}{c}{\textbf{CD Redundancy ($CD_{rd}$)}}  \bigstrut\\
\cline{3-5} \cline{7-9}\cline{11-13}   \multicolumn{1}{c}{} & \multicolumn{1}{c}{} &\multicolumn{1}{p{.5cm}}{\textbf{Popescu et al.}}&\multicolumn{1}{p{.7cm}}{\textbf{Yue et al.}}&\multicolumn{1}{p{.9cm}}{\textbf{Our approach}}&\multicolumn{1}{p{.01cm}}{\textbf{ }}&\multicolumn{1}{p{.5cm}}{\textbf{Popescu et al.}} & \multicolumn{1}{p{.7cm}}{\textbf{Yue et al.}} & \multicolumn{1}{p{.9cm}}{\textbf{Our approach}}&\multicolumn{1}{p{.01cm}}{\textbf{ }} & \multicolumn{1}{p{.5cm}}{\textbf{Popescu et al.}} & \multicolumn{1}{p{.7cm}}{\textbf{Yue et al.}} & \multicolumn{1}{p{.9cm}}{\textbf{Our approach}} \bigstrut\\
    \hline \bigstrut
		~1     & AGV Move to Station & 0.55  & 0.34  & 0.90  &       & 0.48  & 0.50  & 1.00  &       & 0.35  & 0.50  & 0.00 \\
    2     & Arena Announce Tournament & 0.57  & 0.51  & 0.98  &       & 0.27  & 0.87  & 0.96  &       & 0.00  & 0.31  & 0.00 \\
    3     & ATM QueryAccount & 0.70  & 0.83  & 0.90  &       & 0.50  & 0.68  & 1.00  &       & 0.00  & 0.14  & 0.00 \\
    4     & ATM TransferFund & 0.60  & 0.62  & 0.93  &       & 0.44  & 0.82  & 1.00  &       & 0.20  & 0.25  & 0.00 \\
    5     & ATM ValidatePIN & 0.52  & 0.52  & 0.93  &       & 0.52  & 0.70  & 1.00  &       & 0.25  & 0.49  & 0.00 \\
    6     & ATM Withdraw Funds & 0.20  & 0.65  & 0.95  &       & 0.17  & 0.72  & 1.00  &       & 0.50  & 0.31  & 0.00 \\
    7     & ATM Withdrawl Transaction UCDA & 0.55  & 0.73  & 0.96  &       & 0.60  & 0.83  & 1.00  &       & 0.37  & 0.15  & 0.00 \\
    8     & Elevator Dowser & 0.59  & 0.75  & 0.95  &       & 0.50  & 0.44  & 0.82  &       & 0.44  & 0.29  & 0.00 \\
    9     & Elevator Request Elevator & 0.41  & 0.60  & 0.93  &       & 0.51  & 0.92  & 1.00  &       & 0.17  & 0.30  & 0.00 \\
    10    & Elevator Select Destination & 0.42  & 0.58  & 0.91  &       & 0.41  & 1.00  & 1.00  &       & 0.20  & 0.19  & 0.00 \\
    11    & EMS Generate Alarm & 0.19  & 0.68  & 0.95  &       & 0.63  & 0.75  & 1.00  &       & 0.64  & 0.32  & 0.00 \\
    12    & EMS Generate Monitoring Data & 0.38  & 0.58  & 0.90  &       & 0.46  & 0.63  & 1.00  &       & 0.25  & 0.25  & 0.00 \\
    13    & EMS View Alarms & 0.38  & 0.66  & 0.93  &       & 0.63  & 0.64  & 1.00  &       & 0.42  & 0.31  & 0.00 \\
    14    & EMS View Monitoring Data & 0.39  & 0.20  & 0.74  &       & 0.50  & 0.50  & 1.00  &       & 0.35  & 0.50  & 0.20 \\
    15    & iCoot Browse Index & 0.48  & 0.75  & 0.90  &       & 0.50  & 0.50  & 0.88  &       & 0.29  & 0.25  & 0.00 \\
    16    & iCoot Cancel Reservation & 0.43  & 0.47  & 0.93  &       & 0.41  & 0.73  & 1.00  &       & 0.42  & 0.33  & 0.00 \\
    17    & iCoot Change Password & 0.70  & 0.65  & 0.93  &       & 0.48  & 0.80  & 1.00  &       & 0.13  & 0.38  & 0.00 \\
    18    & iCoot Log Off & 0.30  & 0.85  & 0.93  &       & 0.33  & 0.50  & 1.00  &       & 0.50  & 0.25  & 0.00 \\
    19    & iCoot LogOn & 0.59  & 0.85  & 0.85  &       & 0.56  & 0.39  & 1.00  &       & 0.17  & 0.25  & 0.00 \\
    20    & iCoot Make Reservations & 0.40  & 0.65  & 0.93  &       & 0.53  & 0.72  & 1.00  &       & 0.31  & 0.46  & 0.00 \\
    21    & iCoot Search & 0.70  & 0.66  & 0.96  &       & 0.90  & 0.90  & 1.00  &       & 0.00  & 0.08  & 0.00 \\
    22    & iCoot View Car Model Details & 0.40  & 0.41  & 0.88  &       & 0.20  & 0.65  & 1.00  &       & 0.38  & 0.29  & 0.00 \\
    23    & iCoot View Member Details & 0.30  & 0.64  & 0.82  &       & 0.25  & 0.67  & 1.00  &       & 0.00  & 0.29  & 0.00 \\
    24    & iCoot View Rentals. & 0.38  & 0.16  & 0.87  &       & 0.50  & 0.25  & 1.00  &       & 0.42  & 0.50  & 0.00 \\
    25    & iCoot View Results & 0.30  & 0.64  & 0.83  &       & 0.25  & 0.38  & 1.00  &       & 0.50  & 0.30  & 0.00 \\
    26    & Internet Book Store Show Book Details & 0.33  & 0.33  & 0.88  &       & 0.33  & 0.50  & 0.82  &       & 0.33  & 0.50  & 0.00 \\
    27    & Internet Book Store Write Review & 0.67  & 0.61  & 0.97  &       & 0.38  & 0.65  & 0.96  &       & 0.17  & 0.40  & 0.00 \\
    28    & JEWEL Zoom Map & 0.19  & 0.84  & 0.97  &       & 0.28  & 0.81  & 0.95  &       & 0.90  & 0.18  & 0.00 \\
    29    & MyTrip ExecuteTrip & 0.34  & 0.52  & 0.74  &       & 0.44  & 0.67  & 1.00  &       & 0.50  & 0.29  & 0.10 \\
    30    & MyTrip PlanTrip & 0.30  & 0.82  & 0.92  &       & 0.32  & 0.61  & 1.00  &       & 0.48  & 0.35  & 0.00 \\
    31    & OSS Browse Catalog & 0.30  & 0.42  & 0.98  &       & 0.10  & 0.90  & 1.00  &       & 0.50  & 0.17  & 0.00 \\
    32    & OSS Process Delivery Order & 0.69  & 0.57  & 0.80  &       & 0.50  & 0.69  & 0.94  &       & 0.13  & 0.20  & 0.00 \\
    33    & Print Pack Types & 0.39  & 0.63  & 0.91  &       & 0.50  & 0.75  & 1.00  &       & 0.25  & 0.23  & 0.00 \\
    34    & QVS Perform Verification & 0.43  & 0.44  & 0.95  &       & 0.36  & 0.68  & 1.00  &       & 0.39  & 0.48  & 0.00 \\
    35    & Ticket Distributor PurchaseTicket & 0.52  & 0.80  & 0.82  &       & 0.37  & 0.93  & 0.86  &       & 0.25  & 0.15  & 0.12 \\
    36    & TTMS Monitor Train Systems & 0.46  & 0.46  & 0.94  &       & 0.33  & 0.64  & 1.00  &       & 0.29  & 0.38  & 0.00 \\
    37    & TTMS Route Train & 0.32  & 0.23  & 0.94  &       & 0.48  & 0.65  & 0.95  &       & 0.64  & 0.76  & 0.00 \\
    38    & VTS Cancel Approved Request & 0.48  & 0.43  & 0.96  &       & 0.45  & 0.61  & 0.96  &       & 0.31  & 0.25  & 0.00 \\
    39    & VTS Edit Pending Request & 0.48  & 0.44  & 0.92  &       & 0.46  & 0.63  & 1.00  &       & 0.21  & 0.29  & 0.00 \\
    40    & VTS Withdraw Request & 0.48  & 0.48  & 0.86  &       & 0.51  & 0.64  & 1.00  &       & 0.29  & 0.50  & 0.00 \\ \hline \bigstrut
          & \textbf{Average} & \textbf{0.48} & \textbf{0.58} & \textbf{0.91} & \textbf{} & \textbf{0.43} & \textbf{0.67} & \textbf{0.98} & \textbf{} & \textbf{0.32} & \textbf{0.32} & \textbf{0.01} \\ \hline
    \end{tabular}%
   \label{tab:ExperimentResults}%
\end{table}%

\noindent \textbf{\emph{Hypothesis Testing}}: To choose the right statistical test for testing experimental hypotheses we applied \textsl{Kolmogorov-Smirnov} test~\citep{massey1951kolmogorov} to check the normality of the obtained data, the results of the test give no evidence for the data to be normal. Hence we decided to apply a non parametric test for testing experimental hypotheses, we chose to apply \textsl{Friedman} test as our experimental design, procedure applied for the execution of experiment and the data collection methods fulfilled all the conditions for applying \textsl{Friedman} test on the data~\citep{corder2009nonparametric,sheskin2003handbook}. \\

\noindent \textbf{\emph{Testing of hypotheses 1, 2 and 3}}: We applied the \textsl{Friedman} test on the data obtained from the experiment for correctness, completeness and redundancy of the analysis class diagrams of the three approaches. The results of the test as shown in Table~\ref{tab:FriedmanTestCrCmRd} show that the significant value (or p-value) for class diagram correctness ($CD_{cr}$), class diagram completeness ($CD_{cm}$) and class diagram redundancy ($CD_{rd}$) of the analysis class diagrams of the three approaches obtained from the test are 0.000, 0.000 and 0.000, respectively which are all less than the 0.01 (p$<0.01$), which gives clear evidence (with a confidence level of 99\%) for the rejection of the null hypotheses H01, H02 and H03, respectively. Hence the analysis class diagrams of the three approaches significantly differ in terms of correctness (alternate hypothesis H11 hold), significantly differ in terms of completeness (alternate hypothesis H12 hold) and significantly differ in terms of redundancy (alternate hypothesis H13 hold).

To find which approaches are significantly better than the other approaches in terms of correctness, completeness and redundancy of their analysis class diagrams, we applied the post hoc \textsl{Friedman} tests for pair wise comparison of the three approaches. The results of the post hoc tests for correctness, completeness and redundancy are shown in Tables~\ref{tab:FriedmanPostHocTestCr}, \ref{tab:FriedmanPostHocTestCm} and \ref{tab:FriedmanPostHocTestRd}, respectively. The results showed that there is no significant difference between the correctness of analysis class diagrams of~\cite{popescu2008reducing} and~\cite{simula12,yue2015atoucan} approaches, whereas the correctness of the analysis class diagrams of our approach were more than those of Popescu et al. approach and Yue et al. approach. The completeness of analysis class diagrams generated by Yue et al. approach were more than those generated Popescu et al. approach, whereas the completeness of analysis class diagrams generated by our approach were more than those by Popescu et al. approach and Yue et al. approach. There was no significant difference between the redundancy in analysis class diagrams generated by Popescu et al. approach and Yue et al. approach, whereas the redundancy in the analysis class diagrams generated by our approach were less than those by Popescu et al. approach and Yue et al. approach. Hence the results clearly showed that for the forty UCSs the analysis class diagrams generated by our approach were more correct, more complete and less redundant than those generated by the other two approaches. Specifically, from Table~\ref{tab:ExperimentResults} one can note that the analysis class diagrams generated by the proposed approach were 91\% correct, 98\% complete and 01\% redundant, whereas those generated by Yue et al. approach were 58\% correct, 67\% complete and 32\% redundant, and those generated by Popsecu et al. approach were 45\% correct, 43\% complete and 32\% redundant.
\begin{table}[htb]
  \centering
	\scriptsize
  \caption{The Friedman test for the class diagram correctness ($CD_{cr}$), completeness ($CD_{cm}$) and redundancy ($CD_{rd}$) of the three approaches}
    \begin{tabular}{p{.35in}p{2.4in}p{.1in}p{.22in}p{.15in}p{.67in}p{1.18in}}
    \hline
    \textbf{Response variable} & \textbf{Null hypothesis} & \textbf{N} & \textbf{$\chi^2$} & \textbf{p value} & \textbf{Test Result} & \textbf{Conclusion} \bigstrut\\
\hline
    \multirow{2}[4]{*}{$CD_{cr}$} & H01: There are no significant differences between the correctness of the analysis class diagrams generated by the three approaches. & \multicolumn{1}{c}{\multirow{2}[4]{*}{40}} & \multirow{2}[4]{*}{60.843} & \multirow{2}[4]{*}{\textbf{.000}} & \multicolumn{1}{p{.67in}}{\multirow{2}[4]{.67in}{p $<$ 0.01, Enough evidence to reject null hypothesis}} & \multicolumn{1}{p{1.18in}}{\multirow{2}[4]{1.18in}{The correctness of the analysis class diagram generated by the three approaches differ significantly}} \bigstrut\\
          & $CD_{cr}$(Popescu et al.)=$CD_{cr}$(Yue et al.)=$CD_{cr}$(Our approach) & \multicolumn{1}{c}{} &       &       & \multicolumn{1}{l}{} & \multicolumn{1}{l}{} \bigstrut\\

    \multicolumn{1}{l}{\multirow{2}[4]{*}{$CD_{cm}$}} & H02: There are no significant differences between the completeness of the analysis class diagrams generated by the three approaches.  & \multicolumn{1}{c}{\multirow{2}[4]{*}{40}} & \multirow{2}[4]{*}{72.731} & \multirow{2}[4]{*}{\textbf{.000}} & \multicolumn{1}{p{.67in}}{\multirow{2}[4]{.67in}{p $<$ 0.01, Enough evidence to reject null hypothesis}} & \multicolumn{1}{p{1.18in}}{\multirow{2}[4]{1.18in}{The completeness of the analysis class diagram generated by the three approaches differ significantly}} \bigstrut\\
    \multicolumn{1}{l}{} & $CD_{cm}$(Popescu et al.)=$CD_{cm}$(Yue et al.)=$CD_{cm}$(Our approach) & \multicolumn{1}{c}{} &       &       & \multicolumn{1}{l}{} & \multicolumn{1}{l}{} \bigstrut\\

    \multicolumn{1}{l}{\multirow{2}[4]{*}{$CD_{cr}$}} & H03: There are no significant differences between the redundancy in the analysis class diagrams generated by the three approaches.  & \multicolumn{1}{c}{\multirow{2}[4]{*}{40}} & \multirow{2}[4]{*}{56.219} & \multirow{2}[4]{*}{\textbf{.000}} & \multicolumn{1}{p{.67in}}{\multirow{2}[4]{.67in}{p $<$ 0.01, Enough evidence to reject null hypothesis}} & \multicolumn{1}{p{1.18in}}{\multirow{2}[4]{1.18in}{The redundancy of the analysis class diagram generated by the three approaches differ significantly}} \bigstrut\\
    \multicolumn{1}{l}{} & $CD_{rd}$(Popescu et al.)=$CD_{rd}$(Yue et al.)=$CD_{rd}$(Our approach) & \multicolumn{1}{c}{} &       &       & \multicolumn{1}{l}{} & \multicolumn{1}{l}{} \bigstrut\\
    \hline
		\multicolumn{7}{l}{${\chi^2}$ Friedman's chi-square, ${\alpha}$ = 0.01} \bigstrut
    \end{tabular}%
  \label{tab:FriedmanTestCrCmRd}%
\end{table}%
\begin{table}[htb]
  \centering
	\scriptsize
  \caption{The post hoc Friedman test for the pairwise comparison of class diagram correctness ($CD_{cr}$) of the three approaches}
    \begin{tabular}{lcccccp{2.5cm}p{4.5cm}}
		\hline
    \textbf{Pairwise comparison} & \multirow{2}[2]{*}{\textbf{N}} & \multicolumn{2}{c}{\textbf{Test Statistics}} & \multicolumn{2}{c}{\textbf{Mean Rank}} & \multirow{2}[2]{*}{\textbf{Test Result}} & \multirow{2}[2]{*}{\textbf{Conclusion}} \\
\cline{3-6}    \textbf{( A vs B)} &       & \textbf{$\chi^2$} & \textbf{p value} & \textbf{A} & \textbf{B} &       & \\
    \hline
    Popescu et al. vs Yue et al. & \bigstrut 40    & 3.600 & .058  & 1.35  & 1.65  & p\textgreater0.01, No enough evidence for statistical difference & No statistical significant difference between the Class diagram correctness ($CD_{cr}$) of the two approaches \bigstrut\\

    Popescu et al. vs Our approach & \bigstrut 40    & 40.000 & \textbf{.000} & 1.00  & 2.00  & p$<$0.01, Enough evidence for statistical difference & mean rank for our approach \textgreater mean rank for Popescu et al., hence $CD_{cr}$(Our approach) \textgreater $CD_{cr}$(Popescu et al.) \bigstrut\\

    Yue et al. vs  Our approach & \bigstrut 40    & 39.000 & \textbf{.000} & 1.01  & 1.99  & p$<$0.01, Enough evidence for statistical difference & mean rank for our approach \textgreater mean rank for Yue et al., hence $CD_{cr}$(Our approach) \textgreater $CD_{cr}$(Yue et al.) \bigstrut\\
    \hline
    \end{tabular}%
  \label{tab:FriedmanPostHocTestCr}%
\end{table}%

\begin{table}[htb]
  \centering
  	\scriptsize
  \caption{The post hoc Friedman test for the pairwise comparison of class diagram completeness ($CD_{cm}$) of the three approaches}
    \begin{tabular}{lcccccp{2.5cm}p{4.5cm}}
    \hline
     \textbf{Pairwise comparison} & \multirow{2}[2]{*}{\textbf{N}} & \multicolumn{2}{c}{\textbf{Test Statistics}} & \multicolumn{2}{c}{\textbf{Mean Rank}} & \multirow{2}[2]{*}{\textbf{Test Result}} & \multirow{2}[2]{*}{\textbf{Conclusion}} \\
\cline{3-6}    \textbf{( A vs B)} &       & \textbf{$\chi^2$} & \textbf{p value} & \textbf{A} & \textbf{B} &       &  \\
    \hline
    Popescu et al. vs Yue et al. & \bigstrut 40    & 25.973 & \textbf{.000} & 1.11  & 1.89  & p$<$0.01, Enough evidence for statistical difference & mean rank for Yue et al. \textgreater mean rank for Popescu et al., hence $CD_{cm}$(Yue et al.) \textgreater $CD_{cm}$(Popescu et al.) \bigstrut\\

    Popescu et al. vs Our approach & \bigstrut 40    & 40.000 & \textbf{.000} & 1.00  & 2.00  & p$<$0.01, Enough evidence for statistical difference & mean rank for our approach \textgreater mean rank for Popescu et al., hence $CD_{cm}$(Our approach) \textgreater $CD_{cm}$(Popescu et al.) \bigstrut\\

    Yue et al. vs  Our approach & \bigstrut 40    & 39.000 & \textbf{.000} & 1.01  & 1.99  & p$<$0.01, Enough evidence for statistical difference & mean rank for our approach \textgreater mean rank for Yue et al., hence $CD_{cm}$(Our approach) \textgreater $CD_{cm}$(Yue et al.) \bigstrut\\
    \hline
    \end{tabular}%
  \label{tab:FriedmanPostHocTestCm}%
\end{table}%

\begin{table}[htb]
  \centering
 \scriptsize
  \caption{The post hoc Friedman test for the pairwise comparison of class diagram redundancy ($CD_{rd}$) of the three approaches}
    \begin{tabular}{lcccccp{2.5cm}p{4.5cm}}
    \hline
     \textbf{Pairwise comparison} & \multirow{2}[2]{*}{\textbf{N}} & \multicolumn{2}{c}{\textbf{Test Statistics}} & \multicolumn{2}{c}{\textbf{Mean Rank}} & \multirow{2}[2]{*}{\textbf{Test Result}} & \multirow{2}[2]{*}{\textbf{Conclusion}} \\
\cline{3-6}    \textbf{( A vs B)} &       & \textbf{$\chi^2$} & \textbf{p value} & \textbf{A} & \textbf{B} &       &  \\
    \hline
    Popescu et al. vs Yue et al. & \bigstrut 40    & .231  & .631  & 1.46  & 1.54  & p\textgreater0.01, No enough evidence for statistical difference & No statistical significant difference between the Class diagram redundancy ($CD_{rd}$) of the two approaches \bigstrut\\

    Popescu et al. vs Our approach & \bigstrut 40    & 36.000 & \textbf{.000} & 1.95  & 1.05  & p$<$0.01, Enough evidence for statistical difference & mean rank for our approach \textless~mean rank for Popescu et al., hence $CD_{rd}$(Our approach) \textless $CD_{rd}$(Popescu et al.) \bigstrut\\

    Yue et al. vs  Our approach & \bigstrut 40    & 40.000 & \textbf{.000} & 2.00  & 1.00  & p$<$0.01, Enough evidence for statistical difference & mean rank for our approach \textless~mean rank for Yue et al., hence $CD_{rd}$(Our approach) \textless $CD_{rd}$(Yue et al.) \bigstrut\\
    \hline
    \end{tabular}%
  \label{tab:FriedmanPostHocTestRd}%
\end{table}%

\subsection{Validity consideration}
\label{subsec:ValidityConsideration}
Here we present the threats to validity in our experimental study and discuss the strategies used to deal with those threats~\citep{wohlin2003empirical,sjoberg2005survey,wohlin2012experimentation}\\
\\
\textbf{\emph{Internal validity}}: One of the threats to internal validity was: the knowledge, understanding and experience of the subjects in object oriented concepts and analysis modeling. The subjects (or participants) in the experiment were the forty students of Computer Science and Engineering discipline from Indian Institute of Information Technology Design \& Manufacturing Jabalpur, India. The subjects were a mix of Ph.D. scholars,  M.Tech. final year students and B.Tech final year students. All the subjects have studied at least two courses in \textsl{Object Oriented Software Engineering}, one of which is \textsl{Object Oriented Analysis and Design}, they have also done a course project involving \textsl{object oriented analysis}. Additionally, we organised training sessions to brush up their analysis modeling concepts; and tested their concepts and understanding through some exercises.

As the design of the experiment was a complete block design so each subject had evaluated all the three approaches for the UCS given to him/her, hence the threat to selection bias for the comparison groups was nullified. The three approaches were compared on the basis of the analysis class diagrams generated by them for the same set of forty UCSs and using the same quality measures. Hence there was no comparison bias.

The blocking variable was the order in which the approaches were evaluated by the subjects, so to nullify the ordering affect we divided the subjects into six groups and the subjects were asked to answer the questionnaires for the three approaches in the order shown in Table~\ref{tab:OrderOfEvaluation}.\\
\\
\textbf{\emph{Construct validity}}: The threats to construct validity was: Whether the measures used for the evaluation of the class diagrams generated by the tools are able to depict the quality of the class diagrams? We minimized this threat by using the quality metrics presented in Section~\ref{subsec:Metrics} for estimating the quality of analysis class diagrams generated by the tools. These quality metrics are based on a systematic literature review on model quality~\citep{mohagheghi2009definitions} and on the quality measures which were previously defined and used for estimating the correctness, completeness and redundancy of the class diagrams in~\cite{yue2013facilitating}. The quality measures presented in~\cite{yue2013facilitating} were based on comparing an analysis class diagram with some reference class diagram, but it is difficult to get the reference class diagrams for every UCS. Moreover, different persons (or experts) may derive different analysis class diagram for the same UCS and all of them may be correct, so it is difficult to get a single reference class diagram for a given UCS. Therefore we modified and re-formulated those quality measures, so that the measures can be used to assess the quality of analysis class diagrams even if the reference class diagrams are not available. We present 15 sub metrics used for estimating the quality (correctness, completeness and redundancy) of the analysis class diagrams. Out of these 15 sub metrics, 13 are newly defined by us and only 2 sub metrics (correctly named and correctly stereotyped for class correctness) are taken from~\cite{yue2013facilitating}. These modifications were done based on the model quality goals presented in~\cite{mohagheghi2009definitions} and the methodology a human analyst uses to identify problem level objects and relationships from the vocabulary of the problem domain~\citep{booch2006object}.\\
\\
\textbf{\emph{Conclusion validity}}: We minimized the threats to statistical conclusion validity by applying the suitable statistical tests on the data. First we check the normality of the data obtained from the experiment by applying \textsl{Kolmogorov-Smirnov} test, and found that the data was not normal. Then we applied \textsl{Friedman} test (a non parametric test) for hypothesis testing as our experimental design, procedure applied for the execution of experiment and the data collection methods fulfilled all the conditions for applying \textsl{Friedman} test on the data. We tested all hypotheses considering the significance level of 0.01 (p \textless 0.01). Hence, we are confident that the correct statistical tests were applied, as the assumptions of the statistical tests were not violated.  \\
\\
\textbf{\emph{External validity}}: The external validity of an experiment is related to the ability to generalize the results. In the reported experiment the students were used as subjects for comparing the approaches. One of the threats to external validity is whether the results obtained through the answers from the students generalize to software professionals. The subjects of the experiment were well trained in object oriented analysis and modeling through the courses in object oriented software engineering and the course projects done involving analysis and design. Additionally, we conducted training sessions to brush up their analysis modeling concepts. Their concepts and understanding were then tested through different exercises involving analysis modeling. Studies like~\citep{host2000using,holt1987mental,arisholm2004evaluating} reported no significant difference between the students and software professionals used as subjects in the experiments. Porter et al. in their work~\citep{porter1998comparing} reported identical outcomes for the two experiments conducted by them, one involving students as subjects and the other replicated experiment that involved software professionals as subjects. Moreover, studies that involves object oriented design and modeling with UML like~\cite{arisholm2006impact} suggested that the students are better trained representatives than most professionals, specifically those who have not been taught OO modeling with UML in detail. A survey of controlled experiment in software engineering~\citep{sjoberg2005survey} reported that 81\% percent of the subjects were students in a total of 113 experiments investigated in the study.

Other threats to external validity in the experiments like ours' are the size of the study and the number of the representative of the population. In our study we used the forty UCSs from various domains such as control system, real time system, web applications etc. All these UCS were from the standard software engineering books such as~\cite{booch2010object},~\cite{bruegge1999object},~\cite{gomaa2011software}, Rosenberg et al.~\cite{doug2008use,rosenberg2007use} etc. and research works~\cite{liu2004natural,popescu2008reducing,simula12,yue2015atoucan,deeptimahanti2011semi}. Although, these UCSs were not the true representative of the industry level UCS, but are fairly good. Moreover, the same UCSs were used in all the three treatments so their effect if any would had been be same in the evaluation of all the approaches. Regarding the number of representative of the population, in the same survey~\citep{sjoberg2005survey} as reported above in the total of 113 experiments investigated  the number of subjects per experiment ranged from 04 to 266, with a mean value of 48.6. Hence we think the 40 subjects in our study were fair in number.

\section{Discussion}\label{sec:Discussion}
The proposed approach for generating analysis class diagrams from UCSs is a fully automated approach which is supported by a GUI based tool support named \textsl{AutoAMG}. The approach provides in-place visualization of the generated analysis class diagrams. The results of the experimental study conducted for the evaluation of the approach showed that the analysis class diagrams generated by our approach are of better quality. Specifically the results has shown that the analysis class diagrams generated by our approach were 46\% more correct, 55\% more complete and 31\% less redundant than those generated by Popescu et al. approach, and were 33\% more correct, 31\% more complete and 31\%  less redundant than those generated by Yue et al. approach.

Our approach has shown significant improvement over the other two existing approaches because i) The language model (or the set of sentence patterns) used in our approach is comprehensive enough to interpret and identify class diagram elements from possibly all the simple sentences and a few complex sentences (sentences specifying conditions and sentences containing that clause and conjunctive clause) in Engish whereas the language model used in Popescu et al. can not interpret common simple sentences like those containing infinitives (e.g. ``The system prompts to enter the password", ``The system commands the motor to start"), present participles (e.g. ``The system prints the receipt showing transaction number and date"), past participles (e.g. ``The system validates the record entered by the customer"), gerunds (e.g. ``The system starts printing the document"), etc. (Table~\ref{tab:IssuesExistingApproaches}), and the language model used in Yue et al. approach can not interpret common complex sentences like those containing that clauses (e.g. ``The system checks that the password is correct") and conjunctive clauses (e.g. ``The system stops the motor when the tank is full", ``The motor stops when the tank is full."), etc. (Table~\ref{tab:IssuesExistingApproaches}). ii) The transformation rules in our approach takes into account both the sentence structures of the sentences and the semantic relationships between the words of the sentences obtained from TDs to disambiguate the extraction of the desired elements of the analysis class diagrams (Section~\ref{sec:NLPModels}) whereas the Yue et al. approach uses parse trees that do not depict any semantic relationship between the words in the sentences to extract the elements of the analysis class diagram. iii) Our approach applies various heuristics on POS-tags and TDs of the sentences to identify entity terms (Step~4.1, Section~\ref{sec:WorkingOfProposedApproach}), and to identify  domain classes and attributes from entity terms (Step~4.2-4.9, Section~\ref{sec:WorkingOfProposedApproach}) whereas Yue et al. approach is unable to identify the classes and attributes that are documented as group of words in the sentences. Their approach identify the entity classes only from those noun phrases in the sentences that either contains a single noun (e.g. ``customer"), or a single noun with a determiner (e.g. ``The customer") , or  a possessive noun (e.g. ``customer's address") (Rule B1.1 and B1.2~\cite{simula12,yue2015atoucan}). .

Table~\ref{tab:ComparativeApproaches} presents a comparison of the three approaches on the basis of input, output, language model and NLP constructs used. From the existing approaches in literature for the generation of analysis class diagrams, we selected the two approaches viz.~\cite{popescu2008reducing} and~\cite{simula12,yue2015atoucan} for comparison with our approach due to the following reasons:

\begin{enumerate}
\item These are the latest approaches in the literature.
\item These approaches up to certain extent are able to identify most of the elements for generating analysis class diagrams that the proposed approach identifies such as classes, their attributes, their operations, association, generalization and aggregation relationships. Whereas the other approaches are unable to identify all these element, such as the approach CM-Builder2~\citep{harmain2003cm} does not identifies class operations, aggregation and generalization relationships, the approach~\citep{liu2004natural} does not identify attributes of the classes,  aggregation and generalization relationships between the entity classes, and the approach~\citep{ilieva2006models} does not identify attributes, operations, relationship names and relationship types.
\item These approaches require only the UCSs as input, they do not require additional information along with UCSs like glossary files as required by the approach~\citep{liu2004natural} for identifying the candidate classes and as required by the approach~\citep{deeptimahanti2009automated,deeptimahanti2009innovative} for eliminating the redundant classes and attributes.
\item They do not require human intervention for eliminating irrelevant classes as required in~\cite{deeptimahanti2009automated,deeptimahanti2009innovative}.
\end{enumerate}

\begin{table}[htb]
  \centering
	\scriptsize
 \caption{Comparison with existing automated approaches}
    \begin{tabular}{p{.8cm}p{1.2cm}p{.5cm}p{.5cm}p{1cm}p{.5cm}p{4cm}p{1cm}p{1cm}p{1cm}}
			\hline
		 \multicolumn{1}{p{.7cm}}{\multirow{2}[6]{.8cm}{\bigstrut \textbf{Approach}}} & \multicolumn{1}{p{1.2cm}}{\multirow{2}[6]{1.2cm}{\textbf{Input}}} & \multicolumn{4}{p{3.5cm}}{\textbf{Sentence patterns recognized / Language model}}& \multicolumn{1}{p{4cm}}{\multirow{2}[4]{4cm}{\textbf{Theoretical foundation of sentence patterns / language model}}} & \multicolumn{1}{p{1cm}}{\multirow{2}[2]{1cm}{\bigstrut \textbf{No. of Transformation rules}}} & \multicolumn{1}{p{1cm}}{\multirow{2}[2]{1cm}{\textbf{\bigstrut NLP constructs used}}} & \multicolumn{1}{p{1cm}}{\multirow{2}[2]{1cm}{\textbf{Output}}} \\
    \cline{3-6}
    \multicolumn{1}{c}{} & \multicolumn{1}{c}{} & \textbf{Simple} & \textbf{complex} & \multicolumn{1}{p{1cm}}{\textbf{Keyword specific}} & \textbf{special} & \multicolumn{1}{c}{} & \multicolumn{1}{c}{} & \multicolumn{1}{c}{} & \multicolumn{1}{c}{} \bigstrut\\
		\hline
   \cite{popescu2008reducing} & Textual specifications & 2     & 1     & NIL   & 5     &  5 sentence patterns based on constraining grammar proposed in~\cite{juristo2000use} & 12    & link types & Analysis class diagram \bigstrut \\ \\
    \cite{simula12,yue2015atoucan} & Textual specifications & 12    & NIL   & 9 & NIL   & 5 sentence patterns proposed in English grammar book~\citep{greenbaum1996oxford} & 14    & parse tree & Analysis class diagram \bigstrut \\ \\
    Our   & Textual specifications & 26    & 8     & 4     & 4     & 25 sentence patterns proposed in Oxford Advanced Learner's Dictionary of Current English~\citep{OALD1974english,OALD2000english} and in~\cite{hornby1975english} that are successively followed by various researchers in linguistic domain~\citep{hanks2008lexical} & 54    & TDs and POS-tags & Analysis class diagram  \\
    \hline
		    \end{tabular}%
  \label{tab:ComparativeApproaches}%
\end{table}%

The sentence structure rules used in our approach were crafted using the twenty five verb patterns originally proposed by A.S.Hornby et al. in Oxford Advanced Learner's Dictionary of Current English~\citep{OALD1974english,OALD2000english} and in~\cite{hornby1975english}, and successively followed by various researchers in linguistic domain~\citep{hanks2008lexical}. In a grammatical analysis of 152 UCSs, ~\cite{cox2002heuristics} identified 27 most common grammar structures that are used in documenting the UCSs. Our sentence structure rules include all these grammar structures. The proposed sentence structure can recognize possibly all the simple sentences (including sentences containing participles, infinitives and gerunds) as well as a few complex sentences viz. sentences specifying conditions, sentences containing that clause and conjunctive clause. The proposed sentence structure rules when applied on 2000 simple sentences in English taken from Tanaka corpus\footnote{http://tatoeba.org/eng/} were able to correctly recognize 92\% sentences (which is \textless 100\% because of incorrect POS-tags and TDs generated by the parser for some sentences).
\\
\\
\noindent To find the appropriate quality metrics for evaluating the analysis class diagrams we investigated the existing quality metrics, we found that a number of quality metrics to validate design class diagrams (or design models) have been proposed in literature~\citep{rosenberg1997software,genero2002empirical,manso2003no,genero2005survey,mcquillan2007application,ma2010hybrid} but the quality metrics for validating analysis class diagrams (or analysis models) have been rarely explored. As the analysis class diagram is the first structural representation of the problem level objects and the relationships between them. The quality metrics for validating analysis class diagrams must be able to estimate how much correctly and how much completely the transformation from the requirements to the analysis models has been done~\citep{mohagheghi2009definitions}. In~\cite{yue2013facilitating} Yue et al. defined and used a set of measures to compare the quality of analysis class diagrams developed by the subjects in their experiment with the reference class diagrams (the reference class diagrams were taken from software engineering books, and those developed by the student experts). But, it is difficult to get the reference analysis class diagrams for every UCS. Moreover, different experts may derive different analysis class diagrams from the same set of requirements so it is also difficult to get a single reference class diagram for a given UCS. Therefore we modified and re-formulated those quality measures, so that the measures can be used to assess the quality of analysis class diagrams even if the reference class diagrams are not available. We present 15 sub metrics used for estimating the quality (correctness, completeness and redundancy) of the analysis class diagrams. Out of these 15 sub metrics, 13 are newly defined by us and only 2 sub metrics (correctly named and correctly stereotyped for class correctness) are taken from~\cite{yue2013facilitating}. These modifications are done based on the model quality goals obtained through a systematic literature review on model quality~\citep{mohagheghi2009definitions} and the methodology a human analyst uses to identify problem level objects and relationships from the vocabulary of the problem domain~\citep{booch2006object}.\\
\\
\noindent Some limitations of the proposed approach are as follows:
\begin{enumerate}
	\item  The approach requires the UCSs to be written using a few restrictions shown in Table~\ref{tab:RestrictionRules}. The restrictions are required for handling the issues that are associated with the natural language such as ambiguity, variety of sentence types, anaphora (or pronoun) resolution problem and inconsistency~\citep{kamsties2000taming,nuseibeh2001making,fabbrini2001linguistic,yang2010extending}. At present the approach can process and interpret all the simple sentences (including sentences containing participles, infinitives and gerunds) as well as a few complex sentences viz. sentences specifying conditions, the sentences containing that clause and conjunctive clause. In a grammatical analysis of 152 UCSs,~\cite{cox2002heuristics} identified 27 most common grammar structures that are used in documenting the UCSs. These grammar structures are the subset of the sentence structures that our approach can process and interpret.
\item As the approach uses NL parser to process the sentences in UCS, hence the accuracy of the approach is inherently bounded by the accuracy of the NL parser.  The Stanford NL parser used in our approach generates POS tags with accuracy of about 97\%~\citep{manning2011part} and TDs with accuracy of about 84.2\%~\citep{cer2010parsing}, these figures are for all the type of sentences, the accuracy may be more for the simple sentences and a few complex sentences that our approach handles at present. However, the parser accuracy in generating TDs can further be  improved from 84.2\% to 89.1\% by using Charniak-Johnson re-ranking parser~\citep{cer2010parsing,mcclosky2006effective} for generating the dependencies and converting them to Stanford TDs.
\item The proposed approach does not uses entity disambiguation techniques such as misspelling identification, abbreviation identification, alias identification etc.~\citep{misra2013entity}
\item The current implementation of AutoAMG supports the input of one UCS at a time, and the generation of analysis class diagram for one UCS at a time. But, the interactions of the given UCS with other UCSs, as specified in the UCS using INCLUDE, EXTEND keywords and Parent Use Case Name field of the UCS, are shown in the generated analysis class diagram with the help of INCLUDE, EXTEND and generalization relationships between the control class representing given UCS and the control classes representing the other UCSs respectively. In future we will extend our tool to take a set of UCS for a given problem as input, and to generate the analysis class diagram for the problem.
\end{enumerate}	

\noindent The experimental study reported in this paper is to the best of our knowledge the first one to be conducted for comparing different automated approaches used to generate analysis class diagrams. We made the following experimental material along with other details available at\footnote{http://serg.iiitdmj.ac.in/tools/AutoAMG/} for other researchers of this domain to replicate such experiments in future viz. the forty UCSs used in the experimental study, the analysis class diagrams generated by the three approaches (our approach, Popescu et al. approach and Yue et al. approach) for the forty UCSs, the questionnaires for the evaluation of the analysis class diagrams and the tool \textsl{AutoAMG}.

\section{Related Work}
\label{sec:RelatedWork}
This section presents the existing approaches in the literature to automate the process of generating analysis models from software requirements. The available approaches can be classified into semi automated approaches and automated approaches. The semi automated approaches require human intervention for identifying the elements for the generation of the analysis class diagrams whereas the automated approaches do not require any human intervention.

\cite{mich1996nl,mich2002nl} proposed a semi automated approach supported by a CASE tool NL-OOPS (Natural Language - Object Oriented Production System) for generating object oriented models from unrestricted natural language requirements documents. The tool uses as a core the semantic network of a Natural Language Processing System LOLITA (Large-scale Object-based Language Interactor, Translator and Analyser) for identifying classes and relationships. The knowledge of the requirements document is stored in the knowledge base of LOLITA, which adds new nodes to its semantic network. The class model which includes classes and associations is derived from the nodes in the semantic network. The tool requires user intervention for deleting the extra nodes representing spurious classes and to set the level of details for class hierarchy.

\cite{overmyer2001conceptual} proposed a semi automated approach supported by a tool named Linguistic Assistant for Domain Analysis (LIDA) that helps an analyst in deriving analysis models. The tool first reads the textual requirements, then it presents the user with list of nouns, adjectives and verbs from which user has to identify and mark the candidate classes, attributes and the methods. The tool requires user intervention at every step to identify the elements for generating the class diagram.

\cite{harmain2003cm} proposed a semi automated approach supported by a tool named CM-Builder1 for the generation of analysis class diagrams. The approach first reads the textual requirements and provides the lists of candidate classes, attributes and relationships to the user. It also assigns the frequency of reference in the text to each class. The classes having more frequency of references are highly suggested classes. The user has to select the required classes their attributes and relationships between them from the corresponding candidate lists.

\cite{samarasinghe2005generating} proposed a semi automated approach supported by a tool named UCEd to generate domain models from use case models. The tool takes as input a UCS written using the grammar provided by them. The tool then makes every possible combinations of the words in the sentences of the UCS and present them to the user. The user has to select the required classes, their operations, their attributes etc. from the the presented word combinations.\\
\\
\cite{harmain2003cm} proposed an approach supported by a tool named CMBuilder2 that obtains a first cut domain models from the NL requirements automatically. The tool considers all the nouns as candidate classes, all the non copular verbs as candidate relationships; it obtains attributes from possessive relationships and adjectives. The candidate classes having low frequency in the requirements and also do not participating in any relationships are discarded by the tool. The tool fails to identify class operations, aggregation and generalization relationships. Moreover, the generated models contain many unconnected components.

\cite{liu2004natural} proposed an automated approach named UCDA for generating analysis class diagrams from UCS written using restricted grammar. Their approach processes the sentences based on the classification of sentences as transitive, intransitive, ditransitive, intensive, complex transitive, prepositional and non-finite given in~\cite{roberts1956patterns}. To identify domain classes, the approach uses POS-tags generated by a NL parser and a glossary that defines specific terms of the domain. The approach fails to identify attributes of the classes, and aggregation and generalization relationships between the entity classes.

\cite{ilieva2006models} proposed an approach that obtains domain models from unrestricted NL requirements. The approach uses a POS tagger to mark the words in the NL requirements with parts of speech. Then it identifies three roles from the sentences: subject, predicate and object. Using these roles, a semantic network of the words is created. The semantic network is then transformed into a hybrid activity model and a domain model. The domain model represents only the identified classes and the relationship between them, it lacks the attributes, operations, relationship names and relationship types. Any tool supporting their approach was not presented by the authors.

\cite{popescu2008reducing} proposed an approach supported by a prototype tool named \textsl{Dowser} to identify inconsistencies in requirement specifications with the help of automatically generated domain models. The approach uses constraining grammar proposed in~\cite{juristo2000use} that allows only five sentence structures: classification (bottom up, top down and multiple) that represents generalization relationship, composition (component and content) that represents aggregation relationship, identification that represents attribute, complement enumeration and adjacent complements that represent association relationship. The tool  first parses the requirements based on the constraining grammar. Then using a NL parser it generates a link grammar parse of the sentences. It then uses link types to identify the classes, methods, variables and associations, and generates a textual object-oriented analysis model of the specified system. The textual model is then visualized using GraphViz (this needs human involvement). The approach is unable to identify many relationships between the classes that results in many unconnected components in the class diagram.

\cite{deeptimahanti2009automated,deeptimahanti2009innovative} proposed an approach supported by tools named UML Model Generator from analysis of Requirements (UMGAR) to generate class models from restricted natural language requirements. It identifies classes from nouns, attributes from adjectives, and methods from verbs. With the help of a glossary it eliminates the redundant classes.  Attribute classes are eliminated by using a text file which contains a list of words that takes values or indicate status. For eliminating irrelevant classes it requires human intervention. The generated class diagram has many unconnected component. The approach requires the requirements to be stated in either \textsl{subject-verb} or \textsl{subject-verb-object} format. It also does not allow passive voice sentences.

\cite{simula12,yue2015atoucan} proposed an automated approach to derive analysis models from use case models. The approach first uses a NL parser to obtain the parse tree and TDs of the sentences. The approach then uses parse tree and TDs to identify sentence structures. It identifies the sentences on the basis on sentence structures formed using five basic English sentence patterns proposed in~\cite{greenbaum1996oxford}. It then uses the parse tree to identify the elements for generating the class diagram. The approach fails to recognize domain objects and attributes which are documented as a group of words (or nouns). It also fails to distinguish between a class and an attribute, resulting in many incorrectly identified relationships. The generated class diagrams have many unconnected components. Moreover the approach dumps most of the operations in a single control class, whereas most of the other classes are assigned no operations at all, hence the division of responsibilities among the classes is not properly done by the approach.\\
\\
To summarize the earlier efforts done in the direction of obtaining the analysis models from software requirements: the semi-automatic approaches in literature~\citep{overmyer2001conceptual,samarasinghe2005generating} highly relies on users for identifying the elements to derive the analysis models. Due to a very limited set of grammar rules the approaches~\citep{samarasinghe2005generating,popescu2008reducing} fail to analyze the diverse set of NL sentences resulting in an incomplete transformation of software requirements into analysis class diagrams. The analysis class diagram obtained by the approaches~\citep{ilieva2006models,subramaniam2004ucda} lacks either attributes, operations, relationship names,  generalization or aggregation. Many relationships identified between the objects by approaches~\citep{simula12,yue2015atoucan,yue2010automatically,fliedl2007deriving} are incorrect. The approach~\citep{simula12,yue2015atoucan,yue2010automatically} fails to recognize problem level objects and attributes which are documented as a group of words, it fails to distinguish between a class and an attribute, and also fails to distribute responsibilities among the classes.

In comparison the proposed approach attempts to handle the diverse set of sentences in the requirements by using a set of comprehensive sentence structure rules and a set of comprehensive transformation rules. These comprehensive sentence structure rules are based on the twenty five verb patterns proposed by A.S.Hornby in~\cite{hornby1975english} and in Oxford Advanced Learner’s Dictionary of Current English~\citep{OALD1974english,OALD2000english}. To correctly extract potential classes, their attributes, operations and the relationships, the comprehensive transformation rules take into account the sentence structure of the sentences as well as the syntactic and semantic relationships between the words of the sentences. These relationships between the words are found with the help of type dependencies (TDs) and part of speech tags (POS-tags) generated using the Stanford NL parser API.

\section{Conclusion and future work}
\label{sec:Conclusion}
In this paper we have proposed an automated approach for the generation of analysis class diagrams from software requirements documented as use case specifications (UCSs). The approach takes as input a UCS written in English language with a few restrictions, and then parses it using the Stanford NL parser APIs to generate TDs and POS-tags from the sentences in the UCS. To systematically process and interpret the sentences the approach first identifies their sentence structure by applying the proposed set of comprehensive sentence structure rules on their TDs and POS-tags. Based on the identified sentence structures it then applies the proposed set of comprehensive transformation rules on TDs and POS-tags of the sentences to identify the elements for the generation of analysis class diagram. The approach finally generates the analysis class diagram, and visualizes it using GraphViz APIs. We have implemented the proposed approach in a GUI based tool support named \textsl{AutoAMG}.

For the validation of the proposed approach we reported the outcome of a controlled experiment that we have conducted to compare the analysis class diagrams generated by the approach with those generated by the two existing automated approaches, one proposed by~\cite{popescu2008reducing} and the other proposed by~\cite{simula12,yue2015atoucan}. In the experiment forty subjects evaluated the correctness, completeness and redundancy of the analysis class diagrams generated by the three approaches for forty UCSs. The results of the experiment clearly showed that the analysis class diagrams generated by the proposed approach were significantly better in terms of correctness, completeness and redundancy than those generated by the other approaches. Specifically, the results for the forty UCSs showed that the analysis class diagrams generated by our approach were 46\% more correct, 55\% more complete and 31\% less redundant than those generated by Popescu et al. approach, and were 33\% more correct, 31\% more complete and 31\%  less redundant than those generated by Yue et al. approach. \\
\\
{
\normalsize
The presented work has many future directions such as to generalize the approach so that along with the simple sentences and a few complex sentences that the approach can currently interpret and transform, it can also interpret and transform other complex and compound sentences. Other issues to deal with are to handle anaphora or pronoun, to apply entity disambiguation techniques~\citep{misra2013entity} such as misspelling identification, abbreviation identification, alias identification etc. The approach can be extended to further generate platform specific models (PSM) from the platform independent models (PIM), and to generate the template code.
The validation study of the approach can be done in the industrial settings that may help in reenforcing the outcomes of the approach.
}


\vspace{10ex}

\begin{appendices}
\scriptsize
\section{Sentence Structure Rules}
\label{app:SentenceStructureRules}
\scriptsize
Table~\ref{tab:CompleteSSRules} presents the sentence structure rules.

  \begin{longtable}{|p{.5cm}|p{2.6cm}|p{2.4cm}|p{1.9cm}|p{6.6cm}|}
  \caption{\textbf{Sentence Structure Rules}} \label{grid_lllll} \\
	\hline
  \textbf{Rule \#} & \textbf{Antecedent (If the sentence contains TDs:)} & \textbf{Consequent (then the identified sentence structure is:)} & \textbf{Example sentence} & \textbf{Type dependencies of Example sentence} \bigstrut\\
    \hline
		\endfirsthead
\multicolumn{5}{m{0.8\linewidth}}{\textbf{Sentence Structure Rules (Appendix A) --continued from previous page}} \\ \hline
   \textbf{Rule \#} & \textbf{Antecedent (If the sentence contains TDs:)} & \textbf{Consequent (then the identified sentence structure is:)} & \textbf{Example sentence} & \textbf{Type dependencies of Example sentence} \bigstrut\\
\endhead
\hline
		 \textbf{SSR1} & \textit{\textbf{nsubj}(A,B), \textbf{iobj}(A,C), \textbf{dobj}(A,D)} & SVIODO (Subject-Verb-IndirectObject-DirectObject) & The system sends the user an email. & \textit{[det(system-2, The-1), \textbf{nsubj}(sends-3, system-2), root(ROOT-0, sends-3), det(user-5, the-4), \textbf{iobj}(sends-3, user-5), det(email-7, an-6), \textbf{dobj}(sends-3, email-7)]} \bigstrut\\
    \hline \bigstrut
		\textbf{SSR2} & \textit{\textbf{nsubj}(A,B), \textbf{dobj}(A,C), \textbf{complm}(D,E), \textbf{nsubj}(D,F)} & SVDOThatClause (Subject-Verb-DirectObject-ThatClause) & The system informs the user that the battery is full & \textit{[det(system-2, The-1), \textbf{nsubj}(informs-3, system-2), root(ROOT-0, informs-3), det(user-5, the-4), \textbf{dobj}(informs-3, user-5), \textbf{complm}(full-10, that-6), det(battery-8, the-7), \textbf{nsubj}(full-10, battery-8), cop(full-10, is-9), ccomp(informs-3, full-10)]} \bigstrut\\
    \hline \bigstrut
		\textbf{SSR3} & \textit{\textbf{nsubj}(A,B), \textbf{complm}(C,D), \textbf{nsubj}(C,E)} & SVThatClause (Subject-Verb-ThatClause) & The system validates that the password is correct & \textit{[det(system-2, The-1), \textbf{nsubj}(validates-3, system-2), root(ROOT-0, validates-3), \textbf{complm}(correct-8, that-4), det(password-6, the-5), \textbf{nsubj}(correct-8, password-6), cop(correct-8, is-7), ccomp(validates-3, correct-8)]} \bigstrut\\
    \hline \bigstrut
		 \textbf{SSR4} & \textit{\textbf{nsubj}(A,B), \textbf{dobj}(A,C), \textbf{neg}(D,E), \textbf{aux}(D,F), \textbf{infmod}(C,D)} & SVDONotToInf (Subject-Verb-DirectObject-Not-To-Infinitive) & The system warns the user not to restart the system. & \textit{[det(system-2, The-1), \textbf{nsubj}(warns-3, system-2), root(ROOT-0, warns-3), det(user-5, the-4), \textbf{dobj}(warns-3, user-5), \textbf{neg}(restart-8, not-6), \textbf{aux}(restart-8, to-7), \textbf{infmod}(user-5, restart-8), det(system-10, the-9), dobj(restart-8, system-10)]} \bigstrut\\
    \hline \bigstrut
		 \textbf{SSR5} & \textit{\textbf{nsubj}(A,B), \textbf{neg}(C,D), \textbf{aux}(C,E), \textbf{xcomp}(A,C), \textbf{dobj}(C,F)} & SVNotToInf (Subject-Verb-Not-To-Infinitive) & The customer selects not to fill the tank & \textit{[det(customer-2, The-1), \textbf{nsubj}(selects-3, customer-2), root(ROOT-0, selects-3), \textbf{neg}(fill-6, not-4), \textbf{aux}(fill-6, to-5), \textbf{xcomp}(selects-3, fill-6), det(tank-8, the-7), \textbf{dobj}(fill-6, tank-8)]} \bigstrut\\
    \hline \bigstrut
		 \textbf{SSR6} & \textit{\textbf{nsubj}(A,B), \textbf{nsubj}(C,D), \textbf{aux}(C,E), \textbf{cop}(C,F), \textbf{xcomp}(A,C)} & SVDOtobeComp (Subject-Verb-DirectObject-to-be-Complement) & The system marks the errors to be red. & \textit{[det(system-2, The-1), \textbf{nsubj}(marks-3, system-2), root(ROOT-0, marks-3), det(errors-5, the-4), \textbf{nsubj}(red-8, errors-5), \textbf{aux}(red-8, to-6), \textbf{cop}(red-8, be-7), \textbf{xcomp}(marks-3, red-8)]} \bigstrut\\
    \hline \bigstrut
		\textbf{SSR7} & \textit{\textbf{nsubj}(A,B), \textbf{dobj}(A,C), \textbf{aux}(D,E), \textbf{infmod}(C,D)} & SVDOToInf (Subject-Verb-DirectObject-To-Infinitive) & The system commands the motor to start. & \textit{[det(system-2, The-1), \textbf{nsubj}(commands-3, system-2), root(ROOT-0, commands-3), det(motor-5, the-4), \textbf{dobj}(commands-3, motor-5), \textbf{aux}(start-7, to-6), \textbf{infmod}(motor-5, start-7)]} \bigstrut\\
    \hline \bigstrut
		 \textbf{SSR8} & \textit{\textbf{nsubj}(A,B), \textbf{dobj}(A,C), \textbf{partmod}(C,D) and \textbf{POS-tag(D)}==``VBG"} & SVDOPresentPart (Subject-Verb-DirectObject-PresentParticiple) & The system keeps the user waiting. & \textit{[det(system-2, The-1), \textbf{nsubj}(keeps-3, system-2), root(ROOT-0, keeps-3), det(user-5, the-4), \textbf{dobj}(keeps-3, user-5), \textbf{partmod}(user-5, \textbf{waiting}-6)].} \bigstrut\\
    \hline \bigstrut
		\textbf{SSR9} & \textit{\textbf{nsubj}(A,B), \textbf{dobj}(A,C), \textbf{partmod}(C,D) and \textbf{POS-tag(D)}==``VBN"} & SVDOPastPart (Subject-Verb-DirectObject-PastParticiple) & The system validates the record entered by the customer. & \textit{[det(system-2, The-1), \textbf{nsubj}(validates-3, system-2), root(ROOT-0, validates-3), det(record-5, the-4), \textbf{dobj}(validates-3, record-5), \textbf{partmod}(record-5, \textbf{entered}-6), prep(entered-6, by-7), det(customer-9, the-8), pobj(by-7, customer-9)]} \bigstrut\\
    \hline \bigstrut
	\textbf{SSR10} & \textit{\textbf{nsubj}(A,B), \textbf{nsubj}(C,D), \textbf{xcomp}(A,C) and \textbf{POS-tag(C)}==``JJ"} & SVDOAdj (Subject-Verb-DirectObject-Adjective-Complement) & The system keeps the door open & \textit{[det(system-2, The-1), \textbf{nsubj}(keeps-3, system-2), root(ROOT-0, keeps-3), det(door-5, the-4), \textbf{nsubj}(open-6, door-5), \textbf{xcomp}(keeps-3, \textbf{open}-6)]} \bigstrut\\
    \hline
		 \textbf{SSR11} & \textit{nsubj(A,B), nsubj(C,D), xcomp(A,C) and POS-tag(C)==``NN"} & SVDONoun (Subject-Verb-DirectObject-NounComplement) & The system makes the user an administrator. & \textit{[det(system-2, The-1), nsubj(makes-3, system-2), root(ROOT-0, makes-3), det(user-5, the-4), nsubj(administrator-7, user-5), det(administrator-7, an-6), xcomp(makes-3, administrator-7)]} \bigstrut\\
    \hline
    \textbf{SSR12} & \textit{nsubj(A,B), dobj(C,D), advmod(D,E), aux(D,F), xcomp(A,D)} & SVDOConjToInf (Subject-Verb-DirectObject-Conjunctive-To-Infinitive) & The system tells the user where to go. & \textit{[det(system-2, The-1), nsubj(tells-3, system-2), root(ROOT-0, tells-3), det(user-5, the-4), dobj(tells-3, user-5), advmod(go-8, where-6), aux(go-8, to-7), xcomp(tells-3, go-8)]} \bigstrut\\
    \hline
		\textbf{SSR13} & \textit{nsubj(A,B), dobj(A,C), advmod(D,E),nsubj(D,F)} & SVDOConjClause (Subject-Verb-DirectObject-ConjunctiveClause) & The system stops the motor when the tank is full & \textit{[det(system-2, The-1), nsubj(stops-3, system-2), root(ROOT-0, stops-3), det(motor-5, the-4), dobj(stops-3, motor-5), advmod(full-10, when-6), det(tank-8, the-7), nsubj(full-10, tank-8), cop(full-10, is-9), dep(motor-5, full-10)]} \bigstrut\\
    \hline
		\textbf{SSR14} & \textit{nsubj(A,B), dobj(A,C), advmod(A,D)} & SVDOAdverbial (Subject-Verb-DirectObject-Adverbial) & The server responds the query quickly & \textit{[det(server-2, The-1), nsubj(responds-3, server-2), root(ROOT-0, responds-3), det(query-5, the-4), dobj(responds-3, query-5), advmod(responds-3, quickly-6)]} \bigstrut\\
    \hline
		\textbf{SSR15} & \textit{nsubj(A,B), aux(A,C), dobj(A,D) } & SAuxVDO (Subject-AuxiliaryVerb-DirectObject) & The system will eject the ATM card. & \textit{[det(system-2, The-1), nsubj(eject-4, system-2), aux(eject-4, will-3), root(ROOT-0, eject-4), det(card-7, the-5), nn(card-7, ATM-6), dobj(eject-4, card-7)]} \bigstrut\\
    \hline
		\textbf{SSR16} & \textit{nsubj(A,B), dobj(A,C),  prep(A,D), pobj(D,E)} & SVDOPO (Subject-Verb-DirectObject-PrepositionObject) & The system sends the message to the customer. & \textit{[det(system-2, The-1), nsubj(sends-3, system-2), root(ROOT-0, sends-3), det(message-5, the-4), dobj(sends-3, message-5), prep(sends-3, to-6), det(customer-8, the-7), pobj(to-6, customer-8)]} \bigstrut\\
    \hline
		\textbf{SSR17} & \textit{nsubj(A,B), advmod(C,D), aux(C,E), xcomp(A,C)} & SVConjToInf (Subject-Verb-Conjunctive-To-Infinitive) & The system guides where to go. & \textit{[det(system-2, The-1), nsubj(guides-3, system-2), root(ROOT-0, guides-3), advmod(go-6, where-4), aux(go-6, to-5), xcomp(guides-3, go-6)]} \bigstrut\\
    \hline
		\textbf{SSR18} & \textit{nsubj(A,B), advmod(C,D), nsubj(C,E), advcl(A,C)} & SVConjClause (Subject-Verb-ConjunctiveClause) & The motor stops when the tank is full. & \textit{[det(motor-2, The-1), nsubj(stops-3, motor-2), root(ROOT-0, stops-3), advmod(full-8, when-4), det(tank-6, the-5), nsubj(full-8, tank-6), cop(full-8, is-7), advcl(stops-3, full-8)]} \bigstrut\\
    \hline
		  \textbf{SSR19} & \textit{nsubj(A,B), aux(C,D), xcomp(A,C)} & SVToInf (Subject-Verb-To-Infinitive) & The system starts to fill the tank. & \textit{[det(system-2, The-1), nsubj(starts-3, system-2), root(ROOT-0, starts-3), aux(fill-5, to-4), xcomp(starts-3, fill-5), det(tank-7, the-6), dobj(fill-5, tank-7)]} \bigstrut\\
    \hline
		 \textbf{SSR20} & \textit{nsubj(A,B), xcomp(A,C)} & SVGerund (Subject-Verb-Gerund) & The printer starts printing the document & \textit{[det(printer-2, The-1), nsubj(starts-3, printer-2), root(ROOT-0, starts-3), xcomp(starts-3, printing-4), det(document-6, the-5), dobj(printing-4, document-6)]} \bigstrut\\
    \hline
		\textbf{SSR21} & \textit{nsubj(A,B), advmod(A,C)} & SVAdverbialAdjunct (Subject-Verb-AdvervialAdjunct) & The elevator moves up or down. & \textit{[det(elevator-2, The-1), nsubj(moves-3, elevator-2), root(ROOT-0, moves-3), advmod(moves-3, up-4), cc(up-4, or-5), conj(up-4, down-6)].} \bigstrut\\
    \hline
		 \textbf{SSR22} & \textit{nsubj(A,B), cop(A,C)} & SVPredicative (Subject-Verb-Predicative) & The customer is employee. & \textit{[det(customer-2, The-1), nsubj(employee-4, customer-2), cop(employee-4, is-3), root(ROOT-0, employee-4)].} \bigstrut\\
    \hline
		 \textbf{SSR23} & \textit{nsubj(A,B), prep(A,C), num(D,E), pobj(C,D)} & SVForComp (Subject-Verb-For-Complement) & The system waits for 5 seconds. & \textit{[det(system-2, The-1), nsubj(waits-3, system-2), root(ROOT-0, waits-3), prep(waits-3, for-4), num(seconds-6, 5-5), pobj(for-4, seconds-6)].} \bigstrut\\
    \hline
		  \textbf{SSR24} & \textit{nsubjpass(A,B), auxpass(A,C), prep(A,D), pobj(D,E)} & SVPassPO (Subject-VerbPassive-PrepositionObject) & The ATM card is ejected by the system. & \textit{[det(card-3, The-1), nn(card-3, ATM-2), nsubjpass(ejected-5, card-3), auxpass(ejected-5, is-4), root(ROOT-0, ejected-5), prep(ejected-5, by-6), det(system-8, the-7), pobj(by-6, system-8)]} \bigstrut\\
    \hline
		\textbf{SSR25} & \textit{nsubjpass(A,B), aux(A,C), auxpass(A,D), prep(A,E), pobj(E,F)} & SAuxVPassPO (Subject-AuxiliaryVerbPassive-PrepositionObject) & The ATM card will be ejected by the system. & \textit{[det(card-3, The-1), nn(card-3, ATM-2), nsubjpass(ejected-6, card-3), aux(ejected-6, will-4), auxpass(ejected-6, be-5), root(ROOT-0, ejected-6), prep(ejected-6, by-7), det(system-9, the-8), pobj(by-7, system-9)]} \bigstrut\\
    \hline
		\textbf{SSR26} & \textit{nsubj(A,B), prep(A,C), pobj(C,D)} & SVPO (Subject-Verb-PrepositionObject)   & The system prompts for password. & \textit{[det(system-2, The-1), nsubj(prompts-3, system-2), root(ROOT-0, prompts-3), prep(prompts-3, for-4), pobj(for-4, password-5)]} \bigstrut\\
    \hline		
		\textbf{SSR27} & \textit{nsubj(A,B), dobj(A,C)} & SVDO (Subject-Verb-DirectObject)   & ATM customer enters the withdrawal amount. & \textit{nn(customer-2,ATM-1), nsubj(enters-3,customer-2), root(ROOT-0,enters-3), det(amount-6,the-4), nn(amount-6,withdrawal-5), dobj(enters-3,amount-6)} \bigstrut\\
    \hline
     \textbf{SSR28} & \textit{mark(A,B)} & Conditional & If the ATM card is valid. & \textit{[mark(valid-6, If-1), det(card-4, the-2), nn(card-4, ATM-3), nsubj(valid-6, card-4), cop(valid-6, is-5), root(ROOT-0, valid-6)]} \bigstrut\\
    \hline
	  \textbf{SSR29} & \textit{nsubj(A,B)} & SV (Subject-Verb)     & The system restarts & \textit{[det(system-2, The-1), nsubj(restarts-3, system-2), root(ROOT-0, restarts-3)]} \bigstrut\\
    \hline
    \textbf{Rule \#} & \textbf{Antecedent (If the sentence starts with the keyword:)} & \textbf{Consequent (then the identified sentence structure is:)} & \textbf{Example sentence} & \textbf{} \bigstrut\\
    \hline
		\textbf{SSR30} & \textit{Include} & Include & Include UCS Validate PIN & \textit{N/A} \bigstrut\\
    \hline
    \textbf{SSR31} & \textit{Extend} & Extend & Extend Handle Card Jam & \textit{N/A} \bigstrut\\
    \hline
    \textbf{SSR32} & \textit{Resume} & Resume & Resume M2. & \textit{N/A} \bigstrut\\
    \hline
    \textbf{SSR33} & \textit{Repeat} & Repeat & Repeat M4-M8. & \textit{N/A} \bigstrut\\
    \hline
    \textbf{SSR34} & \textit{Except all above} & UnIdentified & N/A    & \textit{N/A} \bigstrut\\
 \hline \bigstrut\\
		\multicolumn{5}{p{0.8\linewidth}}{Note: N/A= Not Applied, M4-M8 represents the step Nos. M4 to M8 of UCS.}
   \label{tab:CompleteSSRules}%
\end{longtable}
\scriptsize

\vspace{2ex}

\section{Transformation Rules}
\label{app:TransformationRules}
\noindent \textbf{1) Rule TR1 is used to identify entity terms}\\
 \\
\textbf{\textsl{Rule-TR1}}: For each sentence in the UCS\\
\setlength\parindent{55pt}
\indent Concatenate the consecutive words in the sentences whose POS-tags starts with ``NN". \\
\indent Parse the modified sentences again and Update the TDs and POS-tags of the sentences to reflect the changes.\\
\setlength\parindent{35pt}
\indent
EndFor\\
\\
\emph{Example:} For sentence ``The system ejects the ATM card.", the POS-tags generated by the parser are shown in Table \ref{tab:POSandTD}, here the consecutive words whose POS-tag starts with ``NN" are ``ATM" and ``card", hence they are combined or concatenated using the transformation rule TR1 to get ``ATMcard" representing a single entity or a single attribute. \\
\\
The modified sentence is : ``The system ejects the ATMcard."\\
\\
The updated TDs are: \textsl{[det(system-2, The-1), nsubj(ejects-3, system-2), root(ROOT-0, ejects-3), det(ATMcard-5, the-4), dobj(ejects-3, ATMcard-5)]}\\
\\
The updated POS-tags are: \textsl{[The/DT, system/NN, ejects/VBZ, the/DT, ATMcard/NN, ./.]}\\
 \\
\\
\noindent \textbf{2) Rules TR2-TR3 are used to identify control and boundary classes}\\
 \\
\noindent\textbf{\textsl{Rule-TR2}}: \textsl{createClass}(UCS.name,``${<<}$control class${>>}$");\\
\\
\emph{Example:} From UCS shown in Table~\ref{tab:UseCaseWithdrawFund} the approach identifies the control class named \textsl{``WithdrawFunds \textless \textless control class \textgreater \textgreater"}, where the stereotype \textsl{``\textless \textless control class \textgreater \textgreater"} denotes that the class is control class. 
\\ \\
\textbf{\textsl{Rule-TR3}}: For each actor of the UCS, \textsl{createClass}(actor.name,``${<<}$boundary class${>>}$"); \\
\\
\emph{Example:} From UCS shown in Table~\ref{tab:UseCaseWithdrawFund} the approach identifies the boundary class named \textsl{``ATMcustomer \textless \textless boundary \textgreater \textgreater \textless \textless primary \textgreater \textgreater"}, where the stereotype \textsl{``\textless \textless boundary \textgreater \textgreater \textless \textless primary \textgreater \textgreater"} denotes that the class is boundary class of primary actor.\\
\\
\\
\textbf{3) Rules TR4-TR9 are used to identify entity classes and their attributes}\\
 \\
\noindent\textbf{\textsl{Rule-TR4}}: POS-tags of each sentence in the UCS are scanned, and all the nouns are stored in a list named ListOfNouns.\\
\setlength\parindent{45pt}
\indent For every two nouns n1 and n2 in ListOfNouns\\
\setlength\parindent{55pt}
\indent If n2 startsWith n1 then\\
\setlength\parindent{65pt}
\indent class = \textsl{createClass}(A,``${<<}$entity class${>>}$");  class.addAttribute(B); \\
\setlength\parindent{55pt}
\indent EndIf\\
\setlength\parindent{15pt}
\setlength\parindent{45pt}
\indent EndFor \\
\\
\emph{Example:}
Let ListOfNouns contains two nouns say n1=``Withdrawal" and  n2=``withdrawalAmount" then by the rule TR4 ``Withdrawal" is identified as a class and ``withdrawalAmount" is added as an attribute of class ``Withdrawal".
\\
\\
 \\
\textbf{\textsl{Rule-TR5}}: If TDs of the sentence contain TDs nsubj(has,A) and dobj(has, B) then\\ \setlength\parindent{55pt}
\indent class = \textsl{createClass}(A,``${<<}$entity class${>>}$");  class.addAttribute(B);\\
\setlength\parindent{45pt}
\indent EndIf\\
\\
\emph{Example:}
{\label{ex:7}}
The TDs generated by parser for sentence M8 of UCS-Withdraw\-Funds are:\\
\textsl{[det(system-2, The-1), nsubj(validates-3, system-2), root(ROOT-0, validates-3), complm(has-7, that-4), det(ATM-6, the-5), nsubj(has-7, ATM-6), ccomp(validates-3, has-7), amod(funds-9, enough-8), dobj(has-7, funds-9)]}\\
\\
Applying the rule TR5 on TDs \textsl{nsubj(has-7, ATM-6)} and \textsl{dobj(has-7, funds-9)}, the approach identifies ``ATM" as class and ``funds" as attribute of class ``ATM".\\
 \\
\textbf{\textsl{Rule-TR6}}: If TDs of the sentence contain TDs prep(A,``in") and pobj(``in",B) \\ \setlength\parindent{45pt}
\indent If(isNoun(B)) then \\ \setlength\parindent{55pt}
\indent class = \textsl{createClass}(B,``${<<}$entity class${>>}$");  class.addAttribute(A);\\
\indent For all TDs of type conj(A, C) before prep(A,``in")  \\ \setlength\parindent{65pt}
\indent    class.addAttribute(C);\\
\setlength\parindent{45pt}
\indent EndFor \\
\indent else (isAdjective(B)) \\ \setlength\parindent{55pt}
\indent class = \textsl{createClass}(A,``${<<}$entity class${>>}$");  class.addAttribute(B);\\
\setlength\parindent{45pt}
\indent EndIf \\
\\
\emph{Example:}
For sentence \textsl{``The system prints the transactionNumber and balance on the receipt."} The TDs generated by the parser are: \\
\\
\textsl{[det(system-2, The-1), nsubj(prints-3, system-2), root(ROOT-0, prints-3), det(transactionNumber-5, the-4), dobj(prints-3, transactionNumber-5), cc(transactionNumber-5, and-6), conj(transactionNumber-5, balance-7), prep(transactionNu-mber-5, on-8), det(receipt-10, the-9), pobj(on-8, receipt-10)]}
\\
\\
From the TDs \textsl{prep(transactionNumber-5, on-8)} and \textsl{pobj(on-8, receipt-10)}, \textsl{receipt} is identified as class and \textsl{transactionNumber} is identified as attribute of class \textsl{receipt.} Also from TDs \textsl{conj(transactionNumber-5,balance-7)} and \textsl{prep(transactionNumber-5, in-8)}, \textsl{balance} is also identified as attribute of class \textsl{receipt}.
\\

\noindent \textbf{\textsl{Rule-TR7}}: If TDs of the sentence contain TDs prep(A,``of") and pobj(``of",B) then \\ \setlength\parindent{55pt}
\indent class = \textsl{createClass}(B,``${<<}$entity class${>>}$");  class.addAttribute(A); \\
\indent For all the consecutive TDs of type conj(A, C) before prep(A,``in") \\ \setlength\parindent{65pt}
\indent class.addAttribute(C);\\
\setlength\parindent{55pt}
\indent EndFor \\
\setlength\parindent{45pt}
\indent EndIf\\
\\
\emph{Example:} For sentence \textsl{``The system prompts for the userName and password of the customer."} the TDs generated by the parser are:\\
\\
\textsl{[det(system-2, The-1), nsubj(prompts-3, system-2), root(ROOT-0, prompts-3), prep(prompts-3, for-4), det(userName-6, the-5), pobj(for-4, userName-6), cc(userName-6, and-7), conj(userName-6, password-8), prep(userName-6, of-9), det(customer-11, the-10), pobj(of-9, customer-11)]
}\\
\\
From the TDs \textsl{prep(userName-6, of-9)} and \textsl{pobj(of-9, customer-11)},
\textsl{customer} is identified as class and \textsl{userName} is identified as attribute of class \textsl{customer}. Also from TDs \textsl{conj(userName-6, password-8)} and \textsl{prep(userName-6, of-9)}, \textsl{password} is also identified as attribute of class \textsl{customer}\\
 \\

\noindent\textbf{\textsl{Rule-TR8}}: If TDs of the sentence contain TDs poss(A, B) then \\
\setlength\parindent{55pt}
\indent class = \textsl{createClass}(B,``${<<}$entity class${>>}$");  class.addAttribute(A);\\
\setlength\parindent{45pt}
\indent EndIf\\
\\
\emph{Example:} For sentence \textsl{``The system prompts for customer's address."} the TDs generated by the parser are: \\
\\
\textsl{[det(system-2, The-1), nsubj(prompts-3, system-2), root(ROOT-0, prompts-3), prep(prompts-3, for-4), poss(address-7, customer-5), possessive(customer-5, 's-6), pobj(for-4, address-7)]}\\
\\
From TD \textsl{poss(address-7, customer-5)}, \textsl{customer} is identified as class and \textsl{address} is identified as the attribute of class \textsl{customer}\\
\\
\textbf{\textsl{Rule-TR9}}: If TDs of the sentence contain TD amod(A, B) then \\
\setlength\parindent{55pt}
\indent class = \textsl{createClass}(A,``${<<}$entity class${>>}$");  class.addAttribute(B); \\
\setlength\parindent{45pt}
\indent EndIf\\
\\
\emph{Example:} For sentence \textsl{``The system informs the interested user."} the TDs generated by the parser are:\\
\\
 \textsl{[det(system-2, The-1), nsubj(informs-3, system-2), root(ROOT-0, informs-3), det(user-6, the-4), amod(user-6, interested-5), dobj(informs-3, user-6)]}\\
\\
From TD \textsl{amod(user-6, interested-5)} \textsl{user} is identified as an entity class and \textsl{interested} is identified as attribute of class \textsl{user}.\\
\\
\\
\\
\textbf{4) Rules TR10-TR42 are used to identify class operations and more class attributes, shown in Table~\ref{tab:AppendixB}}\\
		{
		\scriptsize
\begin{longtable}{|c|p{2cm}|p{5.5cm}|p{6cm}|}
  \caption{\textbf{Transformation Rules to identify class operations and more class attributes}} \label{grid_llll} \\
	\hline
& \textbf{Antecedent ( If A)} & \multicolumn{2}{l|}{\textbf{Consequent ( then use B to identify C )}} \bigstrut\\
		\cline{2-4} \textbf{Rule \#}  & \textbf{A (Sentence Structure of sentence is:)}  & \textbf{B (TDs of the sentence:)} & \textbf{C (operations/attributes:)} \bigstrut\\
		\endfirsthead
\multicolumn{4}{m{0.9\textwidth}}{\textbf{Transformation Rules to identify class operations and more class attributes (Appendix B) --Continued from previous page}} \\ \hline
 & \textbf{Antecedent ( If A)} & \multicolumn{2}{l|}{\textbf{Consequent ( then use B to identify C )}} \bigstrut\\
		\cline{2-4} \textbf{Rule \#}  & \textbf{A (Sentence structure of sentence:)}  & \textbf{B (TDs of the sentence:)} & \textbf{C (operations/attributes:)} \bigstrut\\
\endhead
\hline

    \textbf{TR11} & SVIODO & \textit{\textbf{nsubj}(A,B), \textbf{iobj}(A,C), \textbf{dobj}(A,D)} & op.SourceEntityTerm=B, op.DestEntityTerm=D, op.name=A \bigstrut\\
    \hline
		\textbf{TR12} & SVDOThatClause & \textit{\textbf{nsubj}(A,B), \textbf{dobj}(A,C), \textbf{complm}(D,E), \textbf{nsubj}(D,F)} & op.SourceEntityTerm=B, op.DestEntityTerm=C, op.name=A \bigstrut\\
    \hline
		\textbf{TR13} & SVThatClause & \textit{\textbf{nsubj}(A,B), \textbf{complm}(C,D), \textbf{nsubj}(C,E)} & op.SourceEntityTerm=B, op.DestEntityTerm=E, op.name=A \bigstrut\\
    \hline
		\textbf{TR14} & SVDONotToInf & \textit{\textbf{nsubj}(A,B), \textbf{dobj}(A,C), \textbf{neg}(D,E), \textbf{aux}(D,F), \textbf{infmod}(C,D)} & op.SourceEntityTerm=B, op.DestEntityTerm=C, op.name=A \bigstrut\\
    \hline
		 \textbf{TR15} & SVNotToInf & \textit{\textbf{nsubj}(A,B), \textbf{neg}(C,D), \textbf{aux}(C,E), \textbf{xcomp}(A,C), \textbf{dobj}(C,F)} & op.SourceEntityTerm=B, op.DestEntityTerm=F, op.name=A \bigstrut\\
    \hline
		 \textbf{TR16} & SVDOtobeComp & \textit{\textbf{nsubj}(A,B), \textbf{nsubj}(C,D), \textbf{aux}(C,E), \textbf{cop}(C,F), \textbf{xcomp}(A,C)} & op.SourceEntityTerm=B, op.DestEntityTerm=D, op.name=A \bigstrut\\
    \hline
		 \multirow{4}[8]{*}{\textbf{TR17}} & \multirow{4}[8]{*}{SVDOToInf} & \multirow{4}[8]{4cm}{\textit{\textbf{nsubj}(A,B), \textbf{dobj}(A,C), \textbf{aux}(D,E), \textbf{infmod}(C,D)}} & op.SourceEntityTerm=B, op.DestEntityTerm=D, op.name=A \bigstrut\\
           &        &        & If (TDs of the sentence contains TD dobj(D,F)) then  \bigstrut\\
           &        &        & op.SourceEntityTerm2=C, op.DestEntityTerm2=F, op.name2=D  \bigstrut\\
          &        &        & EndIf \bigstrut\\
    \hline
		\multirow{7}[14]{*}{\textbf{TR18}} & \multirow{7}[14]{*}{SVDOPresentPart} & \multirow{7}[14]{*}{\textit {\textbf{nsubj}(A,B), \textbf{dobj}(A,C), \textbf{partmod}(C,D) \textbf{dobj}(D,E)}} & op.SourceEntityTerm=B, op.DestEntityTerm=D, op.name=A
 \bigstrut\\
          &        &        & If (TDs of the sentence contains TD dobj(D,E)) then \bigstrut\\
           &        &        & op.DestEntityTerm.addAttribute(E) \bigstrut\\
           &        &        & For each TD=conj(X,Y) and (X==E) after dobj(D,E)  \bigstrut\\
           &        &        & destClass.addAttribute(Y) \bigstrut\\
           &        &        & EndFor \bigstrut\\
          &        &        & EndIf  \bigstrut\\
    \hline
		\textbf{TR19} & SVDOPastPart & \textit{\textbf{nsubj}(A,B), \textbf{dobj}(A,C), \textbf{partmod}(C,D)} & op.SourceEntityTerm=B, op.DestEntityTerm=C, op.name=A \bigstrut\\
    \hline
		\textbf{TR20} & SVDOAdj & \textit{\textbf{nsubj}(A,B), \textbf{nsubj}(C,D), \textbf{xcomp}(A,C)} & op.SourceEntityTerm=B, op.DestEntityTerm=D, op.name=A \bigstrut\\
    \hline
		 \textbf{TR21} & SVDONoun & \textit{nsubj(A,B), nsubj(C,D), xcomp(A,C)} & op.SourceEntityTerm=B, op.DestEntityTerm=D, op.name=A \bigstrut\\
    \hline
		\textbf{TR22} & SVDOConjToInf & \textit{nsubj(A,B), dobj(A,C), advmod(D,E), aux(D,F), xcomp(A,D)} & op.SourceEntityTerm=B, op.DestEntityTerm=C, op.name=A \bigstrut\\
    \hline
		\textbf{TR23} & SVDOConjClause & \textit{nsubj(A,B), dobj(A,C), advmod(D,E),nsubj(D,F)} & op.SourceEntityTerm=B, op.DestEntityTerm=C, op.name=A \bigstrut\\
    \hline
		 \textbf{TR24} & SVDOAdverbial & textit{nsubj(A,B), dobj(A,C), advmod(A,D)} & op.SourceEntityTerm=B, op.DestEntityTerm=C, op.name=A \bigstrut\\
    \hline
		\textbf{TR25} & SAuxVDO & \textit{nsubj(A,B), aux(A,C), dobj(A,D) } & op.SourceEntityTerm=B, op.DestEntityTerm=D, op.name=A \bigstrut\\
    \hline
		\multirow{3}[6]{*}{\textbf{TR26}} & \multirow{3}[6]{*}{SVDOPO} & \multirow{3}[6]{*}{\textit{nsubj(A,B), dobj(A,C),  prep(A,D), pobj(D,E)}} & op.SourceEntityTerm=B, op.name=A,   \bigstrut\\
\cline{4-4}           &        &        & If(E=='to' or 'from' or 'on' or 'in' or 'into' or 'through' or 'of') then op.DestEntityTerm=E \bigstrut\\
\cline{4-4}           &        &        & else op.DestEntityTerm=C \bigstrut\\
    \hline
		 \textbf{TR27} & SVConjToInf & \textit{nsubj(A,B), advmod(C,D), aux(C,E), xcomp(A,C)} & op.SourceEntityTerm=B, op.DestEntityTerm=B, op.name=A \bigstrut\\
    \hline
		  \textbf{TR28} & SVConjClause & \textit{nsubj(A,B), advmod(C,D), nsubj(C,E), advcl(A,C)} & op.SourceEntityTerm=B, op.DestEntityTerm=B, op.name=A \bigstrut\\
    \hline
		 \multirow{3}[6]{*}{\textbf{TR29}} & \multirow{3}[6]{*}{SVToInf} & \multirow{3}[6]{4cm}{\textit{nsubj(A,B), aux(C,D), xcomp(A,C)}} & op.SourceEntityTerm=B, op.name=A,   \bigstrut\\
\cline{4-4}           &        &        & If found(dobj(C,E)) then op.DestEntityTerm=E \bigstrut\\
\cline{4-4}           &        &        & Else op.DestEntityTerm=B \bigstrut\\
    \hline
		\multirow{3}[6]{*}{\textbf{TR30}} & \multirow{3}[6]{*}{SVGerund} & \multirow{3}[6]{*}{\textsl{nsubj(A,B), xcomp(A,C)}} & op.SourceEntityTerm=B, op.name=A,  \bigstrut\\
\cline{4-4}           &        &        & If found(dobj(C,D)) then op.DestEntityTerm=D  \bigstrut\\
\cline{4-4}           &        &        & else op.DestEntityTerm=B \bigstrut\\
    \hline
		 \textbf{TR31} & SVAdverbialAdjunct & \textsl{nsubj(A,B), advmod(A,C)} & op.SourceEntityTerm=B, op.DestEntityTerm=B, op.name=A \bigstrut\\
    \hline
		 \multirow{2}[4]{*}{\textbf{TR32}} & \multirow{2}[4]{*}{SVPredicative} & \multirow{2}[4]{*}{\textsl{nsubj(A,B), cop(A,C)}} & If(isAdjective(C)) then B.addAttribute(C)  \bigstrut\\
\cline{4-4}           &        &        & else if(isNoun(C)) then parentClass=C, childClass=B createRelationship(parentClass, childClass, ``generalization") \bigstrut\\
    \hline
		\multirow{3}[6]{*}{\textbf{TR33}} & \multirow{3}[6]{*}{SVForComp} & \multirow{3}[6]{*}{\textit{nsubj(A,B), prep(A,C), num(D,E), pobj(C,D)}} & op.SourceEntityTerm=B, op.name=A,  \bigstrut\\
\cline{4-4}           &        &        & If(A==E) then op.DestEntityTerm=B  \bigstrut\\
\cline{4-4}           &        &        & else op.DestEntityTerm=E \bigstrut\\
    \hline
		\textbf{TR34} & SVPassPO & \textit{nsubjpass(A,B), auxpass(A,C), prep(A,D), pobj(D,E)} & op.SourceEntityTerm=E, op.DestEntityTerm=B, op.name=A \bigstrut\\
    \hline
		\textbf{TR35} & SAuxVPassPO & \textit{nsubjpass(A,B), aux(A,C), auxpass(A,D), prep(A,E), pobj(E,F)} & op.SourceEntityTerm=F, op.DestEntityTerm=B, op.name=A \bigstrut\\
    \hline
		 \textbf{TR36} & SVPO   & \textsl{nsubj(A,B), pobj(A,C)} & op.SourceEntityTerm=B, op.DestEntityTerm=C, op.name=A \bigstrut\\
    \hline
		  \textbf{TR37} & SVDO   & \textsl{nsubj(A,B), dobj(A,C)} & op.SourceEntityTerm=B, op.DestEntityTerm=C, op.name=A \bigstrut\\
    \hline
		\textbf{TR38} & Conditional & \textsl{mark(A,B)} &  \bigstrut\\
    \hline
		\textbf{TR39} & SV     & \textsl{nsubj(A,B)} & op.SourceEntityTerm=B, op.DestEntityTerm=B, op.name=A, op.Para="" \bigstrut\\
    \hline
    \multirow{6}[12]{*}{\textbf{TR40}} & \multirow{6}[12]{1.9cm}{Include or Extend} & \multirow{6}[12]{*}{\textsl{nn(A,B), nn(C,D)}} & op.SourceEntityTerm="system" \bigstrut\\
\cline{4-4}           &        &        & op.name=``\textless\textless B  \textgreater\textgreater" \bigstrut\\
\cline{4-4}           &        &        & op.DestEntityTerm="" \bigstrut\\
\cline{4-4}           &        &        & for(each TD td(X,Y) of the sentence after first TD nn(A,B)) \bigstrut\\
\cline{4-4}           &        &        & op.DestEntityTerm = op.DestEntityTerm + Y \bigstrut\\
\cline{4-4}           &        &        & endFor \bigstrut\\
    \hline
    \textbf{TR41} & Resume & \textsl{nn(A,B)} & op.SourceEntityTerm="system", op.DestEntityTerm="system", op.name=B, op.Para=A \bigstrut\\
    \hline
    \textbf{TR42} & Loop   & \textsl{nn(A,B)} & op.SourceEntityTerm="system", op.DestEntityTerm="system", op.name=B, op.Para=A \bigstrut\\
    \hline
    \multirow{4}[8]{*}{\textbf{TR43}} & \multirow{4}[8]{2cm}{All except SV, SVDO, Conditional, Include, Extend, Resume and Loop} & \multirow{4}[8]{*}{\textsl{root()}} & op.Para="" \bigstrut\\
\cline{4-4}           &        &        & for(each TD td(X,Y) of the sentence after TD root(A,B)) \bigstrut\\
\cline{4-4}           &        &        &     op.Para = op.Para + Y \bigstrut\\
\cline{4-4}           &        &        & endFor \bigstrut\\
    \hline
		\label{tab:AppendixB}%
\end{longtable}
}
\scriptsize
\noindent \textbf{5) Rules TR44-TR45 are used to identify more entity classes}\\
 \\
\textbf{\textsl{Rule-TR44}}: If op.SourceEntityTerm is not present in \textsl{ClassDiagram\_Instance} then class = \textsl{createClass}(op.SourceEntityTerm,``${<<}$entity class${>>}$"); \\
\\
\textbf{\textsl{Rule-TR45}}: For each class C in ClassDiagram\_Instance \\		
\setlength\parindent{65pt}
\indent If (op.DestEntityTerm.name==C.name)AND(C does not contain operation op.name(op.Para)) then\\
\setlength\parindent{80pt}
 \indent	 C.addOperation(op.name(op.Para));\\		
\setlength\parindent{65pt}
\indent EndIf\\
\setlength\parindent{55pt}
\indent EndFor\\
\indent If no such class is found then \\
\setlength\parindent{65pt}
\indent For each class C in ClassDiagram\_Instance \\
\setlength\parindent{80pt}
\indent If(op.DestEntityTerm.name==a.name for some attribute a of class C)AND
(C does not contains operation op.name(op.Para)) then\\
\setlength\parindent{95pt}
\indent	  C.addOperation(op.name(op.Para));\\	
\setlength\parindent{80pt}	
\indent		EndIf\\		
\setlength\parindent{65pt}	
\indent EndFor\\
\setlength\parindent{55pt}
\indent EndIf \\
\indent If no such class is found then \\
\setlength\parindent{65pt}	
\indent	C=createClass(op.DestEntityTerm.name,``${<<}$entity class${>>}$"); C.addOperation(op.name(op.Para));\\
\setlength\parindent{55pt}
\indent EndIf\\
 \\
\textbf{6) Rule TR46 is used to identify association relationships}\\
 \\
\textbf{\textsl{Rule-TR46}}:
For each operation op\\
\setlength\parindent{55pt}
\indent For each relationship r in ClassDiagram\_Instance \\
\setlength\parindent{65pt}
\indent If(op.SourceEntityTerm==r.class1 and op.DestEntityTerm==r.class2)AND(r.name does not contains op.name) \\
\setlength\parindent{80pt}
\indent     append op.name to r.name\\
\setlength\parindent{65pt}
\indent		EndIf \\
\setlength\parindent{55pt}
\indent EndFor \\
\indent If no such relationship found then \\
\setlength\parindent{65pt}
\indent For each relationship r in ClassDiagram\_Instance \\
\setlength\parindent{75pt}
\indent If(op.SourceEntityTerm==r.class1 and op.DestEntityTerm==a.name for some attribute a of class r.class2)AND(r.name does not contains op.name) \\
\setlength\parindent{85pt}
\indent     append op.name to r.name\\
\setlength\parindent{75pt}
\indent		EndIf \\
\setlength\parindent{65pt}
\indent EndFor \\
\setlength\parindent{55pt}
\indent EndIf\\
\indent If no such relationship found then \\
\setlength\parindent{65pt}
\indent   rName=op.name; createRelationship(op.SourceEntityTerm, op.DestEntityTerm, rName, ``association"); \\
\setlength\parindent{55pt}
\indent EndIf\\
\setlength\parindent{45pt}
\\
\vspace{-4ex}
\\
\textbf{7) Rules TR47-TR50 are used to identify generalization relationships}\\
 \\
\textbf{\textsl{Rule-TR47}}: For each sentence of type \textit{Child-\textit{GenSubString}-Parent}, the POS-tags of the sentence are scanned and \\  \setlength\parindent{60pt}
\indent    parentClass=createClass(noun nr on the right of \textit{GenSubString},``${<<}$entity class${>>}$");\\
\indent		 For each noun nl on the left of \textit{GenSubString} \\ \setlength\parindent{70pt}
\indent    childClass=createClass(nl,``${<<}$entity class${>>}$");\\
\indent		 createRelationship(parentClass,childClass,``generalization"); \\
  \setlength\parindent{60pt}
\indent    EndFor\\
\setlength\parindent{40pt}
\indent EndFor			
\\
\vspace{-4ex}
\\
\emph{Example:} For sentence \textsl{``The withdrawal, deposit, transfer and query are types of transaction."} POS-tags generated by the parser are: \\
\textsl{[The/DT, withdrawal/NN, ,/,, deposit/NN, ,/,, transfer/NN, and/CC, query/NN, are/VBP, types/NNS, of/IN, transaction/NNS, ./.]}\\
\\
As this sentence contains \textit{GenSubString}=``types of" hence it is \textit{Child-\textit{GenSubString}-Parent} sentence, therefore rule TR47 is applied. From POS-tags the nouns to the left of \textit{GenSubString} \textsl{``types of"} are \textsl{withdrawal}, \textsl{deposit}, \textsl{transfer} and \textsl{query}, a child class is created for each of these nouns. And the noun to the right of \textit{GenSubString} \textsl{``types of"} is \textsl{transaction}, a parent class is created for this noun. Generalization relationship is established between the identified child classes (\textsl{withdrawal, deposit, transfer and query}) and the identified parent class (\textsl{transaction}) \\
 \\		
\textbf{\textsl{Rule-TR48}}: For each sentence of type \textit{Parent-\textit{GenSubString}-Child}, the POS-tags of the sentence are scanned and \\  \setlength\parindent{60pt}
\indent    parentClass=createClass(noun nl on the left of \textit{GenSubString},``${<<}$entity class${>>}$");\\
\indent		 For each noun nr on the right of \textit{GenSubString} \\ \setlength\parindent{70pt}
\indent    childClass=createClass(nr,``${<<}$entity class${>>}$");\\
\indent		 createRelationship(parentClass,childClass,``generalization");\\
  \setlength\parindent{60pt}
\indent    EndFor\\
\setlength\parindent{40pt}
\indent EndFor	\\
\\
\emph{Example:} For sentence ``Memory has types RAM and ROM.", the POS-tags generated by the parser are: \\
\\
 \textsl{[Memory/NN, has/VBZ, types/NNP, RAM/NNP, and/CC, ROM/NNP, ./.]}\\
\\
As this sentence contains \textit{GenSubString}=``has types" hence it is \textit{Parent-\textit{GenSubString}-Child} sentence, therefore rule TR48 is applied. A parent class is created for noun \textsl{Memory} on the left of \textit{GenSubString} \textsl{has types} and for each noun \textsl{RAM} and \textsl{ROM} a child class is created. A generalization relationship is established between parent class (\textsl{Memory}) and the child classes (\textsl{RAM and ROM})\\
 \\
\textbf{\textsl{Rule-TR49}}: The use case generalization relationships are used to identify the generalization relationships between their control classes.\\
\\
\emph{Example:} For UCS Withdraw Fund the generalization relationship between control classes of Withdraw Fund and ATM is directly obtained from the Parent-UseCase-Name field of UCS.\\
\\
\textbf{\textsl{Rule-TR50}}: The actors generalization relationships are used to identify the generalization relationships between the actors boundary classes.\\
\\
\emph{Example:} For UCS Withdraw Fund the generalization relationship between boundary classes of ATM Customer and User is directly obtained from the Parent-Actor-Name field of UCS.\\
\\
\\
\textbf{8) Rules TR51-TR53 are used to identify aggregation relationships}\\
 \\
\textbf{\textsl{Rule-TR51}}: For each of the two classes c1 and c2 in \textsl{ClassDiagram\_Instance}\\		
\setlength\parindent{60pt}
\indent       If c2 is attribute of c1 then\\		
\setlength\parindent{80pt}	
\indent             aggregateClass=c1;partClass=c2;\\
\indent createRelationship(aggregateClass,partClass,``aggregation"); \\			
\setlength\parindent{60pt}
\indent       EndIf\\		
\setlength\parindent{40pt}	
\indent EndFor\\
\\
The sentences in flows and description sections of UCS are scanned and the sentences containing sub strings of types ``part of", ``consists of" and all their synonyms are used to identify aggregation relationships. We call such sub strings as \textit{AggSubString} (or Aggregation Sub String). We categories the sentences containing \textit{AggSubString} into two kinds:
\begin{enumerate}
	\item \textit{Part-\textit{AggSubString}-Whole} sentences: The sentences containing sub strings ``part of", ``unit of", ``member of" and all their synonyms are referred as \textit{Part-\textit{AggSubString}-Whole} sentences because in these sentences the part class/classes is/are present on the left of \textit{AggSubString} and the whole class is present on the right of \textit{AggSubString}. \\
	\item \textit{Whole-\textit{AggSubString}-Part} sentences: The sentences containing sub strings ``consist of", ``made of", ``contains" and their all synonyms are referred as \textit{Whole-\textit{AggSubString}-Part} sentences because in these sentences the whole class is present on the left of \textit{AggSubString} and the part class/classes is/are present on the right of \textit{AggSubString}.
\end{enumerate}
The following two transformation rules (TR52 and TR53) are applied to identify aggregation relationships:\\
 \\
\textbf{\textsl{Rule-TR52}}: For each sentence of type \textit{Part-\textit{AggSubString}-Whole}, the POS-tags of the sentence are scanned and \\  \setlength\parindent{50pt}
\indent    wholeClass=createClass(noun nr on the right of \textit{AggSubString},``${<<}$entity class${>>}$");\\
\indent		 For each noun nl on the left of \textit{AggSubString} \\ \setlength\parindent{70pt}
\indent    partClass=createClass(nl,``${<<}$entity class${>>}$");\\
\indent		 createRelationship(wholeClass,partClass,``aggregation");\\
  \setlength\parindent{60pt}
\indent    EndFor\\
\setlength\parindent{40pt}
\indent EndFor\\
\\
\emph{Example:} For sentence ``CardReader, CashDispenser and ReceiptPrinter are parts of ATM." the POS-tags generated by the parser are: \\
\textsl{[CardReader/NNP, ,/,, CashDispenser/NNP, and/CC, ReceiptPrinter/NNP, are/VBP, parts/NNS, of/IN, ATM/NNP, ./.]}\\
\\
As the sentence contains \textit{AggSubString}=``parts of" hence it is \textit{Part-\textit{AggSubString}-Whole} sentence therefore rule TR52 is applied. A class representing the part is created for each nouns \textsl{CardReader}, \textsl{CashDispenser} and \textsl{ReceiptPrinter} to the left of \textit{AggSubString} ``parts of"and a class representing the whole is created for the noun \textsl{ATM} on the right of \textit{AggSubString}. An aggregation relationship is established between the classes representing the parts (\textsl{CardReader}, \textsl{CashDispenser} and \textsl{ReceiptPrinter}) and the class representing the whole (\textsl{ATM}). \\
 \\		
\textbf{\textsl{Rule-TR53}}: For each sentence of type \textit{Whole-\textit{AggSubString}-Part}, the POS-tags of the sentence are scanned and \\  \setlength\parindent{50pt}
\indent    wholeClass=createClass(noun nl on the right of \textit{AggSubString},``${<<}$entity class${>>}$");\\
\indent		 For each noun nr on the left of \textit{AggSubString} \\ \setlength\parindent{70pt}
\indent    partClass=createClass(nr,``${<<}$entity class${>>}$"); \\
\indent		 createRelationship(wholeClass, partClass, ``aggregation"); \\
  \setlength\parindent{60pt}
\indent    EndFor\\
\setlength\parindent{40pt}
\indent EndFor	\\
\\ 			
\emph{Example:} For sentence ``A Computer is composed of Hardwares and Softwares." The POS-tags generated by the parser are: \\
\textsl{[A/DT, Computer/NN, is/VBZ, composed/VBN, of/IN, Hardwares/NNPS, and/CC, Softwares/NNPS, ./.]}\\
\\
As the sentence contains \textit{AggSubString}=``composed of" hence it is \textit{Whole-\textit{AggSubString}-Part} sentence therefore rule TR53 is applied. A class representing the whole is created for the noun \textsl{Computer} on the left of \textit{AggSubString} ``composed of" and a class representing the part  is created for each noun \textsl{Hardwares} and \textsl{Softwares} on the right of \textit{AggSubString} ``composed of". An aggregation relationship is created between the class representing the whole (\textsl{Computer}) and the classes representing the parts (\textsl{Hardwares} and \textsl{Softwares})\\
\\
\textbf{9) The rule TR54 is used to eliminate extra classes}\\
 \\
\textbf{\textsl{Rule-TR54}}: For each class c in ClassDiagram\_Instance \\
\setlength\parindent{65pt}
\indent  If c is not present as EndClass in any relationship then\\
\setlength\parindent{80pt}	
\indent     c is deleted from ClassDiagram\_Instance\\
\setlength\parindent{65pt}
\indent       EndIf\\	
\setlength\parindent{50pt}		
\indent EndFor

\section{Types of sentences in English}
\label{app:TypesOfSentnecesInEnglish}
\noindent There are 4-types of sentences in English\footnote{http://www.oxforddictionaries.com/words/sentences}
\begin{enumerate}
	\item \textbf{Simple Sentence} A simple sentence consists of only one independent clause, that has subject and predicate, and expresses a complete thought. \\
Example:\\
i)   “The ATM Customer enters the withdrawal amount.” \\
ii)  “The system commands the motor to start.” \\
iii) “The system validates the record entered by the customer.”\\
iv)  “The ATM card is ejected by the system.”\\
\item \textbf{Compound Sentence} A compound sentence consists of two independent clauses joined by a coordinator (conjunction).\\
Example:\\
i)   “The Card reader ejects the ATM card, and the printer prints the receipt.”\\
ii)  “The system debits the customer’s account, but the cash is not dispensed.”\\
\item \textbf{Complex Sentence} A complex sentence consists of an independent and one or more dependent clauses. \\
Example: \\
i)   “The system checks that the card is valid.”\\
ii)  “While the cash dispenser is counting the cash, the card reader ejects the card.”\\
iii) “The ATM card which the customer inserted was invalid.”\\
\item \textbf{Compound Complex Sentences} A compound complex sentence consists of a complex and a simple sentence joined by coordinator (conjunction).\\
Example:\\
i)  “The Customer inserts the card, and the system checks that the card is valid.”\\
ii) “While the cash dispenser is counting the cash, the card reader ejects the card, and the system starts displaying the message to collect the card and cash.”\\
\end{enumerate}

\section{Sample questionnaires and data collection}
\label{app:sampleQuestionaires}
\noindent Following are the sample questionnaires for the analysis class diagram generated by one of the approach for WithdrawFund UCS. Each subject in the experiment was given 3 sets of such questionnaires one for the analysis class diagram generated by each approach.

The questionnaires were so designed that the data collection from the answers to the questionnaires is straight forward, and the correctness, completeness and redundancy of the analysis class diagram can be easily calculated by simply putting the collected data in the formulas for correctness, completeness and redundancy of analysis class diagram presented in Section~\ref{subsec:Metrics}

\subsection{Questionnaires for class diagram correctness}
\noindent The sample questionnaires for class correctness and relationship correctness are shown in Table~\ref{tab:QClassCorrectness} and Table~\ref{tab:QRelationCorrectness} respectively.
\begin{table}[htb]
\vspace{-4ex}
  \centering
	\tiny
  \caption{Questionnaire for class correctness}
    \begin{tabular}{|p{.2cm}|p{4cm}|p{.8cm}|p{.8cm}|p{1.4cm}|p{3cm}|p{3cm}|}
		\hline
    S.No. & Class & Correctly identified as class ($C_{cr1}$) & Correctly named ($C_{cr2}$) & Correctly stereotyped (entity, boundary or control) ($C_{cr3}$) & Proportion of correctly identified attributes ($C_{cr4}$)=No. of correctly identified attributes in the class /
Total no. of identified attributes in the class & Proportion of correctly identified operations ($C_{cr5}$)=No. of correctly identified operations in the class / Total no. of identified operations in the class \\
		\hline

    1     & WithdrawFunds \textless\textless control class\textgreater\textgreater &     &      &     &      &  \\ \hline
    2     & Message &      &      &      &      &  \\ \hline
    3     & ATMcustomer \textless\textless boundary\textgreater\textgreater\textless\textless primary\textgreater\textgreater &      &      &      &      &  \\ \hline
    4     & Receipt &      &      &      &      &  \\ \hline
    5     & CashAmount &     &      &      &      &  \\ \hline
    6     & Withdrawal &      &      &      &      &  \\ \hline
    7     & Account &      &     &      &      &  \\ \hline
    8     & Transaction &     &      &      &      &  \\ \hline
    9     & ATM   &      &      &      &      &  \\ \hline
    10    & USECASEValidatePIN \textless\textless control class\textgreater\textgreater &      &      &      &      &  \\
							\hline
\multicolumn{7}{l}{Note: In column 1, 2 and 3, write 1 if the element is correctly identified, 0 otherwise} \\
    \end{tabular}%
  \label{tab:QClassCorrectness}%
\vspace{-4ex}
\end{table}%

\begin{table}[htb]
\vspace{-4ex}
  \centering
	\tiny
  \caption{Questionnaire for relationship correctness}
    \begin{tabular}{|p{.2cm}|p{2cm}|p{2cm}|p{1.5cm}|p{1cm}|p{1cm}|p{1cm}|p{.8cm}|p{1cm}|p{1cm}|}
		\hline
    S.No. & End Class1 & End Class2 & Relationship Name & Correctly assigned End-Class1 ($R_{cr1}$) & Correctly assigned End-Class2 ($R_{cr2}$) & Correctly identified as relationship ($R_{cr3}$) & Correctly named ($R_{cr4}$) & Correctly identified relationship type ($R_{cr5}$) & Correctly assigned navigability ($R_{cr6}$) \\
		\hline
    1    & WithdrawFunds \textless\textless control class\textgreater\textgreater & Message & displays &      &      &      &      &      &  \\ \hline
    2     & WithdrawFunds \textless\textless control class\textgreater\textgreater & ATMcustomer \textless\textless boundary\textgreater\textgreater\textless\textless primary\textgreater\textgreater & validates &      &      &      &      &      &  \\ \hline
    3     & ATMcustomer \textless\textless boundary\textgreater\textgreater\textless\textless primary\textgreater\textgreater & WithdrawFunds \textless\textless control class\textgreater\textgreater & enters\_selects &      &      &      &      &      &  \\ \hline
    4     & WithdrawFunds \textless\textless control class\textgreater\textgreater & Receipt & prints &      &      &      &      &      &  \\ \hline
    5     & WithdrawFunds \textless\textless control class\textgreater\textgreater & CashAmount & dispenses &      &      &      &      &      &  \\ \hline
    6     & WithdrawFunds \textless\textless control class\textgreater\textgreater & Withdrawal & enters\_selects\_validates &      &      &      &      &      &  \\ \hline
    7     & WithdrawFunds \textless\textless control class\textgreater\textgreater & Account & selects\_validates &      &      &      &      &      &  \\ \hline
    8     & WithdrawFunds \textless\textless control class\textgreater\textgreater & Transaction & cancels &      &      &      &      &      &  \\ \hline
    9     & WithdrawFunds \textless\textless control class\textgreater\textgreater & ATM   & ejects\_validates &      &      &      &      &      &  \\ \hline
  10      & WithdrawFunds \textless\textless control class\textgreater\textgreater & USECASEValidatePIN \textless\textless control class\textgreater\textgreater &  \textless\textless INCLUDE USE CASE\textgreater\textgreater &      &      &      &      &      &  \\
					\hline
\multicolumn{10}{l}{Note: Write 1 if the element is correctly identified, 0 otherwise}
    \end{tabular}%
  \label{tab:QRelationCorrectness}%
	\vspace{-4ex}
\end{table}%

\subsection{Questionnaires for class diagram completeness}
\noindent 1) No. of sentences in the functional requirements whose functionalities are assigned as operations to some class (or classes) in the class diagram ($N_{sf}$) = ?
\\
\\
2) No. of separate groups of classes and relationships in the class diagram ($N_{sg}$) = ?

\subsection{Questionnaires for class diagram redundancy}
\noindent 1) No. of redundant or extra classes in the class diagram ($N_{rc}$) = ? \\
(Note: A class is considered as a redundant class if it does not participate in any relationship with other classes in the class diagram or if it is an incorrectly identified class)
\\
\\
2) No. of redundant or extra relationships in the class diagram ($N_{rr}$) = ? \\
(Note: More than one relationships of same relationship type and navigability between the two same end classes are considered as redundant relationships. An incorrect relationship is also considered as redundant relationship)
\end{appendices}

\end{document}